\documentclass[12pt]{article}
\newsavebox{\ns}
\newsavebox{\dbrane}
\newsavebox{\dbshort}

\usepackage{epsfig}
\usepackage{amssymb}

\def\appendix{{\newpage\section*{Appendix}}\let\appendix\section%
        {\setcounter{section}{0}
        \gdef\thesection{\Alph{section}}}\section}

\def\be{\begin{equation}}
\def\ee{\end{equation}}
\def\ba{\begin{eqnarray}}
\def\ea{\end{eqnarray}}

\newcommand{\nn}{\nonumber}

\newcommand{\ft}[2]{{\textstyle\frac{#1}{#2}}}
\newcommand{\eqn}[1]{(\ref{#1})}

\def\Dslash{\,\,{\raise.15ex\hbox{/}\mkern-12mu D}}
\def\Dbarslash{\,\,{\raise.15ex\hbox{/}\mkern-12mu {\bar D}}}
\def\delslash{\,\,{\raise.15ex\hbox{/}\mkern-9mu \partial}}
\def\delbarslash{\,\,{\raise.15ex\hbox{/}\mkern-9mu {\bar\partial}}}
\def\pslash{\,\,{\raise.15ex\hbox{/}\mkern-9mu p}}
\def\calDslash{\,\,{\raise.15ex\hbox{/}\mkern-12mu {\cal D}}}
\newcommand\Bprime{B${}^\prime$}
\newcommand{\sign}{{\rm sign}}
\newcommand\re{{\rm Re}}
\newcommand\im{{\rm Im}}

\begin{document}

\begin{titlepage}

\begin{center}
February, 2002
%\today
\hfill hep-th/0202126\\
\hfill HUTP-01/A067\\
\hfill ITEP-TH-06/02\\
\hfill MIT-CTP-3239\\

\vskip 1.5 cm
{\large \bf D-Brane Probes of Special Holonomy Manifolds, and}\\
\vskip 0.16cm
%{\large \bf And}\\
%\vskip 0.16cm
{\large \bf Dynamics of ${\cal N}=1$ Three-Dimensional Gauge Theories}

\vskip 1 cm 
{Sergei Gukov$^1$ and David Tong$^2$}\\
\vskip 1cm

$^1${\sl Jefferson Physical Laboratory, Harvard University, \\ 
Cambridge, MA 02138, U.S.A. \\ {\tt gukov@democritus.harvard.edu}\\}
\vskip 0.5cm
$^2${\sl Center for Theoretical Physics, 
Massachusetts Institute of Technology, \\ Cambridge, MA 02139, U.S.A.
\\ {\tt dtong@mit.edu}\\}

\end{center}

\vskip 0.5 cm
\begin{abstract}
Using D2-brane probes, we study various properties of M-theory 
on singular, non-compact manifolds of $G_2$ and $Spin(7)$ holonomy. 
We derive mirror pairs of ${\cal N}=1$ supersymmetric three-dimensional 
gauge theories, and apply this technique to realize exceptional 
holonomy manifolds as both Coulomb and Higgs branches of the D2-brane 
world-volume theory. We derive a ``$G_2$ quotient construction'' 
of non-compact manifolds which admit a metric of $G_2$ holonomy.  
We further discuss the moduli space of such manifolds, 
including the structure of geometrical transitions in each case.
For completeness, we also include familiar examples
of manifolds with $SU(3)$ and $Sp(2)$ holonomy, where some of 
the new ideas are clarified and tested.

\end{abstract}

\end{titlepage}

\pagestyle{plain}
\setcounter{page}{1}
\newcounter{bean}
\baselineskip16pt
\tableofcontents

%\newpage

\section{Introduction and Summary} 
\label{intro}

D-brane probes of non-compact Calabi-Yau manifolds have 
proven to be powerful tools in understanding the dynamics 
of both string theory and supersymmetric gauge theories \cite{DRM}.
The purpose of this paper is to extend some of these results
to cases with less supersymmetry, where the background manifold 
has exceptional holonomy. 

When M-theory is compactified on a smooth manifold $X$ of 
exceptional holonomy, with all typical scales much larger than the 
Planck scale, 
the supergravity approximation is valid and one can simply derive the 
low-energy effective theory by the familiar rules of the Kaluza-Klein 
reduction \cite{paptown}. This leads to a rather simple effective 
theory containing, in particular, only abelian gauge fields. To 
get more interesting physical phenomena, such as non-abelian 
gauge symmetry and phase transitions, we must take a limit where 
the manifold $X$ develops a singularity.
Since the physics associated with singularity is local,
{\it i.e.} does not depend very much on the details of
the smooth part of the manifold, one can study such 
phenomena by isolating the singular region of $X$ and
studying M-theory or string theory on a non-compact model of $X$.
For this reason, it is interesting to understand M-theory dynamics
on non-compact manifolds of special holonomy.

\begin{figure}
\setlength{\unitlength}{0.9em}
\begin{center}
\begin{picture}(25,15)
\put(11,15){M-theory on X}
\put(0,9){IIA with D6-branes}\put(3,7.5){on $L$}
\put(22,9){IIA on X}
\put(0,2){IIB with D5-branes}\put(3,0.5){on $L$}
\put(20,2){IIB with NS5-branes}\put(23.5,0.5){on $L$}

\put(12,14.5){\line(-1,-1){4}}
\put(17,14.5){\line(1,-1){4}}
\put(4,7){\line(0,-1){3.8}}\put(4.5,4.5){T}
\put(24,8.5){\line(0,-1){5.3}}\put(24.5,4.5){T}
\put(10,1.5){\line(1,0){9}}\put(14.5,2){S}
%\multiput(11.4,4)(0.6,0){3}{\circle*{.2}}
%\put(11.2,2){\line(1,1){8}}
%
%\put(15.6,5.5){$\updownarrow$}
%
%\put(-1,0){\vector(1,0){3}}\put(-1,0){\vector(0,1){3}}
%\put(-1,0){\vector(1,1){3}}
%\put(2.3,-0.2){$x^{6}$}\put(-1.2,3.2){$x^{3,4,5}$}\put(2.3,3.2){$x^{7,8,9}$}
\end{picture}\end{center}
\caption{From M-theory to type II strings and back: the web of dualities.}
\label{figa}
\end{figure}
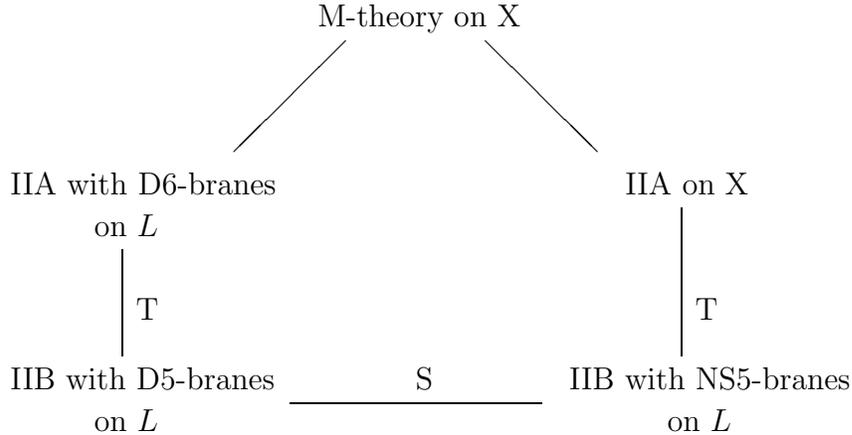

In this paper we will consider M-theory on a non-compact manifold $X$
of holonomy $SU(3)$, $G_2$, $Sp(2)$ and $Spin(7)$, such that 
$X$ has real dimension ${\rm dim}(X)=6, 7, 8$ and 8 
respectively (see Table 1).
\begin{table}\begin{center}
\begin{tabular}{|c|c|c|c|}
\hline
$Hol(X)$ & dim$(X)$ & SUSY on $D2$-brane & Examples of $X$ \\
%$Hol(X)$ & Supersymmetry on  & Examples of $X$ \\
%& D2-brane probe  &  \\
\hline
\hline
$SU(3)$ & 6 & $\mathcal{N}=2$ & The conifold \\
\cline{1-4}
&&& The cone over ${\bf \mathbb{C}P}^3$ \\
\cline{4-4}
&&& The cone over $SU(3)/U(1)^2$ \\
\cline{4-4}
$G_2$ & 7 & $\mathcal{N}=1$ & New manifold with $h_2 + h_3=2$  \\
\cline{4-4}
&&& New manifold with $h_2 + h_3=3$  \\
\cline{4-4}
&&& The cone over ${\bf S}^3 \times {\bf S}^3$  \\
\cline{1-4}
$Sp(2)$ & 8 & $\mathcal{N}=3$ & $T^* {\bf \mathbb{C}P}^2$, $T^* \mathcal{B}_n$  \\
\cline{1-4}
$Spin(7)$ & 8 & $\mathcal{N}=1$ & New manifolds with $h_2+h_3 \ge 3$ \\
\hline
\end{tabular}\end{center}
\caption{A list of models analyzed in this paper.}
\end{table}
To determine the physics of M-theory on $X$, it 
is often useful to reduce to IIA on a spatial ${\bf S}^1$. 
There are essentially two possibilities: 
we may either choose the ${\bf S}^1$ to be transverse to $X$, or to 
be embedded within $X$. These are depicted in Figure \ref{figa}, 
where, starting at the top of the figure, we move clockwise or 
anti-clockwise respectively. In the former case, we simply end up with 
IIA string theory compactified on $X$. In the latter case however, 
the resulting IIA background, $X/U(1)$, depends strongly on the 
choice of the circle action. A particularly convenient choice of 
embedding ${\bf S}^1$ --- which we call L-picture,
following \cite{aw,GS} --- occurs when  
the resulting IIA space-time geometry is topologically flat:
\be
X/U(1)\cong \mathbb{R}^n,\quad\quad n={\rm dim}(X)-1.
\ee 
If such a quotient exists, all the information about the topology of 
$X$ is encoded in the fixed point locus, which we denote $L$ (thus 
giving the name to the L-picture). From 
the IIA perspective, these fixed points are the positions of 
D6-branes lying in the directions transverse to $X/U(1)$:
\be
M^{10-n} \times L \subset M^{10-n} \times \mathbb{R}^n
\ee
This ensures that the locus $L$ has dimension
\be
{\rm dim}(L)={\rm dim}(X)-4
\ee
The task of identifying the geometry of $L$ directly is rather hard.
It has been undertaken recently for some examples of
$G_2$ \cite{aw} and $Spin(7)$ \cite{GS} manifolds.
However, as explained in \cite{GS}, under the assumption that such 
an L-picture exists,
the homology of the fixed point set can be easily determined from
the homology of $X$ by means of the following general formulas,
\begin{eqnarray}
h_0 (L) & = & h_2 (X) + 1 \nonumber \\
H_i (L; \mathbb{Z}) & \cong & H_{i+2} (X; \mathbb{Z}), \quad i>0
%\nonumber \\
%H^2 (L;\mathbb{Z}) & \cong & H^4 (X;\mathbb{Z}) \nonumber \\
%\nu & \stackrel{\rho}{=} & \lambda
\label{genhom}
\end{eqnarray}
These formulas, which were derived by matching the 
BPS states in the IIA and M-theory picture, allow us to 
simply write down the topology of the fixed point set $L$ in 
many cases of interest. In the following we shall develop the 
L-picture in more detail, determining the explicit curve 
in several cases. We show that much of the physics of M-theory 
on $X$ can be understood from this picture. 

As shown 
in Figure \ref{figa}, we may connect the $L$-picture 
with the manifold $X$ in two ways: either by returning to 
M-theory, or by performing a T-duality to IIB string 
theory. In the latter route, $L$ is interpreted as a locus 
of NS5-branes which describes the T-dual of $X$. Familiar 
examples include the equivalence between parallel NS5-branes  
and $A$-type ALE spaces, and between orthogonal 
NS5-branes and (generalized) Calabi-Yau conifolds \cite{u,sunil}. 
Here we discuss in detail several $G_2$ examples. 

One of the main themes of this paper is the story of brane 
probes on manifolds of exceptional holonomy. For M2-brane 
probes placed transverse to $X$ of holonomy $SU(3)$, $G_2$, 
$Sp(2)$ or $Spin(7)$, the world-volume theory enjoys 
${\cal N}=2,1,3$ or 1 supersymmetry respectively in $d=(2+1)$ 
dimensions. If we reduce on a circle transverse to the 
M2-brane, we have a D2-brane probe of a IIA 
background. For the $SU(3)$ and $G_2$ cases, we have two 
ways of performing this reduction, resulting in a D2-brane 
probe of the $L$ picture, or a D2-brane probe of the manifold 
$X$. The two resulting $d=2+1$ dimensional theories on the probe 
world-volumes are related by mirror symmetry, a duality 
of three-dimensional gauge theories which, among other things, 
exchanges Coulomb and Higgs branches \cite{is}. In 
situations where we can successfully identify both world-volume 
theories, this provides a stringy strategy to derive large classes 
of three dimensional mirror pairs. In fact, it actually provides 
two such strategies, depending on the route we take around the circle in 
Figure \ref{figa}. We may either lift to M-theory, and return 
immediately to IIA on a different circle \cite{pz},  a 
procedure which, in view of politically correct 
sensibilities, we refer to as the "M-theory flip". Alternatively, 
we may choose the IIB route, resulting in Hanany-Witten brane 
configurations \cite{hw}. This latter procedure has been used 
to great effect in deriving mirror symmetry for various 
three dimensional gauge theories $[10-16]$.

In the present paper, we employ a logic that is somewhat reversed 
from the description above; we use mirror symmetry of the 
field theory to derive aspects of the probe world-volume theory. 
We choose this method because, for manifolds of $G_2$ holonomy, 
the ${\cal N}=1$ (two supercharges) theory on the world-volume 
of the D2-brane probe is 
not yet well understood. For $G_2$ orbifolds, and their partial 
resolutions, one may determine the probe theory using the 
techniques of \cite{DRM} (see \cite{tsimpis} for some initial work in 
this direction). However, for the conical singularities of 
(phenomenological) interest, this procedure does not work. A 
new technique is required, and the $L$ picture provides this. 
In contrast to the probe of $X$, the world-volume theory 
of a D2-brane probe of the D6-brane locus $L$ is, in many cases, 
extremely simple. In particular, in the limit in which $X$ 
develops a conical singularity, the D6-brane locus $L$ also 
degenerates into a cone \cite{aw}, and often becomes a collection of 
flat, intersecting, $D6$ branes \cite{BDL}. Each of these 
contributes a hypermultiplet to the D2-brane world-volume theory, 
the precise coupling of which breaks supersymmetry 
to ${\cal N}=1$. The world-volume field theory operators 
corresponding to deformations away from the singular limit can 
then be identified, and the full quantum corrected Coulomb branch 
is conjectured to reproduce the manifold $X$, with the dual 
photon playing the role of the M-theory circle. At this point, 
we can invoke mirror symmetry of the gauge theory --- derived 
using independent techniques --- to write down a putative theory 
for the D2-brane probe of $X$. As we shall see, in many cases of 
interest, we can reconstruct the manifold $X$ as the Higgs branch 
of this dual gauge theory.

Before proceeding to the detailed outline of the paper, it is worth 
making a few cautionary points. In particular, 
while the techniques of brane probes are tried and tested for 
theories with ${\cal N}\geq 2$ supersymmetry in three dimensions, 
one must necessarily be more skeptical when dealing with probes of 
exceptional holonomy manifolds. With only ${\cal N}=1$ supersymmetry 
(2 supercharges), 
the power of holomorphy is lost, and with it our cherished 
non-renormalization 
theorems. Our hopes rest on the discrete parity symmetry,
\ba
P:(x^0,x^1,x^2)\rightarrow (x^0,-x^1,x^2)
\ea
under which both gauge field mass terms, as well as 
real superpotentials \cite{ahw,gk,bks} are odd. Since the 
question of parity invariance of the theory is determined at 
one-loop \cite{red,agw,ch}, we may use this to prohibit the 
lifting of moduli spaces of vacua in some of our models. 
However, no further information is available. In particular, 
we know of no field theoretic reason that even the topology 
of the vacuum moduli space need agree with the space-time 
background although, at least for weakly curved backgrounds, 
we would expect this on physical grounds. Nevertheless, we 
shall see that using the techniques of mirror symmetry, we 
are able to reconstruct both the topology, and the isometries 
of the manifold $X$. This latter statement is particularly  
non-trivial since the quotient from M-theory to the IIA L-picture 
partially destroys the isometries, which are expected to be 
recovered only in the strong coupling limit. The fact that 
mirror symmetry of ${\cal N}=1$ theories yields the full 
isometry group of $X$ is, we believe, vindication of our methods. 
Finally, the most ambitious hope would be to recover the metric 
on $X$ using these techniques. In particular, when applied to 
$d=1+1$ dimensional theories, the $G_2$ quotient construction 
of Section 4.1 (described in more detail below) provides a linear 
sigma model whose target space admits a metric of $G_2$ holonomy. 
In the infra-red, the theory necessarily flows to a (near) Ricci-flat 
metric. However, in the absence of something akin to Yau's theorem, 
we cannot be sure that this is indeed the metric of $G_2$ holonomy.

The plan of the paper is as follows. In 
section 2, we review what is known about abelian mirror symmetry, 
and derive several classes of putative ${\cal N}=1$ mirror pairs, 
using both field theoretic as well as string theory techniques. We 
give several examples that will have useful applications in the 
sequel. Readers interested only in the brane probe theories, 
and not the methods used to derive them, may safely skip this 
section. We have also tried to make each subsequent section 
self-contained. For each manifold $X$ listed in Table 1, 
we follow a simple pattern. Firstly we identify the locus $L$ of 
D6-branes, and write down the theory on a probe D2-brane, 
whose quantum corrected Coulomb branch realizes $X$. Secondly, 
we determine the mirror three-dimensional gauge theory and, thus, 
reconstruct $X$ algebraically as a Higgs branch. In this fashion 
we work our way around the circle of Figure \ref{figa}. 

Section 3 
acts as a warm up exercise, 
where we review our methods as applied to the simplest ${\cal N}=2$ 
example which arises on a D2-brane probe of a Calabi-Yau conifold. 
We describe various aspects of deformations and mirror pairs in 
this case.  

Section 4, describing $G_2$ manifolds,  
contains the main part of the paper. We start by reviewing 
the restrictions on flat special Lagrangian 
planes which, physically, corresponds to the requirement 
that a collection of D6-branes preserves four supercharges. 
Upon lifting to M-theory, this results in a manifold $X$ 
of $G_2$-holonomy. Using the techniques developed in 
Section 2, we write down a linear sigma model whose 
target space is topologically $X$, and therefore 
admits a metric of $G_2$ holonomy. We refer to this 
as the ``$G_2$-quotient construction''. 

The remainder of Section 4 examines various 
examples of the $G_2$ quotient construction, starting with the 
cone on ${\mathbb C}{\bf P}^3$ and the cone on $SU(3)/U(1)^2$, 
both of which were discussed by Atiyah and Witten \cite{aw}. 
We recover some of the results of Acharya and Witten \cite{bobed} 
from a brane probe 
perspective. Moreover, we show that the cone over $SU(3)/U(1)^2$ 
has an extra, non-normalizable, moduli not considered in \cite{aw}.  
We then turn attention to two further examples where the $L$ picture 
consists of three and four orthogonally intersecting D6-branes. We 
determine the homology of the M-theory lift $X$ and, using 
mirror symmetry, derive algebraic descriptions of these manifolds 
as quotient spaces. These manifolds have a rich moduli space 
of (non-normalizable) deformations in which two cycles undergo 
flop transitions, or are replaced by three cycles\footnote{For
a recent discussion of geometric transitions in M-theory
on $G_2$ holonomy manifolds see, for example $[25-38]$.}.
{}From the L-picture it is clear that each such transition
is inherited from the Calabi-Yau conifold discussed in Section 3.
Finally, we turn to the more subtle case where $X$ is a cone over 
${\bf S}^3\times {\bf S}^3$. The theory on a D2-brane probe 
of this model has already been discussed by Aganagic and Vafa 
\cite{av}. We elaborate on their construction and provide the 
mirror gauge theory. 

In Section 5
we discuss D2-brane probes of hyperK\"ahler 
8-manifolds with $Sp(2)$ holonomy. This is partly to 
elucidate some of the issues unique to 8-dimensional 
spaces in preparation for the $Spin(7)$ examples.
We also clarify some outstanding issues about the probe theory
and extend the results of \cite{gvw} to hyperK\"ahler
singularities of the form $T^* \mathcal{B}_n$,
where $\mathcal{B}_n$ is a del Pezzo surface.
%Just in the case of ${\bf \mathcal{C}P}^2$ model,
%we find that for every $\mathcal{B}_n$ there is
%actually a family of supersymmetric compactifications,
%parametrized by the value of the $G$-flux, most of which
%have a single massive vacuum.
In particular, we find that for every $\mathcal{B}_n$
there is a model (with a special value of the $G$-flux),
which has two vacua.

In the final section, we discuss manifolds of $Spin(7)$ holonomy. 
We restrict ourselves to manifolds $X$ whose L-picture consists 
of up to seven, mutually orthogonal D6-branes. This does not 
include any of the examples discussed in \cite{GS}, and the explicit
metric on these $Spin(7)$ holonomy manifolds is not known.
Nevertheless, these manifolds possess an intricate moduli space,
including branches of different topologies in which two-cycles are 
exchanged for three-cycles. To our knowledge, this is the first 
example of a geometric transition in a $Spin(7)$ manifold, albeit 
one which is locally equivalent to the Calabi-Yau conifold transition. 
Unlike for the above discussion of $G_2$ manifolds, there are now 
insufficient dimensions to perform an M-theory flip and string 
theory provides no method of deriving mirror pairs of three-dimensional 
gauge theories probing $Spin(7)$ backgrounds. Nevertheless, our 
field theoretic techniques allow us to derive a mirror theory, 
providing an algebraic description of the manifolds. 

Even though we mainly consider a single brane probe,
we expect that some of our results can be generalized
to non-abelian gauge theories on multiple branes.
In particular, it would be interesting to extend
mirror symmetry to such models, and explore various phases.
We will not pursue this here. Let us just briefly mention
that in the last two cases of $Sp(2)$ and $Spin(7)$ holonomy
it is easy to predict what happens if we place a large
number of membranes at the conical singularity of $X$.
Namely, following the usual ideas of AdS/CFT correspondence \cite{ADS},
it is natural to expect three-dimensional conformal field theories
with $\mathcal{N}=3$ and $\mathcal{N}=1$ supersymmetry, respectively.
In fact, in the case of hyperK\"ahler singularities there are
more reasons to expect that the same happens at finite $N$ \cite{gvw}.

We also include an Appendix, containing a new class of mirror pairs for 
${\cal N}=1$ Maxwell-Chern-Simons theories which, in 
particular, allow for the possibility of compact Coulomb branches. 
We illustrate this phenomenon with a theory which has an ${\bf S}^3$ 
Coulomb branch.

The results of Sections 2.1 and 4.1, together with the example of 
Section 4.3, are summarized in the companion paper \cite{d2lite}.

%%%%%%%%%%%%%%%%%%%%%%%%%%%%%%%%%%%%%%%%%%%%%%%%%%%%%%%%%%%%%%%%%%%%

\section{Mirror Symmetry in Three Dimensional Gauge Theories}
\label{mirrorsec}

%%%%%%%%%%%%%%%%%%%%%%%%%%%%%%%%%%%%%%%%%%%%%%%%%%%%%%%%%%%%%%%%%%%

Mirror symmetry of three dimensional gauge theories refers to 
a conjectured quantum equivalence between a pair of theories 
which, for the remainder of this paper, we shall refer to as 
Theory A and Theory B. The Coulomb branch of Theory A coincides 
with the Higgs branch of Theory B and vice versa. 
We start here with a review of the most general abelian ${\cal N}=4$
mirror pairs \cite{is,berk,ks}, and discuss the
techniques of \cite{ahiss,matt,ks,tong} which allow one to
deform these theories to mirror pairs with less supersymmetry.
The most general abelian mirror pair may be most simply
derived using the Kapustin-Strassler formula \cite{ks}, yielding
\ba
{\bf Theory\ A:} && U(1)^r \mbox{\ with\ } N \mbox{\ hypermultiplets\ }
\nn\\
{\bf Theory\ B:} && \hat{U(1)}{}^{N-r} \mbox{\ with\ } N
\mbox{\ hypermultiplets\ }
\nn\ea
where the ${\cal N}=4$ vector multiplets contain a gauge field and 
a triplet of real scalars $\phi$, together with four Majorana 
spinors. The ${\cal N}=4$ hypermultiplets also contain four 
Majorana spinors, this time paired with a doublet of complex 
scalars $w$, 
\be
w=\left(\begin{array}{l} q \\ \tilde{q}^\dagger \end{array}\right)
\ee
where $q$ and $\tilde{q}$ are in conjugate representations 
of the gauge group. For Theory A we denote the  charge of 
the hypermultiplets as $R_i^a$, while for Theory B it is 
$\hat{R}_i^p$,
$i=1,\cdots N\ ;\ a=1,\cdots r\ ;\ p=1,\cdots N-r$. Each of these
matrices is assumed to be of maximal rank. Mirror symmetry requires, 
\ba
\sum_{i=1}^NR_i^a \hat{R}_i^p = 0 \ \ \ \ \ \forall\ a, p
\label{charge}\ea
We denote the coupling constant of the two theories as
$e^2$ and $\hat{e}^2$ respectively.
For simplicity we shall concentrate on the case where
Theory A lies on its Coulomb branch and Theory B on its
Higgs branch, each a $4r$ real dimensional hyperK\"ahler 
manifold. The  theories include further parameters consistent with
supersymmetry: a triplet of masses, $m_i$,
for each hypermultiplet of Theory A, and a triplet of
FI parameters, $\zeta^p$, for each gauge factor of Theory B.
(Note that including FI parameters for Theory A or mass parameters
for Theory B would partially lift the vacuum moduli space of interest,
and so we set these to zero). Note further that not
all the mass parameters of Theory A are independent; precisely
$k$ of them may be absorbed by suitable shifts of the
vector multiplet scalars. 

The mirror map between operators is given by,
\ba
R_i^a\phi_a+m_i=\hat{w}_i^\dagger\tau\hat{w}_i\quad,\quad
\hat{R}_i^p\hat{\phi}_p+\hat{m}_i={w}_i^\dagger\tau w_i
\label{opmap}\ea
where $\tau$ is the triplet of Pauli matrices. The FI and mass
parameters obey the similar relationship,
\ba
\zeta^p = \sum_i \hat{R}_i^p m_i
\label{massfimap}\ea
With ${\cal N}=4$ supersymmetry, non-renormalization theorems
guarantee that the metric on the Higgs branch is classical,
while that on the Coulomb branch receives no corrections beyond
one-loop\footnote{Non-perturbative corrections, allowed by
supersymmetry, are not present in abelian theories.}. Each is a 
toric hyperK\"ahler manifold, consisting of a torus ${\bf T}^r$ 
fibered over a real $3r$ dimensional base. It is a 
simple matter to explicitly calculate the metric in each case, 
to discover that they coincide in the strong-coupling, infra-red 
limit $e^2\rightarrow\infty$ \cite{is,berk}. This is the statement of
mirror symmetry in these theories.

It is instructive to examine the symmetries of each model. Theory A 
has a $U(1)^{N-r}_F\times U(1)^r_J$ global symmetry group. The F-currents
are flavor symmetries acting on the chiral multiplets, while the
J-currents act transitively on the dual photons $\sigma$, 
defined as $d\sigma={}^\star F$,
\be
U(1)_J:\sigma \rightarrow \sigma + \alpha
\ee
This is to be
contrasted to the $U(1)^r_F\times U(1)_J^{N-r}$ global symmetry
group of Theory B. The mapping is obvious: $F\leftrightarrow J$.
In each
case the $U(1)^{N-r}$ factor is related to the mass and
FI parameters of \eqn{massfimap}, while the $U(1)^r$
factor acts on the ${\bf T}^r$ torus of the moduli space, 
resulting in tri-holomorphic isometries of the metric.
For certain choices of the charges,
the flavor symmetry of either theory may be enhanced to a non-abelian group.
In such circumstances, only the maximal torus is manifest
in the mirror picture as a J-symmetry.
Nevertheless, in the infra-red there is a quantum
non-abelian symmetry enhancement and the symmetry groups of the
two theories once again agree \cite{is}. As well as its
action on the global, tri-holomorphic symmetries, the mirror
map also exchanges the $SU(2)_N\times SU(2)_R$ R-symmetry
currents of the theories. 

The above analysis leads us to view three dimensional
mirror symmetry in the same light as Seiberg duality,
valid only in the extreme infra-red. However, work
by Kapustin and Strassler \cite{ks} suggests that in fact this
need not be the case. They show that there exists a
deformation of Theory B such that the metric on the Higgs
branch coincides with the metric on the Coulomb branch
of Theory A for all values of $e^2$:
\ba
{\bf Theory\ B{}^\prime:} && U(1)^r\times \hat{U(1)}{}^N \mbox{\ with\ }
N \mbox{\ hypermultiplets\ }
\nn\ea
One may understand the deformation from Theory B to Theory B${}^\prime$ as
a two-step process. One firstly gauges the $U(1)^{r}$ flavor symmetry
of Theory B --- it is the extra $\hat{U(1)}{}^r\subset \hat{U(1)}{}^N$
above.
This is subsequently coupled
via a Chern-Simons (CS) interaction to a
further $U(1)^r$ symmetry group. These
$U(1)^r$ fields have no further couplings to hypermultiplets, and
their kinetic terms are normalized as $1/e^2$, the same as the coupling
constants of Theory A.
Examples of the resulting hyperK\"ahler quotient construction were studied, 
for example, in \cite{gr}. The net effect is to squash, by an amount 
$1/e^2$, the asymptotic ${\bf T}^r$ fibers of the Higgs branch 
associated to the 
action of the
flavor group.
Since the action of the flavor group is tri-holomorphic, this squashing
preserves the three complex structures on the Higgs branch. The resulting
metric coincides with the Coulomb branch metric of Theory A at finite
gauge coupling.

To fill in the details, the $\hat{U(1)}{}^N$ naturally splits into
$r+(N-r)$ abelian gauge
fields, under which the hypermultiplets have charge $(R_i^a, \hat{R}_i^p)$.
The hypermultiplets
are neutral under $U(1)^r$. These latter fields couple only to
$\hat{U(1)}{}^r$ via a CS-coupling,
\ba
R_i^aR_i^b\ A^a\wedge \hat{F}{}^b
\label{rr}\ea
together with further terms required by supersymmetry (see below). 
The coupling
constants $e^2$ of Theory A now play the role of coupling constants
for the $U(1)^r$ of Theory B, while those of $\hat{U(1)}{}^N$
are sent to infinity (they may also be made finite if we further
modify Theory A). Let us examine the vacuum moduli space
of Theory B${}^\prime$,
\ba
V_{B^\prime}&=&\sum_{a=1}^r e_a^2 (R_i^aR_i^b \hat{\phi}^b)^2
+\sum_{a=1}^r \hat{e}^2_a\left(R_i^aw_i^\dagger\vec{\tau} 
w_i+R_i^aR_i^b\phi^b\right)^2
\label{bprime}\\ && +\sum_{p=1}^{N-r} \hat{e}_p^2\left( 
\hat{R}_i^p w_i^\dagger \vec{\tau} w_i + \hat{R}_i^pm_i\right)^2
+\sum_{i=1}^N\left(R_i^a\hat{\phi}^a+\hat{R}_i^p\hat{\phi}^p\right)^2
w_i^\dagger w_i
\nn\ea
where $\vec{\tau}$ are the Pauli matrices. The presence of $\phi$ and 
$\hat{\phi}$
in the D-terms is a consequence of the supersymmetric completion of the
CS-coupling above. The Higgs branch of this theory is parameterized by
$w_i$ and $\phi_a$, together with the $r$ dual photons $\sigma^a$ arising from
$U(1)^r$. The D-terms above provide $3N$ constraints which are moment
maps for the $\hat{U(1)}{}^N$ gauge orbits, the action of which
includes a translation of the dual photons,
\ba
w_i &\rightarrow&\exp\left(i R_i^b\hat{\alpha}_b+i\hat{R}_i^p
\hat{\alpha}_p\right) w_i
\nn \\
\sigma^a &\rightarrow& \sigma^a + R_i^aR_i^b\hat{\alpha}^b
\label{shift}\ea
It is useful to examine how Theory B${}^\prime$
reduces to Theory B in the limit $e^2\rightarrow \infty$.
This allows the vector multiplet scalars $\phi$ 
to fluctuate unconstrained by a kinetic term. The $\phi$'s
then appear only in the second term in \eqn{bprime} and their
role is simply to remove this constraint. As for the corresponding
$\hat{U(1)}{}^r$
gauge action, this may be absorbed by its action on the
dual photons \eqn{shift}, leaving the $w_i$ to be constrained
only by $(N-r)$ D-terms, and the corresponding gauge
action $\hat{U(1)}{}^{N-r}$. Note that, for certain
non-minimal choices of charge $R$, there may remain a
discrete remnant of the $\hat{U(1)}{}^r$ gauge symmetry.

%%%%%%%%%%%%%%%%%%%%%%%%%%%%%%%%%%%%%%%%%%%%%%%%%%%%%
\subsection{Deforming Mirrors without Cracking Them}
\label{qftsec}
%%%%%%%%%%%%%%%%%%%%%%%%%%%%%%%%%%%%%%%%%%%%%%%%%%%%%

The agreement of the metrics --- and hence the two-derivative
terms in the low-energy expansion ---  at all values of the
coupling constants suggests a view of mirror symmetry radically
different from that first envisaged.
Rather than being reminiscent of Seiberg duality,
Theory A and Theory B${}^\prime$, may be thought of as
two different descriptions of the same physics {\em at all
energy scales}. This is the conjecture of Kapustin and
Strassler \cite{ks}. Of course, this is a much stronger
statement than mere agreement of the metrics on the vacuum 
moduli space, but nonetheless it has survived at least one 
non-trivial test
\cite{ahkt}. For the purpose of this paper, we shall assume 
the validity of 
this conjecture, and will provide evidence 
that the conclusions we derive from this do indeed hold. 

An important corollary of the Kapustin-Strassler conjecture
is that it may be possible to deform the two theories, breaking
supersymmetry but preserving their equivalence. This procedure was 
described for breaking to ${\cal N}=2$ Maxwell-Higgs theories in 
\cite{ahiss,matt} and to ${\cal N}=2$ Maxwell-Chern-Simons-Higgs 
theories in \cite{tong,dt}. One may attempt to be yet braver, and 
deform to theories with ${\cal N}=1$ supersymmetry. In this case we 
lose much control over our theories, since phase transitions abound. 
Nevertheless, we proceed blindly. In following sections, we shall 
re-derive some of these mirrors from string theory methods, 
giving further evidence of their validity.

Let us explain in more detail the techniques we use,
starting with the theories without CS couplings. 
Suppose we gauge a flavor symmetry of Theory \Bprime. The
mirror deformation is to gauge the corresponding
J-symmetry of Theory A. This is achieved via a CS coupling.
This coupling requires the newly introduced dual photon
to transform transitively under a gauge symmetry and,
in the strong coupling limit, effectively removes this
gauge symmetry. In this manner, gauging a flavor symmetry,
say of Theory B${}^\prime$, is mirror to un-gauging a symmetry
of Theory A. Moreover, since this procedure commutes with the
strong coupling limit (at least in the classical Lagrangian)
we quote only the simpler mirror theories A and B: it is a
trivial matter to re-derive the all-scale mirror ``Theory \Bprime''  
following the prescription described above.

If we perform this procedure in an ${\cal N}=2$
invariant manner --- as first described in \cite{ahiss} ---
we arrive at the following mirror pairs,
\ba
{\bf Theory\ A:}&& U(1)^r \mbox{\ with $k$ neutral chirals
and $N$ charged hypermultiplets} \nn\\
{\bf Theory\ B:} && \hat{U(1)}{}^{N-r} \mbox{\ with $N-k$ neutral chirals
and $N$ charged hypermultiplets}
\nn\ea
where the gauge fields all lie within ${\cal N}=2$ vector
multiplets. The charges of the hypermultiplets in the two
theories, $R_i^a$ and $\hat{R}_i^p$, are once again related  
through \eqn{charge}. 
The neutral chiral multiplets of Theory A, which we denote as 
$\Psi_\alpha$, $\alpha=1,\cdots,k$, couple to the
hypermultiplets, containing $Q_i$ and $\tilde{Q}_i$, 
via the familiar gauge invariant cubic superpotential,
\be
{\cal W}=S_i^\alpha \tilde{Q}_i\Psi_\alpha Q_i
\label{comsup}\ee
with Yukawa couplings $S_i^\alpha$. Theory B has 
analagous interactions with coupling constants 
$\hat{S}_i^\rho$, $\rho=1,\cdots N-k$. These obey the mirror 
map,
\be
\sum_{i=1}^N S_i^\alpha \hat{S}_i^\rho = 0\ \ \ \ \ \ \forall
\ \alpha,\rho
\label{ss}\ee
All four matrices $R$, $\hat{R}$, $S$ and $\hat{S}$ must
be of maximal rank. While the above derivation of ${\cal N}=2$ 
mirrors relied on the mapping of the abelian symmetries, one 
may also arrive at them by use of the explicit operator mapping 
\eqn{opmap} as shown in \cite{matt}. 

We may now continue this procedure to ${\cal N}=1$ theories,
with similar results. Before describing these results, 
let us recall some of the less familiar aspects of the ${\cal N}=1$ 
(two supercharge) theories. The ${\cal N}=2$ vector multiplet 
decomposes into an 
${\cal N}=1$ vector multiplet and an ${\cal N}=1$ scalar multiplet. 
The former contains the gauge field and a single Majorana 
fermion. ${\cal N}=1$ gauge multiplets in three-dimensions contain 
no auxiliary fields, and all potential terms are therefore 
associated to scalar multiplets. These may be combined into a 
real scalar superfield which is a function of a single Majorana 
superspace coordinate (see \cite{ss,ahw} for further details),
\be
\Phi = \phi +\theta\psi-\theta^2D
\ee
where $\phi$ is a real scalar, $\psi$ is a Majorana spinor, and 
$D$ is a real auxiliary field. Each of these fields has a 
natural complexification, resulting in a complex scalar 
superfield $Q$. This is nothing more than the familiar chiral 
superfield of ${\cal N}=2$ theories. 

Since ${\cal N}=1$ supersymmetry in three dimensions provides no 
holomorphic luxuries, interactions between scalar superfields are 
written in terms of a {\em real} superpotential, 
\be
\int d^2\theta\ f(\Phi^a)=\frac{\partial f}{\partial \phi^a} D^a
+\frac{\partial^2 f}{\partial\phi^a\partial\phi^b}\psi^a\psi^b
\ee
With these conventions in mind, we can state our ${\cal N}=1$ 
mirror pairs. They are,
\ba
{\bf Theory\ A:}&& U(1)^r \mbox{\ with $k$ scalar and $N$ 
hypermultiplets} \nn\\
{\bf Theory\ B:} && \hat{U(1)}{}^{N-r} \mbox{\ with $(3N-k)$ 
scalar and $N$ hypermultiplets}
\nn\ea
where the gauge fields, this time, live in ${\cal N}=1$
vector multiplets and, by hypermultiplet, we mean the full 
${\cal N}=4$ matter multiplet, each of which contains 
a doublet of complex scalar superfields. We write,
\be
W=\left(\begin{array}{l}Q \\ \tilde{Q}^\dagger \end{array}\right)
\ee
As in previous cases, the coupling of these hypermultiplets to 
the gauge fields is through the charges $R$ and $\hat{R}$ 
satisfying \eqn{charge}. The scalar multiplets couple only 
to the hypermultiplets through Yukawa couplings. For Theory 
A, the real superpotential is
\ba
f=\sum_{i,\alpha} 
W^\dagger_i\vec{\tau}\,W_i\cdot \vec{T}_i^\alpha\,\Phi_\alpha
\label{realsup}\ea
$\alpha=1,\cdots, k$. As before, $\vec{\tau}$ are Pauli 
matrices, and the couplings are determined by the triplet 
of $k\times N$ matrices, $T$. The Yukawa couplings 
for Theory B are of the same form, only now fixed by the 
triplet of $(3N-k)\times N$ matrices $\hat{T}$. They 
must satisfy,
\be
\sum_{i=1}^N \vec{T}_i{}^\alpha\cdot\vec{\hat{T}}_i{}^\rho=0
\ \ \ \ \ \ \forall\ \alpha, \rho
\label{tt}\ee
This is one of the main results of the paper and we shall employ 
it extensively in applications to D2-brane probes. As we 
have discussed above, the mirror symmetry of ${\cal N}=1$ 
three-dimensional gauge theories discussed here is conjectural, 
and difficult to prove. Nevertheless, we shall see later that 
this conjecture does yield the correct results in places 
where we can test it. For 
completeness, we should also point out that these are not 
the most general ${\cal N}=1$ mirrors, although they are all 
we need for the purposes of this paper. In the Appendix
we describe further ${\cal N}=1$ mirrors with CS couplings.  

Before proceeding, a comment on notation: as we have seen above,
in this paper we shall consider
${\cal N}=1$ supersymmetric theories which include matter
multiplets more usually found in theories with higher
supersymmetry. We shall therefore refer to ${\cal N}=1, 2$
and $4$ matter multiplets as scalar, chiral and hyper-multiplets
respectively. Throughout this paper, the following symbols will be
used to denote various scalar fields:
\ba
\phi && \mbox{Real scalar field} \nn\\
\sigma && \mbox{Real, periodic scalar, dual to the photon} \nn\\
\psi && \mbox{Complex scalar field, neutral under the gauge group} \nn\\
q    && \mbox{Complex scalar field, charged under the gauge group} \nn\\
w    && \mbox{Doublet of complex scalar fields}
\nn\ea
The real scalar field $\phi$ may be found in either a
scalar multiplet, or an ${\cal N}=2$ vector
multiplet. Similarly, the complex scalar ${\psi}$ lives in
either a chiral multiplet, or an ${\cal N}=4$ vector
multiplet. The complex scalar $q$ arises in either a
chiral or hypermultiplet, while the doublet $w$ exclusively
resides within a hypermultiplet. We will often write this
explicitly as $w^\dagger = (q^\dagger,\tilde{q})$.

%%%%%%%%%%%%%%%%%%%%%%%%%%%%%%%%%%%%%%%%%%%%%%%%%%%%%%%%%%%%%%%%%%%
\subsection{${\cal N}=1$ Mirrors from IIB Brane Models}
\label{branesec}
%%%%%%%%%%%%%%%%%%%%%%%%%%%%%%%%%%%%%%%%%%%%%%%%%%%%%%%%%%%%%%%%

A large subclass of the mirror pairs decribed above
have an alternative
derivation in terms of Hanany-Witten type brane configurations
in IIB string theory \cite{hw,gk}. We can construct many abelian ${\cal N}=1$
field theories by considering D3-branes strung between various
combinations of D5 and NS5-branes, lying in the directions,
\ba
D3 && 126 \nn\\
NS5 && 12345 \nn\\
NS5^\prime && 12389 \nn\\
D5 && 12348
\nn\ea
\begin{table}\begin{center}
\begin{tabular}{|c|c|c|c|c|}
\hline
Configuration & Angles & Condition & SUSY & second 5-brane \\
\cline{1-5}
1 & $\theta_4$ & $\theta_4 = 0$ & $\mathcal{N}=4$ & NS5 $(12345)$ \\
\cline{1-5}
2(i) & $\theta_2$, $\theta_3$ &
$\theta_2 = \theta_3$ &
$\mathcal{N}=2$ & NS5 $(123 [48]_{\theta_2} [59]_{\theta_3})$ \\
\cline{1-5}
2(ii) & $\theta_3$, $\theta_4$ &
$\theta_3 = \theta_4$ &
$\mathcal{N}=2$ & $(p,q)5$\ $(1234 [59]_{\theta_3})$ \\
\cline{1-5}
3(i) & $\theta_1$, $\theta_2$, $\theta_3$ &
$\theta_3 = \theta_1 + \theta_2$ &
$\mathcal{N}=1$ & NS5 $(12 [37]_{\theta_1} [48]_{\theta_2} [59]_{\theta_3})$ \\
\cline{1-5}
3(ii) & $\theta_2$, $\theta_3$, $\theta_4$ &
$\theta_3 = \theta_2 + \theta_4$ &
$\mathcal{N}=1$ & $(p,q)5$\ $(123 [48]_{\theta_2} [59]_{\theta_3})$ \\
\cline{1-5}
4(i) & $\theta_1$, $\theta_2$, $\theta_3$, $\theta_4$ &
$\theta_4 = \theta_1 + \theta_2 + \theta_3$ & $\mathcal{N}=1$ &
$(p,q)5$\ $(12 [37]_{\theta_1} [48]_{\theta_2} [59]_{\theta_3})$ \\
\cline{1-5}
4(ii) & $\theta_1$, $\theta_2$, $\theta_3$, $\theta_4$ &
$\theta_1 = - \theta_2$, $\theta_3 = \theta_4$ & $\mathcal{N}=2$ &
$(p,q)5$\ $(12 [37]_{\theta_1} [48]_{\theta_2} [59]_{\theta_3})$ \\
\cline{1-5}
4(iii) & $\theta_1$, $\theta_2$, $\theta_3$, $\theta_4$ &
$\theta_1 = \theta_2 = \theta_3 = \theta_4$ & $\mathcal{N}=3$ &
$(p,q)5$\ $(12 [37]_{\theta_1} [48]_{\theta_2} [59]_{\theta_3})$ \\
%\cline{1-5}
\hline
\end{tabular}\end{center}
\caption{Supersymmetric five-brane configurations in IIB theory.}
\end{table}
An example of this type was discussed by Gremm and
Katz in \cite{gk}. Note that, although we deal exclusively with 
abelian mirrors, this brane set-up suggests many non-abelian 
mirror pairs. Here we extend this class of IIB brane models to 
more general configurations of five-branes. In fact, all models 
discussed by Gremm and Katz \cite{gk}
have one massless scalar field, corresponding
to D3-brane motion in the $x^3$ direction
(common to all of the five-branes involved).
More general five-brane configurations which lead
to $\mathcal{N}=1$ field theory on a D3-brane were
classified in \cite{OT,ohta2} and nicely summarized in \cite{BHKK}.
For completeness we (shamelessly) reproduce here a table from 
\cite{ohta2} wherein all possible five-brane configurations,
together with the amount of supersymmetry on D3-branes 
stretched between them, are cataloged. 
Apart from its charge, each $(p,q)$-fivebrane is specified
by its orientation in $x^3$ -- $x^7$,  $x^4$ -- $x^8$,
and $x^5$ -- $x^9$ planes.
We denote the corresponding angles by $\theta_1$,
$\theta_2$, and $\theta_3$, and refer to such a five-brane as:
\ba
(p,q)5 && 12 [37]_{\theta_1} [48]_{\theta_2} [59]_{\theta_3} 
\nn\ea
As in \cite{BHKK}, it will be convenient for us to
label the charge of the 5-brane also by an angle:
\be
\tan \theta_4 = {p \over q}
\ee
The models discussed by Gremm and Katz \cite{gk} correspond to case $3(ii)$.
%Let's consider the other configurations with $\mathcal{N}=1$ supersymmetry.
Here we consider examples  of other models which will illustrate certain 
points.

\subsubsection*{\rm{\em Example 1}}

\begin{figure}

\setlength{\unitlength}{0.9em}
\begin{center}
\begin{picture}(22,11)

\put(8,5){\line(1,0){11.5}}\put(9.5,5.5){D3}

\put(8,0){\line(0,1){10}}\put(8.5,10){NS5}
\put(17,0){\line(1,2){5}}\put(22.5,10){NS5'}

\put(9,2){\line(1,1){8}}\put(15.5,10){D5}
\multiput(11.4,4)(0.6,0){3}{\circle*{.2}}
\put(11.2,2){\line(1,1){8}}

\put(15.6,5.5){$\updownarrow$}

\put(-1,0){\vector(1,0){3}}\put(-1,0){\vector(0,1){3}}
\put(-1,0){\vector(1,1){3}}
\put(2.3,-0.2){$x^{6}$}\put(-1.2,3.2){$x^{3,4,5}$}\put(2.3,3.2){$x^{7,8,9}$}

\end{picture}\end{center}
\caption{A D-brane model of $\mathcal{N}=1$ theory in IIB string
theory with NS5'-brane of type $3(i)$.}
\label{figb}
\end{figure}
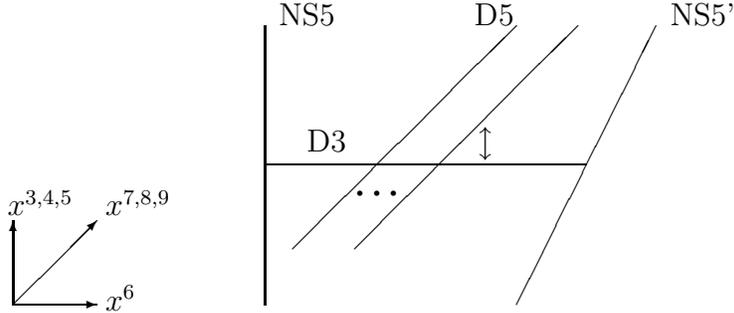

We start with $3(i)$ model
and a single $U(1)$ gauge group. Apart from the first NS5-brane
in directions $x^1, \ldots, x^5$, the configuration involves
an NS5${}^\prime$-brane and $N$ D5-branes as drawn in 
Figure \ref{figb}, and oriented as follows: 
\ba
D3 && 126 \nn\\
NS5 && 12345 \nn\\
NS5^\prime && 12 [37]_{\theta_1} [48]_{\theta_2}
[59]_{\theta_1 + \theta_2} \nn\\
D5 && 12789
\nn\ea
Note, that if one of the angles $\theta_1$, $\theta_2$,
or $\theta_3 = \theta_1 + \theta_2$
is equal to zero, then supersymmetry is enhanced to $\mathcal{N}=2$.
Below we assume a generic situation when this doesn't happen.
The position of the D3-brane stretched between the NS5 and 
NS5${}^\prime$-branes is then fixed, so that the only massless 
mode is a $U(1)$ gauge field. There are also $N$ hypermultiplets
coming from strings stretched between D3-brane and D5-branes.

As well as the massless modes, there are also some low-energy 
massive modes. The position of the D3-brane in $x^3,x^4$ and 
$x^5$ direction yields a triplet of scalars $\vec{\phi}$, 
whose mass $\vec{M}$ is determined by $\theta_{1}$ and $\theta_2$.  
For small angles, $M_1\sim\tan\theta_1$, etc. 
Similarly, the position of the D5-branes in these same directions 
results in a triplet of mass parameters $\vec{m}_i$ which 
vanish when the D5 and D3-branes coincide. Summarizing, we have:
\begin{center}
{\bf Theory A:} $U(1)$ with 3 massive scalar 
and $N$ hypermultiplets
\end{center}
The scalar potential of this theory is given by,
\be
V_A=\sum_{c=1}^3\left(\sum_{i=1}^Nw^\dagger_i{\tau}_c w_i
-M_c\phi_c\right)^2 
+\sum_{i=1}^N |\vec{\phi}-\vec{m}_i|^2w_i^\dagger w_i 
\nn\ee
If hypermultiplets are massless, this theory has a classical
Higgs branch of real dimension $(4N-4)$. It corresponds to 
breaking the D3-brane into $N+1$ segments, each stretched 
between adjacent D5-branes and free to move in directions $x^7$, 
$x^8$, and $x^9$ and with no constraint on the Wilson line 
$A_6$. At the origin of the Higgs branch there is a singularity 
at which the photon becomes massless. Emanating from this is 
the one-dimensional Coulomb branch, which is simply ${\bf S}^1$. 

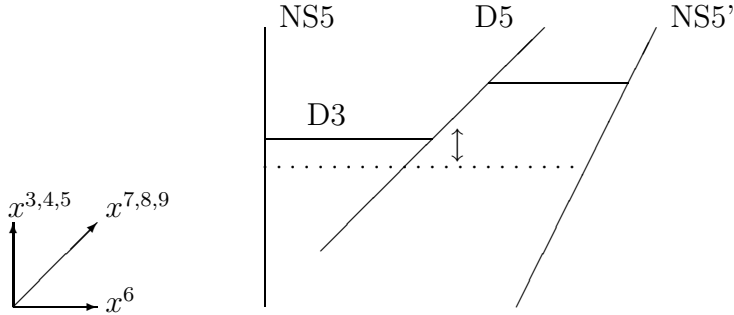
\begin{figure}

\setlength{\unitlength}{0.9em}
\begin{center}
\begin{picture}(22,11)

%\put(8,5){\line(1,0){11.5}}\put(9.5,5.5){D3}
\multiput(8,5)(0.5,0){23}{\circle*{.1}}
\put(8,6){\line(1,0){6}}\put(9.5,6.5){D3}
\put(16,8){\line(1,0){5}}

\put(8,0){\line(0,1){10}}\put(8.5,10){NS5}
\put(17,0){\line(1,2){5}}\put(22.5,10){NS5'}

\put(10,2){\line(1,1){8}}\put(15.5,10){D5}
%\multiput(11.4,4)(0.6,0){3}{\circle*{.2}}
%\put(11.2,2){\line(1,1){8}}

\put(14.6,5.5){$\updownarrow$}

\put(-1,0){\vector(1,0){3}}\put(-1,0){\vector(0,1){3}}
\put(-1,0){\vector(1,1){3}}
\put(2.3,-0.2){$x^{6}$}\put(-1.2,3.2){$x^{3,4,5}$}\put(2.3,3.2){$x^{7,8,9}$}

\end{picture}\end{center}
\caption{``Higgs branch'' of a $U(1)$ gauge theory with
one massive hypermultiplet.}
\label{figc}
\end{figure}

The Higgs branch remains if the hypermultiplets acquire an 
identical mass $\vec{m}_i=\vec{m}$. This can be seen from the 
string picture as the left-most 
segment moves in directions $x^3$, $x^4$, $x^5$ along
the NS5-brane, while the rightmost
segment has to move in directions $x^7$, $x^8$, $x^9$
along D5-brane, and in the same time also in
directions $x^3$, $x^4$, $x^5$ along the NS5${}^\prime$-brane. 
This deformation is sketched in Figure \ref{figc}.

From the field theory perspective, we see that the mass parameter 
becomes promoted to a FI parameter upon setting $\vec{\phi}=\vec{m}$. 
Topologically, the Higgs branch is $T^\star {\bf CP}^{N-1}$, where 
the size of the zero section is determined by $\vec{M}$ and $\vec{m}$. 
The singularity at the origin of the Higgs branch is removed for 
$\vec{m}\neq 0$, but nevertheless the Coulomb branch remains: it 
is simply separated from the Higgs branch in field space. The fact 
that the vacuum moduli space is not connected implies that this 
theory contains domain walls interpolating between the Higgs 
and Coulomb phases. 

The mirror theory can be obtained by S-duality
in IIB string theory, which maps D5-branes into
NS5-branes and vice versa. Specifically, we find:
\begin{center}
{\bf Theory B:} $U(1)^{N-1}$ with $3(N-1)$ scalar 
and $N$ hypermultiplets
\end{center}
This has the same matter content as the mirror theory derived by 
field theory means in the previous section. Moreover, the couplings 
may be easily read from the brane picture, and again agree with the 
field theory analysis. Each vector multiplet pairs up with 3 scalar 
multiplets to act essentially as a ${\cal N}=4$ vector multiplet. 
These couple in an ${\cal N}=4$ invariant fashion to all but the 
final hypermultiplet. The Coulomb and Higgs branches are 
$4(N-1)$ and $1$ dimensional respectively. 

\subsubsection*{\rm{\em Example 2}}

Let us now turn to an example that corresponds to a $G_2$ 
manifold that we discuss in Section \ref{gsecd}. 
It is an elliptic model (meaning the $x^6$ direction is 
compactified) of type 4i) from the table 2. The brane 
configuration has three five-branes of the same type, 
oriented to be mutually orthogonal. We start with,
\ba
D5_1: && 12345 \nn\\
D5_2: && 12389 \nn\\
D5_3: && 12479 \nn\\ 
D3: && 126 
\ea
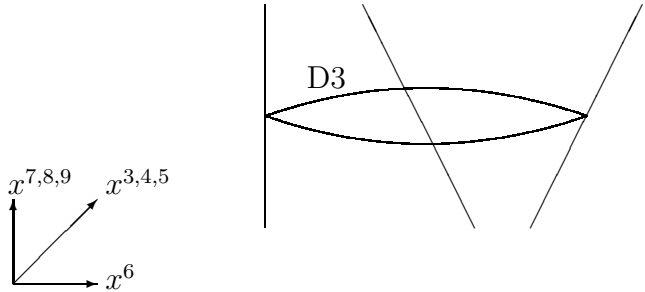
\begin{figure}

\setlength{\unitlength}{0.9em}
\begin{center}
\begin{picture}(22,11)
\qbezier(8,6)(13.7,8)(19.5,6)\put(9.5,7){D3}
\qbezier(8,6)(13.7,4)(19.5,6)
\put(17.5,2){\line(1,2){4}}
%\put(21.5,9){NS5$_2$}
\put(8,2){\line(0,1){8}}
%\put(5.5,10){NS5$_1$}
\put(15.5,2){\line(-1,2){4}}
%\put(12.5,9){NS5$_3$}
%\put(14.6,5.5){$\updownarrow$}
%\put(19,4.8){$\updownarrow$}
\put(-1,0){\vector(1,0){3}}\put(-1,0){\vector(0,1){3}}
\put(-1,0){\vector(1,1){3}}
\put(2.3,-0.2){$x^{6}$}\put(-1.2,3.2){$x^{7,8,9}$}\put(2.3,3.2){$x^{3,4,5}$}

\end{picture}\end{center}
\caption{IIB Brane model for Example 2 with orthogonal 5-branes.}
\label{figd}
\end{figure}
This is drawn in Figure \ref{figd}. The theory on the D3-brane 
has ${\cal N}=1$ supersymmetry and has the (interacting)  
massless field content,
\begin{center}
{\bf Theory A:} $U(1)$ with 6 scalar and 3 hypermultiplets
\end{center}
where the real scalars $\phi_\alpha$, $\alpha=1,\ldots,6$, denote the 
fields corresponding to D3-brane motion along 
$x^{3,4,5,7,8,9}$. These are coupled to the hypermultiplets 
through a superpotential of the form \eqn{realsup}. To 
avoid writing out a triplet of $3\times 6$ matrices, we 
may write the superpotential as,
\be
f=\sum_{i=1}^3 \vec{A}_i\cdot W_i^\dagger \vec{\tau} W_i
\ee
where the triplets $\vec{A}_i$ are suitable combinations 
of the $\Phi_\alpha$,
\be
\vec{A}_1=(\Phi_7,\Phi_8,\Phi_9)\ \ \ \ ,\ \ \ \ 
\vec{A}_2=(\Phi_7,\Phi_4,\Phi_5)\ \ \ \ ,\ \ \ \ 
\vec{A}_3=(\Phi_3,\Phi_8,\Phi_5)
\nn\ee
This theory has a seven dimensional Coulomb branch, parameterized 
by the six scalars $\phi_\alpha$ and the dual photon (hence 
the relevance to $G_2$ manifolds as we shall see). To read off the 
mirror theory, we perform a single S-duality to get,
\ba
NS5_1: && 12345 \nn\\
NS5_2: && 12389 \nn\\
NS5_3: && 12479 \nn\\ 
D3: && 126 
\nn\ea
Since each pair of NS5-branes are mutually orthogonal, 
we may read off the theory on D3-brane world-volume 
by following the the standard rules for theories with 
${\cal N}=2$ supersymmetry \cite{u}, treating each pair of 
NS5-branes in turn. It 
is simple to read off the massless fields from the 
brane set-up. We have,
\begin{center}
{\bf Theory B:} $U(1)^2$ with 3 scalar and 3 hypermultiplets
\end{center}
where we have ignored the overall, free $U(1)$ gauge symmetry. 
The hypermultiplets may be taken to have charges $(+1,-1,0)$ and 
$(0,-1,+1)$ under the gauge group. More subtle are the 
interactions. We expect two types of interactions; the 
usual Yukawa couplings of scalar multiplets  to 
hypermultiplets, and the 
quartic superpotential couplings present in the ${\cal N}=2$ 
theories \cite{kw,u}. Let us start with the Yukawa couplings. 
Recall that for a theory with ${\cal N}=4$ supersymmetry, the 
D- and F-terms are unified in a triplet of auxiliary fields 
which rotate into each other under the $SU(2)_R$ R-symmetry.  
However, this symmetry is lost in the usual ${\cal N}=2$ 
superspace formulation of the theory. It may be made 
manifest if we work in ${\cal N}=1$ superspace, in 
which case the interaction terms are encoded in the 
real superpotential $\vec{\Phi}\cdot W^\dagger \vec{\tau}W$. 
One usually employs the notation that the D-term is 
proportional to $\tau^3$, while the F-term is a complex 
combination of $\tau^1$ and $\tau^2$. In the present 
case, we have only ``D-term'' type interactions (only 
real scalar fields), but the rotation of the branes ensures 
that the scalar couples to a different combination for 
each hypermultiplet. Thus we find the Yukawa couplings,
\ba
f_{\rm Yuk}&=&\Phi_1(W_1^\dagger\tau^3W_1-W_2^\dagger
\tau^3W_2)+ \Phi_2(W_2^\dagger\tau^1W_2-W_3^\dagger\tau^1W_3) 
\nn \\ && +\Phi_3(W_3^\dagger \tau^2W_3-W_1^\dagger\tau^2W_1)
\label{corblimey}\ea
For the quartic superpotential, we assume that it arises from 
pairs on NS5-branes, in which case we may follow the rules 
laid down in \cite{u}. In the case where the D-term is 
proportional to $\tau^3$ (for example, $\Phi_1$ above), 
this quartic superpotential takes the form,
\be
\tilde{Q}_1Q_1\tilde{Q}_2Q_2+{\rm h.c.} = 
W^\dagger_1\tau^1W_1\,W^\dagger_2\tau^1W_2-
W^\dagger_1\tau^2W_1\,W^\dagger_2\tau^2W_2
\ee
Hence, comparing with the indices in equation \eqn{corblimey}, 
we obtain,
\ba
f_4&=&W^\dagger_1\tau^1W_1\,(W^\dagger_2\tau^1W_2-W^\dagger_3\tau^1W_3) 
+W^\dagger_2\tau^2W_2\,(W^\dagger_3\tau^2W_3-W^\dagger_1\tau^2W_1) \nn\\
&&+W^\dagger_3\tau^3W_3\,(W^\dagger_1\tau^3W_1-W^\dagger_2\tau^3W_2)
\label{ffour}\ea
As a simple check that this is the correct superpotential, 
note that it gives no further constraints on the $w_i$ than 
\eqn{corblimey} alone, ensuring that the Higgs branch of this 
model has the requisite dimension seven. The scalar potential 
arising from the superpotential $f=f_{\rm Yuk}+f_4$ yields 
the constraints,
\ba
|q_1|^2 - |\tilde{q}_1|^2 - |q_2|^2 + | \tilde{q}_2|^2 &=& 0 \nn\\
\re(\tilde{q}_2q_2-\tilde{q}_3q_3) &=& 0 \nn\\ 
\im(\tilde{q}_3q_3-\tilde{q}_1q_1) &=& 0
\nn\ea
These give 3 real constraints on the 12 real parameters $w_i$. 
After dividing by the $U(1)^2$ gauge action, we arrive at 
the Higgs branch.
The field content derived from the brane picture agrees 
with that derived in the previous section using 
field theory methods and, most importantly, the Higgs branches  
coincide. However, the brane theory also gave rise 
to the quartic superpotential \eqn{ffour}. As discussed 
above, this higher dimension operator did not alter the 
vacuum moduli space, and the two techniques therefore 
agree. However, as we shall see in future examples, 
there are cases where the field content of the mirror 
theories derived through field theory techniques and 
brane constructions do not agree. In cases with a greater 
number of hypermultiplets, one often finds that the 
field theory mirror has a greater number of neutral 
scalars than the brane mirror. However, in all of 
these cases, the Higgs branch determined by the two 
theories is the same. Essentially, the extra constraints 
arising from the Yukawa couplings in the field theory 
context are exactly equal to the constraints arising from 
the quartic superpotential in the brane context. An example 
of this occurs in Section \ref{gsece}. In fact, this same 
discrepancy between field theory methods and brane methods 
is also seen in mirror pairs with ${\cal N}=2$ supersymmetry 
but, once again, the moduli spaces agree. Presumably in all 
these cases one may give masses to the extra scalar multiplets 
such that, after integrating them out, the quartic superpotentials 
are generated. Such points aside, we stress again that in this 
paper we are interested only in the vacuum moduli space and 
these agree in all cases.

\section{$SU(3)$ Holonomy}
\label{susec}

In this section we discuss some of our techniques in the 
well-studied case of non-compact Calabi-Yau three-folds. 
Much of what we say here is not new, but we shall use 
the opportunity to stress the points which will play an 
important role in the forthcoming sections. We discuss 
separately the small resolution and deformation of the 
conifold and, in each case, attempt to work our way around 
Figure \ref{figa}. For the resolution these results are 
well-known  
and will be used extensively in later examples. However, for 
the deformation of the conifold to $T^\star{\bf S}^3$, the 
mirror probe theory is not well understood. We review the 
difficulties in this case. 

The conifold under consideration is a 
three dimensional, non-compact, singular Calabi-Yau manifold 
$X$, defined by the complex equation,
\be
xy-wz=0
\ee 
As described in the introduction, we start with M-theory 
on $X$ and and pick a ${\bf S}^1\subset X$ on which to reduce,
such that the resulting IIA spacetime is topologically 
$\mathbb{R}^6$. The sole remanant of the conifold is then 
two intersecting D6-branes with world-volumes \cite{u,sunil}, 
\ba
\begin{array}{ll} 
D6 & 123456 \\
D6 & 123678 \nn
\end{array}
\nn\ea
both of which lie at the same point in the $x^9$ direction. 
There exist two ways of smoothening out the singularity at 
the intersection point, $x=y=w=z=0$. They are known as the small 
resolution, and the deformation. We deal with each in turn.

\subsubsection*{\rm{\em The Small Resolution}}

Changing the K\"ahler structure of the conifold results in the 
manifold $X\cong\mathcal{O} (-1) \oplus \mathcal{O} (-1) \rightarrow 
{\bf \mathbb{C}P}^1$, which is topologically  
$\mathbb{R}^4\times {\bf S}^2$. This deformation is non-normalizable: it 
appears as a parameter (as opposed to the VEV of a dynamical field) 
in the low-energy IIA action. 

D-brane probes transverse to the resolved conifold have been well-studied 
in recent years --- see for example, 
\cite{kw,mp,aklm,u,sunil} --- resulting in a full understanding of 
the circle of dualities discussed in the introduction. Let us 
review this chain of dualities, moving anti-clockwise around 
the circle of Figure \ref{figa}.  We start by reducing to the 
L-picture. The equations \eqn{genhom} require the fixed 
point locus, $L$, to consist of two disconnected components, each 
of trivial topology. In other words, $L=\mathbb{C}\cup\mathbb{C}$ 
corresponding to two disconnected D6-branes which, by 
supersymmetry, are necessarily orthogonal. Together with the D2-brane 
probe, the world-volume directions span
\ba
\begin{array}{ll} D2 & 12 \nn\\
D6 & 123456 \label{cond6}\\
D6 & 123678 \nn
\end{array}
\nn\ea
While the world-volume directions of the D6-branes are identical 
to the singular case of the conifold, the singularity has been
resolved by simply separating the D6-branes in the $x^9$ direction. 

It is simple to write down the theory on the D2-brane.
Since we shall be using this technique in later
sections to derive more complicated mirror pairs,
we spell out in detail how the spectrum arises. Firstly,
the 2-2 strings result in the usual ${\cal N}=8$
supersymmetry multiplet in three-dimensions,
consisting of a $U(1)$ gauge field and
seven transverse scalars. The latter parameterise
the motion of the D2-brane along $x^m$, $m=3,\cdots,9$. 
The $U(1)$ gauge field may be dualised into a periodic 
scalar. It parameterizes the M-theory circle.
The 2-6 strings supply a hypermultiplet for each 
D6-brane, whose couplings break supersymmetry to 
${\cal N}=2$. In the current situation, $(x^3+i x^6)$ sits 
in a free chiral multiplet and shall be ignored in the 
following. The remaining four real scalars in the 
vector multiplet naturally form two neutral complex 
scalars, 
\be
\psi_1=x^4+ix^5,\quad \psi_2=x^7+ix^8
\ee
each of which is the lowest component of a chiral multiplet. 
Similarly, the $U(1)$ gauge field combines with the real 
scalar $\phi=x^9$ to form a ${\cal N}=2$ vector multiplet. 
The final theory therefore has matter content,
\begin{center}
{\bf Theory A:} $U(1)$ with 2 neutral chirals and 2 charged
hypermultiplets
\end{center}
The hypermultiplets come from 2-6 strings and
have charge $(+1,-1)$ (note that the overall sign is a matter
of convention). The scalar potential then is given by,
\ba
V_A&=&e^2(|q_1|^2-|q_2|^2-|\tilde{q}_1|^2+|\tilde{q}_2|^2)^2
+e^2|q_1\tilde{q}_1|^2 +e^2|q_2\tilde{q}_2|^2 \nn\\
&&+\left(\phi^2+|\psi_1|^2\right)w^\dagger_1w_1
+\left((\phi-m)^2+|\psi_2|^2\right)w^\dagger_2w_2
\label{bpot}\ea
In this picture, the  K\"ahler class of the 
conifold is given by the real mass parameter $m$, corresponding 
to the separation of the D6-branes in the $x^9$-direction. This 
theory has a six dimensional Coulomb branch, parameterised by
$\phi$ and $\psi_i$, together with the dual photon. 
By construction, this coincides with the resolved conifold. 
Note that, as stressed in the introduction, this is a statement 
only about the topology of the space; the metric on the Coulomb branch 
is not expected to coincide with the Ricci flat metric. When 
$m=0$, there is a two dimensional Higgs branch, parameterised 
by the mesons $q_1\tilde{q_2}$, and $q_2\tilde{q_1}$. This  
corresponds to the D2-brane dissolved inside the intersection of the 
D6-branes. 

Theory A has a $U(1)_J\times U(1)_1\times U(1)_2\times U(1)_F$ global 
symmetry group. The $U(1)_J$ factor acts only on the 
dual photon, while $U(1)_a$ rotates $\psi_a$, for $a=1,2$. In the 
strong coupling limit, we expect the $U(1)_J$ symmetry to be 
enhanced to $SU(2)$, reflecting the full isometry group of the 
conifold. The 
hypermultiplet scalars are neutral under each of the first three 
$U(1)$ factors. In contrast, 
the $U(1)_F$ symmetry does not act on the Coulomb branch, but 
rotates the hypermultiplet scalars: $w_1$ has charge $+1$, and $w_2$ 
charge $-1$. This global symmetry is the manifestation of the 
axial combination of gauge symmetries on the D6-brane which, in 
turn, arises from symmetries of the three-form $C$ in M-theory. 

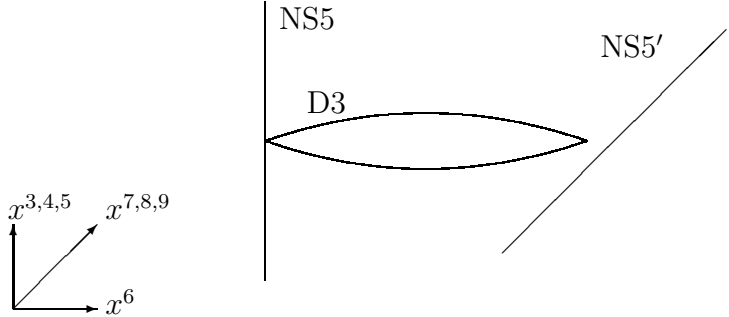
\begin{figure}

\setlength{\unitlength}{0.9em}
\begin{center}
\begin{picture}(22,11)

\qbezier(8,6)(13.7,8)(19.5,6)\put(9.5,7){D3}
\qbezier(8,6)(13.7,4)(19.5,6)
\put(8,1){\line(0,1){10}}\put(8.5,10){NS5}
\put(16.5,2){\line(1,1){8}}\put(20,9){NS5${}^\prime$}
\put(-1,0){\vector(1,0){3}}\put(-1,0){\vector(0,1){3}}
\put(-1,0){\vector(1,1){3}}
\put(2.3,-0.2){$x^{6}$}\put(-1.2,3.2){$x^{3,4,5}$}\put(2.3,3.2){$x^{7,8,9}$}

\end{picture}\end{center}
\caption{IIB Brane model for the deformed conifold.}
\label{fige}
\end{figure}

To derive the mirror of Theory A, we continue to work our way 
around the circle in Figure \ref{figa}. Performing a T-duality 
in the $x^6$ direction, followed by an S-duality leads to 
the brane set-up shown in Figure \ref{fige}. The gauge theory 
on the D3-brane was first discussed in \cite{berk3} (see 
also \cite{u}) and reproduces the well-known 
linear sigma-model of the resolved conifold\footnote{The
non-abelian world-volume theory on multiple D-brane
probes is somewhat more involved, and includes a
quartic superpotential \cite{kw}. This vanishes in the
abelian case of interest.}

\begin{center}
{\bf Theory B:} $U(1)$ + 2 charged hypermultiplets
\end{center}
The hypermultiplets both have charge $+1$ under the $U(1)$ gauge field.
There is a further free vector multiplet,
parameterizing the motion of the D2-brane transverse to the conifold,
and on the M-theory circle. This theory coincides with the 
mirror of Theory A derived in the 
previous section using field theory techniques \cite{berk3,matt}. 
The scalar potential of this theory is given by,
\ba
V_B=e^2\left(|q_1|^2+|q_2|^2-|\tilde{q}_1|^2-|\tilde{q}_2|^2-m\right)^2
+\phi^2\left(w_1^\dagger w_1+w_2^\dagger w_2\right)
\label{apot}\ea
The D-term, together with the action of the
gauge group yields the resolved conifold of interest as the
Higgs branch of the theory where the K\"ahler class is determined
by the FI parameter. When $m=0$, there is a two dimensional
Coulomb branch, corresponding to D2-brane splitting into
fractional branes. Quantum effects cause the Coulomb branch to 
split into two parts, parameterized asymptotically by 
$v_\pm=\exp(\pm \phi\pm i\sigma)$. Under mirror symmetry, 
$v_+ \rightarrow q_1\tilde{q}_2$, while $v_-\rightarrow q_2\tilde{q}_1$. 

Finally, to close the circle of Figure \ref{figa}, we perform one 
further T-duality in the $x^6$ direction. The T-dual of intersecting 
NS5-brane configurations were discussed in \cite{u,sunil}, and 
lead us back to IIA compactified on the resolved conifold. The 
theory on the D2-brane probe is once again Theory B, as may be 
derived through a process of partial resolutions of the 
$\mathbb{C}^3/\mathbb{Z}_2\times \mathbb{Z}_2$ orbifold \cite{mp}.
This completes our journey around the resolved conifold.

\subsubsection*{\rm{\em The Deformation}}

The singularity of the conifold may also be removed by deforming the defining 
equation to,
\be
xy-wz=\rho
\ee
for some complex parameter $\rho$. This also changes the complex structure.  
The resulting Calabi-Yau manifold is $X\cong T^* {\bf S}^3$, and 
has topology $\mathbb{R}^3\times {\bf S}^3$. As for the resolution, 
the deformation is non-normalizable in the IIA effective action. 
Let us try and repeat our discussion of the previous section, 
starting first with the L-picture.
Using equations \eqn{genhom}, we find that topologically 
$L\cong {\bf S}^1\times \mathbb{R}$. Supersymmetry restrictions 
require $L$ to be a complex curve (or, equivalently, a 
Special Lagrangian curve) in $\mathbb{C}^2$, 
\be
\psi_1\psi_2=\rho
\label{ccd}\ee
From 
the U-dual perspective of NS5-branes, these deformations 
were dubbed brane diamonds in \cite{aklm}. 
Our task is to reproduce this curve from the world-volume 
perspective of the D2-brane probe. Since asymptotically 
the locus $L$ becomes the two orthogonal D6-branes 
described in \eqn{cond6} (now at the same $x^9$ position), 
we expect the field content to be the same as Theory A 
above; it remains to ask what further couplings the 
deformation parameter $\rho$ induces in the theory.

Since this issue will play a role in later examples, let us 
spend some time examining the necessary quantum numbers of the 
deformation. We start with the M-theory picture. 
Recall that supersymmetry dictates that the moduli space 
of M-theory on the deformed conifold is parameterized by a 
hypermultiplet (in 
contrast to the resolved conifold which is parameterized by a five 
dimensional vector multiplet). The complex deformation $\rho$ 
provides two of the four scalars in this hypermultiplet. A further 
scalar arises from the M-theory 3-form,
\be
\theta=\int_{{\bf S}^3}C_3
\ee
The global symmetries of M-theory on this background include not only 
the geometrical symmetries of $X$, but also a symmetry arising 
from the gauge symmetry of $C$, which act as translations on 
$\theta$. For finite size ${\bf S}^3$, this symmetry is spontaneously 
broken. However, in the singular limit, the $U(1)_C$ symmetry acting on 
$\theta_1$ is restored. Indeed, we have already identified this 
symmetry in the probe theory of the previous section:
\be
U(1)_C\equiv U(1)_F
\ee
The deformation of the probe theory is therefore expected to break $U(1)_F$.

Before determining the operator deformation of   
probe theory, let us examine how it arises from the D6-brane point 
of view. From this perspective, the deformation parameter $\rho$  
corresponds to a vacuum expectation value (VEV)  
for the hypermultiplet scalars arising from 6-6 strings localized 
at the intersection \cite{aklm}.\footnote{The usual D-term constraints for 
hypermultiplets are suppressed by the ratio of the D6-brane 
world-volume to the volume of the intersection.} This VEV breaks 
the axial combination of the gauge symmetry on the D6-branes, in 
agreement with our comments about $U(1)_C$ above. Moreover, this 
picture also suggests the correct D2-brane field theory deformation 
\cite{aklm},
\be
{\cal W}=M_1\tilde{Q}_1Q_2+M_2\tilde{Q}_2Q_1
\label{defcon1}\ee
which leads to the scalar potential,
\ba
V^\prime_A&=&e^2(|q_1|^2-|q_2|^2-|\tilde{q}_1|^2+|\tilde{q}_2|^2)^2
+e^2|q_1\tilde{q}_1|^2 +e^2|q_2\tilde{q}_2|^2 \nn\\ 
&& +\phi^2w_1^\dagger w_1+|\psi_1\tilde{q}_1+M_2\tilde{q}_2|^2 
+|\psi_1q_1+M_1q_2|^2 \nn\\ 
&& +(\phi-|t|)^2w_2^\dagger w_2 + |\psi_2\tilde{q}_2+M_1\tilde{q}_1|^2 
+|\psi_2q_2+M_2q_1|^2
\nn\ea
This coupling leaves the $U(1)_J\times U(1)_1\times U(1)_2$ 
symmetry of the probe theory unbroken, but is expected to alter 
the enhanced symmetry group at strong coupling. The 
$U(1)_F$ symmetry is broken by this interaction, in 
agreement with expectations. Moreover, for $M_1M_2\neq 0$, 
it lifts the Higgs branch. 

Finally, let us show that the interaction \eqn{defcon1} 
does indeed correspond to the deformation \eqn{ccd}. To see 
this, note that the hypermultiplets 
arising from 2-6 strings, which are expected to become 
massless on the curve \eqn{ccd}. The two hypermultiplets 
of Theory A contain between them four Dirac fermions, 
whose mass matrix is,
\be
{\cal M}_F=\left(\begin{array}{cccc} 0 & \psi_1^\dagger 
& 0 & M_2^\dagger \\ \psi_1 & 0 & M_1 & 0 \\ 
0 & M_1^\dagger & 0 & \psi_2^\dagger \\ 
M_2 & 0 & \psi_2 & 0 \end{array}\right)
\ee
The curve $L$ on which the D6-branes lies is determined by the 
zero locus of ${\cal M}_F$,
\be
\det{\cal M}_F=|\psi_1\psi_2-M_1M_2|^2=0
\nn\ee
which agrees with \eqn{ccd} for the choice of mass parameters 
$M_1M_2=\rho$.

This completes the D-brane probe in the L-picture. However, 
in attempting to move further around the circle of Figure \ref{figa},
we meet an obstacle. In particular, the 
theory on a D-brane probe transverse to $X$ is, to our 
knowledge, undetermined. (For related work, see \cite{aklm}). 
We may attempt to determine the mirror theory using purely 
field theoretical techniques \cite{ahiss}. The mirror of 
the operator \eqn{defcon1} is 
\ba
\Delta{\cal W}=M_1V_++M_2V_-
\label{defcon17}\ea
where $V_\pm$ are the chiral superfields parameterizing the 
quantum corrected Coulomb branch of Theory B. They are related 
to vortex creation operators \cite{ahiss} and are poorly understood. 
It would certainly be interesting check that the Higgs branch 
of Theory B with the above deformation does indeed reproduce the 
deformed conifold.

\section{$G_2$ Holonomy}
\label{gsec}

In this section we consider M-theory on manifolds of
$G_2$ holonomy. We start by recalling the condition on 
D6-branes for the preservation of 
${\cal N}=1$ supersymmetry in four dimensions \cite{BDL}. 
Each of these IIA backgrounds lifts to M-theory compactified 
on a non-compact manifold $X$ of $G_2$-holonomy. It is 
a simple matter to write down the theory on a D2-brane 
probe of these models. Employing our mirror pairs of 
three-dimensional gauge theories discussed in Section 
2, we find an algebraic quotient description of the 
manifold $X$ as the Higgs branch. We stress that this 
procedure does not the provide $G_2$ holonomy metric 
on $X$. It does, however, give a simple construction of 
manifolds admitting a metric of $G_2$ holonomy, and may 
be employed as a linear sigma model for these purposes.

In the remainder of this section we give several examples 
of our construction, showing how it reproduces the 
cones over ${\mathbb C}{\bf P}^3$ and $SU(3)/U(1)^2$ 
discussed in \cite{aw} as well as some of the results 
of \cite{bobed}. We further use our technique to describe 
new $G_2$ manifolds arising from orthogonally intersecting 
D6-branes.

\subsection{Intersecting Special Lagrangian Planes}
\label{gseca}

Consider M-theory compactified on a manifold $X$ of $G_2$ holonomy.
When the M-theory circle, ${\bf S}^1\cong U(1)$, is embedded in $X$,
the resulting IIA string theory background consists of a
six-manifold $X/U(1)$, possibly with RR fluxes. Fixed points of
the $U(1)$ action lead to D6-branes. 
In the present case of a $G_2$ manifold, the condition on $L$ to
preserve four supercharges on the D6-brane is that $L$
must describe a special Lagrangian submanifold in $\mathbb{R}^6$.

In order to write this condition explicitly, let us
define coordinates on the ``interesting'' part of
the space, $\mathbb{R}^6$, parameterized by $x^3, x^4, x^5, x^7, x^8, x^9$.
It is convenient to pair them into complex coordinates:
\be
z_k = x^{2+k} + i x^{6+k}, \quad k=1,2,3
\ee
and introduce a complex structure (a holomorphic
$SU(3)$ invariant 3-form):
\be
\Omega = dz_1 \wedge dz_2 \wedge dz_3
\ee
We also need a K\"ahler form on $\mathbb{R}^6 \cong \mathbb{C}^3$:
\be
J = dz_1 \wedge d \bar z_1 + dz_2 \wedge d \bar z_2
+ dz_3 \wedge d \bar z_3
\ee

Now, the special Lagrangian condition on 3-submanifold
$L \subset \mathbb{R}^6$ implies two conditions:
$a)$ the restriction of $J$ to $L$ vanishes (then $L$ is
said to be Lagrangian), and $b)$ the restriction of
$Im (e^{i \gamma} \Omega)$ to $L$ vanishes for some
real phase $\gamma \in [0,2 \pi)$.
Then,
\be
Re (e^{i \gamma} \Omega)
\ee
restricts on $L$ as
a volume form, and one says that $L$ is {\it calibrated}
with respect to $Re (e^{i \gamma} \Omega)$ \cite{HL}.
In particular, this condition defines an orientation of $L$:
\be
Re (e^{i \gamma} \Omega) \vert_L = {\rm vol} (L) > 0
\label{lorientation}
\ee
The simplest examples of a special Lagrangian curve $L$ 
is a collection of 3-planes linearly embedded into 
$\mathbb{R}^6 \cong X/U(1)$. They correspond to {\it flat} 
D6-branes intersecting at a point, and
lift to singular manifolds $X$ of $G_2$ holonomy.
Up to some obvious change of coordinates,
such a special Lagrangian 3-plane can be described
by a choice of three angles $\theta_k$, $k=1,2,3$,
corresponding to rotations of $L$ in the
$x^4-x^7$, $x^5-x^8$, $x^6-x^9$ plane, respectively.
Then, the special Lagrangian conditions is simply \cite{OT}:
\be
\theta_1 \pm \theta_2 \pm \theta_3 = 0,
\quad \mathrm{mod} \quad 2 \pi
\label{3kings}\ee
There are a few remarks in order here.
First, note that none of the angles can be zero (mod $2 \pi$),
for otherwise supersymmetry would be larger \cite{OT}.
Secondly, a `natural' choice of the orthogonal intersecting
branes breaks all the supersymmetry; indeed it is T-dual
to the well-known non-supersymmetric D0-D6 system.
More generally, the above condition can never be satisfied if all
$\theta_k = \pm \pi /2$.

\subsubsection*{Probe Theory and a $G_2$ Quotient Construction}

Consider a collection of $N$ flat D6-branes, each of which has 
world-volume directions,
\ba
D6_i && 123[47]_{\theta_1^i}[58]_{\theta_2^i}[69]_{\theta_3^i}
\ \ \ \ \ \ \ \ i=1,\ldots,N
\nn\ea
We probe this configuration with a D2-brane with spatial
world-volume in the $x^1-x^2$ plane. This breaks 
the supersymmetry of the D6-branes by a further half, 
resulting in  
${\cal N}=1$ supersymmetry in $d=(2+1)$ dimensions. For the 
singular case of intersecting, flat D6-branes, the theory on 
the D2-brane probe is simple to write down. The 2-2 strings give 
rise to the usual gauge field and seven scalars. Of these, there is 
one free ${\cal N}=1$ scalar multiplet parameterizing motion
in the $x^3$ direction common to all D6-branes. Further fields 
arise from the 2-6 strings. These give rise to $N$ hypermultiplets 
which, as usual, we denote as 
$w^\dagger_i\equiv (q_i^\dagger, \tilde{q}_i)$. Thus, we
have the interacting ${\cal N}=1$ supersymmetric
theory on the probe as
\begin{center}
{\bf Theory A:} $U(1)$ with 6 scalar multiplets and N hypermultiplets
\end{center}
where each hypermultiplet has charge $+1$ under the gauge field. 
The couplings of the hypermultiplets to the scalar multiplets 
are determined by the geometry of the D6-branes: each 
hypermultiplet couples minimally to the three scalar fields 
orthogonal to the corresponding D6-brane. We define the scalar 
fields $\phi_\alpha=x^{\alpha+3}$, $\alpha=1,\ldots,6$.
\be
f=\sum_{i=1}^N\sum_{c=1}^3 \sum_{\alpha=1}^6
W_i^\dagger\tau^c\,W_i\cdot{T}_{c,i}^\alpha 
\phi_\alpha
\ee
where the couplings are determined by the triplet of 
matrices,
\be
{T}^\alpha_{c,i}=-\sin\theta_c^i\,\phi_\alpha\,\delta_{c,\alpha} 
+\cos\theta_c^i\,\phi_{\alpha}\,\delta_{c,\alpha-3}\ \ \ \ \ \ c=1,2,3
\ee
The Coulomb branch of this theory, parameterized by the 
six real scalars $\phi_\alpha$, together with the dual 
photon $\sigma$, is a seven dimensional manifold $X$ that 
admits a metric of $G_2$ holonomy. Since Theory A is of the 
class of theories discussed in Section 2, we 
may simply write down the mirror theory whose Higgs 
branch is conjectured to give the $G_2$ manifold $X$,
\begin{center}
{\bf Theory B:} $U(1)^{N-1}$ with $3(N-2)$ scalar and $N$ 
hypermultiplets 
\end{center}
The $i^{\rm th}$ gauge group acts on the $i^{\rm th}$ 
hypermultiplet with charge $+1$, and the $(i+1)^{\rm th}$ 
hypermultiplet with charge $-1$. All other hypermultiplets are 
neutral. The Yukawa terms are of the same form as above,
\be
f= \sum_{i=1}^N\sum_{c=1}^3 \sum_{\rho=1}^{3N-6}
W_i^\dagger\tau^c\,W_i\cdot\hat{T}_{c,i}^\rho 
\phi_\rho
\nn\ee
where, as in Section 2, 
the triplet of coupling matrices are defined to satisfy,
\be
\sum_{c,i}\hat{T}^\rho_{c,i}T^\alpha_{c,i}=0\ \ \ \ \ \ \ \ 
\forall\ \rho,\alpha
\ee
The Higgs branch of this theory is parameterized by 
$w_i$, the $4N$ real scalars in the hypermultiplets, modulo the 
$(N-1)$ $U(1)$ gauge orbits. To this we add  the $3(N-2)$ 
real constraints coming from the Yukawa interactions,
\be
\sum_{i,c}\hat{T}^{\rho}_{c,i} w_i^\dagger\tau^c w_i = 0\ \ \ \ \ \ 
\rho=1,\ldots,3(N-2)
\label{g2c}\ee
This quotient construction yields a conical manifold which 
admits a metric of $G_2$ holonomy. In some cases the 
conical singularity may be (partially) resolved by adding 
constants to the right-hand side of \eqn{g2c}. This blows 
up two-cycles and, in the IIA picture, corresponds to 
translating the D6-branes. Note that when the Yuakawa 
matrices $\hat{T}$ fall into suitable $SU(2)$ triplets, 
the above method coincides with the toric hyperK\"ahler 
quotient construction, supplemented by a further quotient by a  
tri-holomorphic isometry to yield a manifold of dimension 
seven. This is the construction discussed by Acharya and 
Witten \cite{bobed}. However, in general, our charges differ. 
We now explain, in some detail, how 
this construction works in specific cases.

\subsection{2 D6 Branes: The cone over ${\bf \mathbb{C}P}^3$}
\label{gsecb}

\begin{figure}

\setlength{\unitlength}{0.9em}
\begin{center}
\begin{picture}(22,11)

\put(7,5){\line(1,0){18}}\put(7.5,6){D6$_1$}

%\put(16,-1){\line(0,1){12}}
%\put(13,-1){\line(1,2){6}}

\put(19,-1){\line(-1,2){6}}\put(11,9){D6$_2$}

\qbezier(7,5.3)(16,5)(12.7,11)\qbezier(19.3,-1)(16,5)(25,4.7)

\qbezier(15,7)(17,7)(18,5)
\put(17,7){${2 \pi \over 3}$}
\put(19.5,2.7){$L$}
\put(-1,0){\vector(1,0){3}}\put(-1,0){\vector(0,1){3}}
\put(2.3,-0.2){$x^{3,4,5}$}\put(-1.2,3.2){$x^{7,8,9}$}

\end{picture}\end{center}
\caption{Intersection of special Lagrangian D6-branes dual
to M-theory on $G_2$ holonomy cone over ${\bf \mathbb{C}P}(3)$.
Resolution of the conical singularity corresponds to deforming
intersecting D6-branes into a single smooth surface $L$.}
\label{figf}
\end{figure}
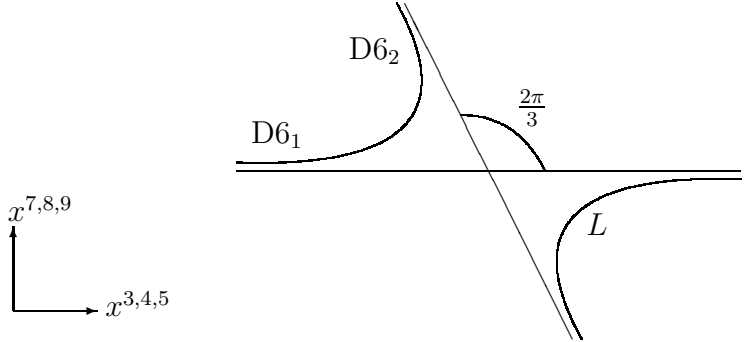

We start with the simplest configuration, consisting of 
only two D6-branes. The natural `symmetric' solution for the 
intersection of two D6-branes is when all the angles are equal:
\be
\theta_1 = \theta_2 = \theta_3 = {2 \pi \over 3}
\ee
More precisely, by this we mean that D6-branes are represented
by 3-planes inside $\mathbb{R}^6 \cong \mathbb{C}^3$, such
that in each $\mathbb{C}$-plane they look like straight lines
intersecting at $2 \pi / 3$ angle, see Figure \ref{figf}.
Put differently, one can take this picture in one copy
of $\mathbb{C}$, say in $x^3 - x^7$ plane, and tensor
it three times.

As we shall see below, this D6-brane intersection 
lifts in M-theory to a cone on ${\bf \mathbb{C}P}(3)$
with $G2$ holonomy metric \cite{aw}.
A resolution of the conical singularity yields a smooth
$G2$ manifold of the homotopy type \cite{gary}:
\be
X \cong \mathbb{R}^3 \times {\bf S}^4
\ee
As we described above, the proposed intersection
of D6-branes has the right amount of supersymmetry,
and as we explain below, it also has the right
structure of global symmetries.

Before we go into the identification of symmetries, let us make a
few general comments one should bear in mind.
M-theory on $G2$-manifold $X$ has certain global symmetries,
some of which come from gauge symmetries of the $C$-field,
while others are geometric symmetries of $X$ itself.
Let us denote the total group of these symmetries by $K_X$.
If $X$ develops a conical singularity the group of
global symmetries is in general enhanced to a group
of the corresponding symmetries of the six-dimensional
base $Y$; we call it $K_Y$, {\it cf.} \cite{aw}
%(In \cite{aw} the reasoning was inverse: it was
%shown that not all of the symmetries of $Y$ extend to symmetries of $X$.)
For example, in the $Y={\bf \mathbb{C}P}^3$ model
we are considering, the corresponding groups are:
\be
K_X = Sp(2) \times \mathbb{Z}_2
\ee
and
\be
K_Y = Sp(2) \times \mathbb{Z}_2 \times U(1)_C
\ee
where index $C$ refers to the fact that $U(1)_C$
comes from the gauge symmetries of the $C$-field.
After we reduce this model to type IIA via
compactification on a circle, the $U(1)_C$ global
symmetry can be understood as a gauge symmetry
on D6-branes (remember, that the diagonal
$U(1)$ decouples) \cite{aw}.

%{\bf As long as we have the right number of intersecting
%D6-branes in the L-picture, and as long as we correctly
%identify the deformations of our models, we should
%be able to automatically reproduce all the $U(1)$
%symmetries associated with $C$-field in M-theory,
%both for singular and smooth $X$.}

Turning to the geometric symmetries, after reduction from 
M-theory to type IIA, a geometric part of $K_Y$ gets broken 
to a subgroup:
\be
Sp(2) \to U(1)_J \times SU(2)
\ee
where $U(1)_J$ is the isometry of the M-theory circle.
%While not seen in IIA string theory, <- this is wrong, I think
It reappears in the probe picture 
as the dual photon. We will discuss this in more detail below. 
The $SU(2)$ factor is  to be identified with a geometric 
symmetry of the D6-brane configuration. Let us see how this 
appears. In the above notations for the spatial coordinates,
one D6-brane can be described by {\it real}
equations in $\mathbb{C}^3 \cong \mathbb{R}^6$:
\be
D6_1 : \quad Im (z_i) = 0
\ee
Once can easily check that the holomorphic 3-form
defined above indeed restricts to a volume form on
this special Lagrangian plane, with a certain orientation.

The second D6-brane can be described by a similar
set of linear equations:
\be
D6_2 : \quad Im (\omega z_i) = 0
\ee
where $\omega = \exp(2 \pi i / 3)$.
Again, the orientation is defined by \eqn{lorientation}.
Written in terms of real coordinates $x^m$,
these equations look like:
\be
D6_1 : \quad x^7 = x^8 = x^9 = 0
\label{d61}\ee
for the first D6-brane, and
\be
D6_2 : \quad
{1 \over 2} x^7 + {\sqrt{3} \over 2} x^3 =
{1 \over 2} x^8 + {\sqrt{3} \over 2} x^4 =
{1 \over 2} x^9 + {\sqrt{3} \over 2} x^5 = 0
\ee
for the second D6-brane. 

These equations, defining the D6-brane locus $L$,
are invariant under $SU(2) \cong SO(3)$ symmetry group,
as expected for the ${\bf S}^4$ model at hand.
To see this, let $a$ be an element of $SO(3)$,
and introduce two 3-vectors:
\be
\vec{\phi}_1 = (x^7, x^8, x^9)^T, \quad
\vec{\phi}_2 = (x^3, x^4, x^5)^T
\ee
Then, the $SO(3)$ action is realized as follows:
\be
SO(3) \colon \quad \vec{\phi}_1 \mapsto a \cdot \vec{\phi}_1,
\quad \vec{\phi}_2 \mapsto a \cdot \vec{\phi}_2
\ee
Since $SO(3)$ acts in the same way on both $\vec{\phi}_1$ and
$\vec{\phi}_2$, the above equations manifestly remain invariant.
Hence, the $SO(3) \cong SU(2)$ is a symmetry of
our D6-brane configuration, in agreement with
proposed relation to M-theory on $G2$ manifold $X$. Note that
any other solution to the special lagrangian equations \eqn{3kings}
would not have this property.

It remains to understand the supersymmetric deformation 
of this model away from the singular limit. The singularity 
of the manifold $X$ may be resolved to have topology 
$\mathbb{R}^3\times {\bf S}^4$. Comparing with equations 
\eqn{genhom}, the D6-branes must deform to lie on a curve 
with topology,
\be
L \cong \mathbb{R} \times {\bf S}^2
\label{l2}\ee
asymptotic to a union of the special Lagrangian 3-planes:
$$
L_0 = \{ \vec \phi_1 =0 \} \cup \{ (\vec \phi_1 + \sqrt{3} \vec \phi_2) = 0 \}
$$
In fact, we may identify this curve $L$. The supersymmetry condition
implies that $L$ is special Lagrangian submanifold in
$\mathbb{C}^3 = \mathbb{R}^3 \times \mathbb{R}^3$,
and $SU(2) \cong SO(3)$ symmetry implies that $L$
is homogeneous in coordinates $\vec \phi_1$, $\vec \phi_2$:
\be
\vec{\phi}_1\cdot\vec{\phi}_2=-|\vec{\phi}_1||\vec{\phi}_2|
\label{antipar}\ee
The special Lagrangian condition gives one extra constraint \cite{joycesymm}:
\be
|\vec \phi_1|~ (3 |\vec \phi_2|^2 - |\vec \phi_1|^2) = \rho
\label{cp3lsurf}
\ee
that completely defines a 3-dimensional variety $L$,
for every value of the real parameter $\rho$.
Most importantly, for non-zero values of the deformation
parameter $\rho$ this equation defines a smooth surface with 
topology \eqn{l2}. In the singular limit $\rho \to 0$, we recover a 
configuration
of two intersecting D6-branes described by $L_0$.
As expected, the boundary of $L$ (the same as boundary of $L_0$)
is a union of two spheres:
$$
F = {\bf S}^2 \cup {\bf S}^2
$$
We therefore see that the moduli space of M-theory on $X$ 
consists of a single parameter (suitably complexified). It 
was further shown in \cite{aw} that this deformation is 
${\bf L}^2$-normalizable; the associated scalar has finite kinetic 
terms. This is in contrast to the conifold case discussed in the 
previous section. It would be interesting to understand this from the 
L-picture.

\subsubsection*{Probe Theory}

Having discussed the various properties of the L-picture, we 
now come to the theory on a D2-brane probe with  spatial
world-volume in the $x^1-x^2$ plane. As discussed at the 
beginning of this section, the theory on the world-volume 
may be easily written down,
\begin{center}
{\bf Theory A:} $U(1)$ with 6 scalar multiplets and 2 hypermultiplets
\end{center}
where the hypermultiplets arise from the 2-6 strings and are taken
to both have charge $+1$ under the
gauge field. 
The couplings between hypermultiplets and scalar multiplets are 
described in term of a real (non-holomorphic) superpotential,
\be
f=\sum_{a,i=1}^2 T^a_i\,W_i^\dagger\,\vec{\tau}\cdot\,\vec{\Phi}_aW_i
\label{yoyomar}\ee
where $W_i$ for each $i=1,2$ is the hypermultiplet,
expressed as a doublet of chiral multiplets;
$\vec{\tau}$ are the Pauli matrices; and $\vec{\Phi}_a$, 
$a=1,2$ are each a triplet of scalar multiplets with lowest components 
$\vec{\phi}_a$. The Yukawa coupling matrix is given by,
\be
T_i^a=\left(\begin{array}{cc} 1 & 0 \\ -\ft12 & -\ft{\sqrt3}2 \end{array}
\right)
\label{annie}\ee
where the hypermultiplet index $i=1,2$ labels the
rows and the scalar multiplet index $a=1,2$ labels the columns. 
Note, in particular, the minus signs in the second row which 
distinguish a brane rotated by $2\pi/3$ from a brane rotated by $\pi/3$.
The latter possibility is equivalent to replacing the brane 
with an anti-brane, and breaks supersymmetry. The bosonic sector 
of the probe theory does not notice the difference between these 
two possibilities. 
However, as we shall see momentarily, the minus signs are crucial for the 
fermionic sector. In terms of
component fields, the superpotential leads to the scalar potential, 
%\begin{eqnarray}
\be
V = e^2 (w_1^{\dagger}\vec\tau w_1+\ft12w_2^{\dagger}\vec\tau w_2)^2
+ \ft34 e^2 (w_2^{\dagger} \vec \tau w_2)^2
+ w_1^{\dagger} w_1 |\vec \phi_1 |^2
%+ |\psi_1|^2) \nn\\ &&
+ {1 \over 4} w_2^{\dagger} w_2 |\vec \phi_1 + \sqrt{3} \vec \phi_2 |^2
%+ |\psi_1 + \sqrt{3} \psi_2|^2\right)
\label{cp3pot}
\ee
%\end{eqnarray}
One could further add FI terms to this theory. They lift
the Coulomb branch of interest and correspond to background
NS-NS B-fields, in the string theory picture. We set them
to zero here. Any diagonal bare mass parameters can be absorbed 
by a shift of one of the scalars. 

The symmetry group of the classical Lagrangian is
$U(1)_J \times SU(2) \times U(1)_F$. As usual the
J-symmetry acts solely on the dual
photon. The Coulomb branch scalars, $\vec{\phi}_1$ and $\vec{\phi}_2$
transform in the ${\bf 3}$ of the $SU(2)$, while the hypermultiplet scalars 
are singlets. Finally the $U(1)_F$ flavor symmetry
acts on the hypermultiplets orthogonally to the gauge symmetry:
we take $w_1$ to have charge $+1$ and $w_2$ charge $-1$.

%\begin{table}\begin{center}
%\begin{tabular}{|c|c|c|c|}
%\hline
%& $U(1)_J$ & $SU(2)$ & $U(1)_F$ \\
%\cline{1-4}
%$\phi_{4,5,6}$ && {\bf 3} & \\
%\cline{1-4}
%$\phi_{7,8,9}$ && {\bf 3} & \\
%\cline{1-4}
%$w_1$ & & & +1 \\
%\cline{1-4}
%$w_2$ & & & -1 \\
%\hline
%\end{tabular}\end{center}
%\caption{Representations of the massless fields in Theory A.}
%\end{table}

Theory A has a moduli space of vacua in which the
gauge group remains unbroken: it is the Coulomb branch.
The D-terms of the potential (the first two terms in
\eqn{cp3pot}) require us to set $w_1=w_2=0$, ensuring that the
scalar potential vanishes for any value of $\vec{\phi}_1$
and $\vec{\phi}_2$. The dual photon supplies the final
moduli, bringing us to the requisite seven. As discussed in the 
introduction, the conservation of parity in this theory ensures that 
the Coulomb branch survives at the quantum level. Since the
dual photon manifestly has a different origin to the
other six scalars,  the Coulomb branch has a natural
decomposition into a ${\bf S}^1$ fiber over $\mathbb{R}^6$.
The fiber degenerates at the positions of the D6-branes, as
suggested by a one-loop computation in the gauge theory.
At finite energies the Coulomb branch
inherits the $U(1)_J\times SU(2)$ isometry from
the field theory Lagrangian. In the strong coupling 
limit, we expect this symmetry to be enhanced to $SO(5)$.

%Let is examine the one-loop effective
%action. The metric on the moduli space is given
%by,
%\be
%ds^2= H_\sigma^{-1}(d\sigma+\vec{\omega}_1\cdot d\vec{\phi}_1
%+\vec{\omega}_2\cdot d\vec{\phi}_2)^2 + H_1 d\vec{\phi}_1^2+
%H_2 d\vec{\phi}_2^2
%\label{g2loop}\ee
%where $\vec{\nabla}\times \vec{\omega_a} = {\vec \nabla} H_a$, for $a=1,2$
%and the $H$'s are given by the functions,
%\ba
%H_\sigma&=&\frac{1}{e^2}+\frac{1}{|\vec{\phi}_1|}
%+\frac{2}{|\vec{\phi}_1+\sqrt{3}\vec{\phi}_2|} \nn\\
%H_1&=&\frac{1}{e^2}+
%\frac{1}{|\vec{\phi}_1|}+\frac{\ft12}
%{|\vec{\phi}_1+\sqrt{3}\vec{\phi}_2|} \label{hsfour} \\
%H_2&=&\frac{1}{e^2}+\frac{\ft32}{|\vec{\phi}_1+\sqrt{3}\vec{\phi}_2|}
%\nn\ea

The above probe theory describes the manifold $X$ at the 
point where it develops a conical singularity. What deformation 
of Theory A smoothens out this singularity? As we shall see, 
the necessary change of the probe theory is more analogous 
to the deformation of the conifold, than to the small 
resolution of the conifold. Our clue is the $U(1)_F$ symmetry 
of the probe theory. This does not act as an isometry on the 
Coulomb branch, arises from a $U(1)_C$ C-field symmetry of 
M-theory on $X$. From the discussion of \cite{aw} 
we know that the $U(1)_C$ symmetry is generically broken. It is 
restored only when $X$ develops a conical singularity. The 
situation is therefore very similar to the deformed conifold picture. 
To describe the probe theory on the resolved locus \eqn{cp3lsurf},
we are looking for a deformation of Theory A which breaks the 
$U(1)_F$ flavor symmetry while simultaneously preserving the existence,
not only of the Coulomb branch, but also of the $SU(2)$
symmetry. We claim that the correct deformation is once again of the 
form \eqn{defcon1} but, without the luxury of ${\cal N}=2$ 
supersymmetry, we need not restrict ourselves to holomorphic 
superpotentials. However, the $SU(2)$ symmetry does place 
severe constraints on the possible couplings. To see this, note that 
while the hypermultiplet scalars do not transform under this global 
symmetry, the same is not true of the hypermultiplet fermions. This is 
apparent if we examine the Yukawa couplings in Theory A. With supersymmetry 
broken to ${\cal N}=1$, it is most natural to work with real Majorana rather 
than complex Dirac fermions. In $d=(2+1)$, each hypermultiplet contains
four Majorana fermions, $\lambda^p$, $p=1,2,3,4$. The couplings to the 
vector multiplet scalars are,
\be
\left(\lambda_1^p\vec{\phi}_1\lambda_1^q
- \lambda_2^p(\ft12\vec{\phi}_1+\ft{\sqrt{3}}{2}\vec{\phi}_2)
\lambda_2^q\right)\cdot\left(\vec{V}_{pq} + i\vec\eta_{pq}\right)
\label{g2yuk}\ee
where the $\vec{\eta}$ are the self-dual $4\times 4$ 
anti-symmetric 't Hooft matrices, and the $\vec{V}$ are 
$4\times 4$ symmetric matrices given by,
\ba
&&\eta_1=i\tau^2\otimes \tau^1,\quad\eta^2=-i\tau^2\otimes\tau^3,\quad 
\eta^3=1\otimes i\tau^2 \nn\\
&&V^1=-\tau^1\otimes \tau^3,\quad V^2=-\tau^1\otimes\tau^1 ,\quad 
V^3=\tau^3\otimes 1
\nn\ea
which satisfy the relations
\be
[\eta^i,\eta^j] = -2\epsilon^{ijk}\eta^k\ \ \ \ ; \ \ \ \ 
[\eta^i,V^j] = -2\epsilon^{ijk}V^k\ \ \ \ ;\ \ \ \ 
[V^i,V^j] = 2\epsilon^{ijk}\eta^k
\ee
Note that the anti-symmetric $\vec{\eta}$ terms in the Yukawa 
coupling vanish in the abelian theory of interest. The relative minus 
sign between the two couplings follows from the superpotential couplings 
\eqn{annie}. 
Under the $SU(2)$ symmetry, $\vec{\phi}_a$ transform 
in a triplet which implies the transformations of the fermions,
\ba
SU(2)&:& \vec{\delta}\lambda_1^p = \vec{\eta}^p_{\ q}\lambda_1^q, 
\quad\quad \vec{\delta}\lambda_2^p = \vec{\eta}^p_{\ q}\lambda_2^q, 
%\nn\\
%U(1)_F&:& \delta\lambda_1^p = \epsilon(\eta^3)^p_{\ q}\lambda_1^q,\quad\quad 
%\delta\lambda_2^p = -\epsilon(\eta^3)^p_{\ q}\lambda_2^q 
\nn\ea
In these conventions the gauge and flavor transformations are 
implemented by use of the anti-symmetric, anti-self dual 
't Hooft matrix, $\bar{\eta}_3=-i\tau^3\otimes\tau^2$, satisfying 
$[\bar{\eta}^3,\vec{\eta}]=[\bar{\eta}^3,\vec{V}]=0$,
\ba
%SU(2)&:& \vec{\delta}\lambda_1^p = \vec{\eta}^p_{\ q}\lambda_1^q, 
%\quad\quad \vec{\delta}\lambda_2^p = \vec{\eta}^p_{\ q}\lambda_2^q, 
%%\nn\\
U(1)_G:& \delta_G\lambda_1^p = (\bar{\eta}^3)^p_{\ q}\lambda_1^q,\quad\quad 
\delta_G\lambda_2^p = (\bar{\eta}^3)^p_{\ q}\lambda_2^q \nn\\
U(1)_F:& \delta_F\lambda_1^p = (\bar{\eta}^3)^p_{\ q}\lambda_1^q,\quad\quad 
\delta_F\lambda_2^p = -(\bar{\eta}^3)^p_{\ q}\lambda_2^q 
\nn\ea
From these transformation laws, we may deduce the correct 
generalization of the coupling \eqn{defcon1}: it is a superpotential 
that results in the real fermion mass terms, 
\be
\lambda_1^pX_{pq}\lambda_2^q
\label{defg2}\ee
where $X$ must satisfy $[X,\vec{\eta}]=[X,\bar{\eta}^3]=0$ for 
the coupling to have the correct quantum numbers. The only 
solution to this equation is
\be
X=M_1 \cdot {\bf 1} + M_2 \cdot \bar{\eta}^3,
\ee 
where $M_1$ and $M_2$ are real parameters, and ${\bf 1}$ is the unit 
$4\times 4$ matrix. The hypermultiplet 
fermion mass matrix arising from the Yukawa coupling \eqn{g2yuk}, 
together with this deformation is an $8\times 8$ real, symmetric 
matrix ${\cal M}_F$ with determinant,
\be
\det{\cal M}_F=\left( M^4+M^2
(\vec{\phi}_1 +\sqrt{3}\vec{\phi}_2)+\ft14
|\vec{\phi}_1|^2|\vec{\phi}_1+\sqrt{3}\vec{\phi}_2|^2
\right)^2
\ee
with $M^2=M_1^2+M_2^2$. The zero locus of the determinant
in the full quantum theory is expected to reproduce
the special Lagrangian locus $L$ given by equations 
\eqn{antipar} and \eqn{cp3lsurf}.
Our (rather conservative) hope is that the semi-classical
analysis gives at least the right topology of $L$. At first 
sight it seems rather difficult to reproduce the two 
equations defining $L$ from the root of an order eight 
polynomial. However, for the fermion mass matrix above, 
the equation $\det{\cal M}_F =0$ only has solutions when 
$\phi_1$ and $\phi_2$ are anti-parallel,
\be
\vec{\phi}_1\cdot\vec{\phi_2}=-|\vec{\phi}_1||\vec{\phi_2}|
\ee
and the roots then satisfy the further constraint
\be
M^2 = |\vec{\phi}_1|(|\sqrt{3}|\vec{\phi}_2|-|\vec{\phi}_1|)\geq 0
\ee
Comparing this locus with the special Lagrangian curve $L$ given 
in equations \eqn{antipar} and \eqn{cp3lsurf}, we see that the 
curves do not precisely agree. As mentioned in the introduction, we 
have no reason to expect exact agreement. Indeed, the zero locus of 
the fermion mass matrix yields the one-loop correction to the 
degeneration of the dual photon fiber, and we have no reason to 
expect higher loop corrections to be negligible near the 
intersection of the D6-branes. Nevertheless, the topology of the 
locus derived from field theory does coincide with the special 
Lagrangian manifold $L$. This is heartening.

Finally, we note that if the minus sign in \eqn{g2yuk} is replaced
with a plus --- corresponding to branes rotated by $\pi/3$,
breaking supersymmetry --- 
the fermion masses vanish on a locus with $\vec{\phi}_1$ and 
$\vec{\phi}_2$ parallel. This is not a special Lagrangian deformation, 
reflecting the breaking of supersymmetry in that case. 

\subsubsection*{IIB Brane Construction and the Mirror Theory}

Let us now continue our way around Figure \ref{figa}, and derive the 
mirror theory whose Higgs branch yields the $G_2$ manifold $X$. We 
have presented two methods to derive the mirror theory:
using field theory techniques and D-brane models.
Here we use both. Firstly, from the 
discussion of Section \ref{qftsec}, it is simple to write down 
the field theory mirror,
\begin{center}
{\bf Theory B:} $U(1)$ with 2 hypermultiplets.
\end{center}
where the hypermultiplets have charge $+1$ and $-1$
respectively. This theory has a seven real-dimensional
Higgs branch $X=\mathbb{C}^4/U(1)$. Since $\mathbb{C}^4$ may
be thought of as the cone over ${\bf S}^7$, and
${\bf S}^7/U(1)\cong {\bf \mathbb{C}P}^3$, the resulting 
Higgs branch is topologically the cone over $Y$, with 
$$
Y = {\bf \mathbb{C}P}^3
$$
in agreement with the claims made in this section. Note that 
this theory yields the singular, conical limit of $X$. 
The above procedure can also be performed for the smooth locus 
$L$ given in \eqn{cp3lsurf}, resulting in brane diamond 
configuration analogous to those of \cite{aklm}, but with lower 
supersymmetry. As with the deformed conifold case, we do not 
well understand the mirror theory whose Higgs branch yields 
the blown up $G_2$ manifold. Presumably, as in the conifold 
case, the mirror of the supersymmetric operator \eqn{defg2} 
is related to the vortex creation operator.

\begin{figure}

\setlength{\unitlength}{0.9em}
\begin{center}
\begin{picture}(22,11)
\qbezier(8,6)(13.7,8)(19.5,6)\put(9.5,7){D3}
\qbezier(8,6)(13.7,4)(19.5,6)
\put(17.5,2){\line(1,2){4}}\put(22,10){NS5$_2$}
\put(8,2){\line(0,1){8}}\put(8.5,10){NS5$_1$}
%\put(14.6,5.5){$\updownarrow$}
%\put(19,4.8){$\updownarrow$}
\put(-1,0){\vector(1,0){3}}\put(-1,0){\vector(0,1){3}}
\put(-1,0){\vector(1,1){3}}
\put(2.3,-0.2){$x^{6}$}\put(-1.2,3.2){$x^{7,8,9}$}\put(2.3,3.2){$x^{3,4,5}$}

\end{picture}\end{center}
\caption{IIB Brane model for Theory B:
$U(1)$ with 2 hypermultiplets.}
\label{figg}
\end{figure}
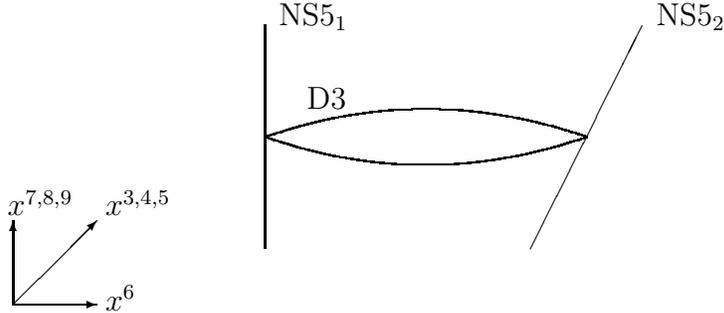

We can also derive this theory using brane configurations by 
T-dualizing to IIB along the $x^6$ direction, common 
to both D6-branes, and transverse to the D2-brane. 
If T-duality is performed in the singular limit of 
flat intersecting D6-branes, the resulting brane configuration 
consists of D5-branes, with a 
single D3-brane, each with world-volume directions,
\ba
D3 && 126 \nn\\
D5_1 && 12789 \nn\\
D5_2 && 12 [37]_{\theta_1} [48]_{\theta_2} [59]_{\theta_3}
\nn\ea
where
\be
\theta_1 = \theta_2 = \theta_3 = {2 \pi \over 3}
\ee
A Wilson line for the flavor symmetry ensures the D5-branes are 
separated along the $x^6$ circle. 
In the low-energy limit, the theory on the D3-brane probe 
of this periodic (``elliptic'') model coincides with Theory A 
described above. More interesting is the D3-brane world-volume 
theory upon performing an S-duality. Following Hanany and 
Witten \cite{hw}, this should result in the mirror theory. 
The IIB brane configuration is depicted in Figure \ref{figg}.
The world-volumes are,
\ba
D3 && 126 \nn\\
NS5_1 && 12789 \nn\\
NS5_2 && 12 [37]_{\theta_1} [48]_{\theta_2} [59]_{\theta_3}
\nn\ea
The theory on the D3-brane probe may be read from the brane 
picture. Each segment yields a $U(1)$ vector field, as well 
as three scalar multiplets. Each of these scalar multiplets 
acquires a mass due to the relative orientation of the NS5-branes. 
Further, there exist 
hypermultiplets arising from strings stretched across each 
NS5-brane. These have charges $(+1,-1)$ and $(-1,+1)$, 
respectively, under the two gauge factors, ensuring that 
the diagonal $U(1)$ decouples. The massless fields are 
simply those of Theory B above. Finally, note that the brane 
picture also has the possibility of a superpotential, arising 
from integrating out massive neutral scalars,
\be
f=\sum_{a,i=1}^2 \tilde{S}^a_i\,W_i^\dagger\,\vec{\tau}\cdot\,
\vec{\Phi}_aW_i +\sum_{a=1}^2 m_a\vec{\Phi}_a\cdot\vec{\Phi}_a
\ee
where the scalars couple as
\be
\tilde{S}^a_i=\left(\begin{array}{cc} +1 & -1 \\ -1 & +1 \end{array}
\right)
\ee
By symmetry the scalar masses are $m_1=-m_2$, ensuring that 
upon integrating out the scalar multiplets there is no 
remaining superpotential. At low-energies the resulting 
dynamics is therefore that of Theory B.

\subsubsection*{Generalizations: The cone over ${\bf W \mathbb{C} P}^3$}

There is an obvious generalization to this model in which 
we add further D6-branes, lying parallel to the ones 
already introduced. Consider the 
IIA background consisting of $p+q+2$ D6-branes with 
orientation,
\ba
D2 && 12 \nn\\
(p+1)\times D6_1 && 126789 \nn\\
(q+1)\times D6_2 && 126 [37]_{\theta_1} [48]_{\theta_2} [59]_{\theta_3}
\nn\ea
where, as before
\be
\theta_1 = \theta_2 = \theta_3 = {2 \pi \over 3}
\ee
We claim that this IIA brane configuration lifts to 
M-theory on the singular $G_2$ manifold $X$ which is the 
cone over the weighted projective space
${\bf W \mathbb{C} P}^3_{p,p,q,q}$. These manifolds were 
discussed in \cite{aw,bobed} in the quest for $G_2$ 
compactifications yielding four dimensional chiral fermions. 
The manifold $X$ has a two co-dimension 4 singularities of ALE type: 
an $A_p$ and an $A_q$. If M-theory is compactified on $X$, these
singularities support a seven dimensional $SU(p)$ and
$SU(q)$ gauge connection respectively. The intersection
of this singularity supports a chiral fermion in the 
$({\bf p},{\bf q})$ representation of $SU(p)\times SU(q)$. 

The theory on a D2-brane probe of this D6-brane 
configuration is,
\begin{center}
{\bf Theory\ A:} $U(1)$ with 6 scalars and $N+2$ hypermultiplets
\end{center}
where $N=p+q$. All hypermultiplets have charge $+1$ under the the 
gauge fields. The 6 scalar multiplets pair up into two triplets 
as described in the previous section \eqn{yoyomar}. 
The first $p+1$ hypermultiplets have Yukawa couplings with the
first triplet, while the remaining $q+1$ chirals couple to both 
triplets with charges \eqn{annie}. 
The singularity where the D6-branes meet may be partially 
resolved by separating the D6-branes. From the probe perspective, 
this corresponds to adding mass terms for the hypermultiplets, 
and results in 
$pq$ singularities of the type we met in the previous 
section. These may further be resolved by introducing 
operators in the probe theory of the form \eqn{defg2}. 

The simplest way to  see that this background indeed lifts 
to the cone over ${\bf W \mathbb{C} P}^3_{p,p,q,q}$
is to examine the mirror three-dimensional gauge theory. Again, 
we perform T- and S-dualities to type IIB, to find a configuration 
of $N$ NS5-branes, and a single D3-brane. In this case we find 
a further subtlety: the theory on the D3-brane depends on the 
relative ordering of the NS5-branes around the circle. Here we 
concentrate on the simplest case where all branes with the same 
orientation are adjacent. (It would be interesting to examine whether 
exchanging brane positions results in Seiberg-like duality 
for these three-dimensional gauge theories \cite{toric}). The 
interacting ${\cal N}=1$ mirror theory may then simply be read from 
the brane picture:
\begin{center}
{\bf Theory\ B:} $U(1)^{N+1}$ with $3N$ scalars and $N+2$ hypermultiplets
\end{center}
Notice that the number of gauge fields is one more than the 
number of scalar triplets. In fact, we may combine the scalar 
multiplets with $N$ of the gauge fields into 
$U(1)^N=U(1)^p\times U(1)^q$ ${\cal N}=4$ vector multiplets. 
These couple to the hypermultiplets in an 
${\cal N}=4$ invariant fashion as follows: the first $p+1$ 
hypermultiplets are charged only under $U(1)^p$, while the 
remaining $q+1$ multiplets are charged under $U(1)^q$. 
The charges of each subset are determined by the quiver diagrams of 
$SU(p+1)$ and $SU(q+1)$ respectively 
(where in each case the overall, free, gauge field is ignored). 
Before taking into account the remaining ${\cal N}=1$ $U(1)$ vector 
multiplet, the resulting Higgs branch is therefore the hyperK\"ahler 
8-manifold given by the direct product  
$\mathbb{C}^2/ \mathbb{Z}_p\times \mathbb{C}^2/ \mathbb{Z}_q$.

The extra $U(1)$ action in Theory B reduces the Higgs branch 
to a seven dimensional manifold. It acts as a $U(1)$ quotient 
of $\mathbb{C}^2/ \mathbb{Z}_p\times \mathbb{C}^2/ \mathbb{Z}_q$, 
which acts with charge $+1$ on the $(p+1)^{\rm th}$
hypermultiplet, and charge $-1$ on the $(p+2)^{\rm th}$,
with all others neutral. This is precisely the construction 
\cite{bobed} where manifolds admitting $G_2$ holonomy 
metrics were constructed by quotienting hyperK\"ahler 
manifolds by a tri-holomorphic isometry. This particular 
example was considered in \cite{bobed} where it was shown that 
the Higgs branch is indeed a cone over 
${\bf W \mathbb{C} P}^3_{p,p,q,q}$.

\subsection{3 D6 Branes: The cone over $SU(3)/U(1)^2$}
\label{gsecc}

We turn now to the second example of Atiyah and Witten \cite{aw}, 
the cone over the flag manifold $Y=SU(3)/U(1)^2$. A resolution 
of this singular space yields a smooth $G_2$ manifold of 
homotopy type \cite{gary}
\be
X\cong \mathbb{R}^3\times \mathbb{C}\bf{P}^2
\nn\ee
Atiyah and Witten \cite{aw} argue that the moduli space 
of M-theory compactified on this space contains three 
components, intersecting at a singular point and rotated 
by a triality symmetry. In the following we shall confirm 
this scenario using the D-brane probe theory.

\begin{figure}

\setlength{\unitlength}{0.9em}
\begin{center}
\begin{picture}(22,11)
\put(7,5){\line(1,0){18}}\put(23,5.5){D6$_1$}
\put(13,-1){\line(1,2){6}}\put(16,9){D6$_3$}
\put(19,-1){\line(-1,2){6}}\put(11,9){D6$_2$}
\qbezier(7,5.3)(16,5)(12.7,11)\qbezier(19.3,-1)(16,5)(25,4.7)

%\put(17,7){${2 \pi \over 3}$}
\qbezier(15,7)(17,7)(17.5,5)
%\qbezier(15.2,3.4)(17,3)(17.3,5)
%\qbezier(17,7)(18,6)(18,5)
\put(18,6){${2 \pi \over 3}$}

\put(-1,0){\vector(1,0){3}}\put(-1,0){\vector(0,1){3}}
\put(2.3,-0.2){$x^{3,4,5}$}\put(-1.2,3.2){$x^{7,8,9}$}

\end{picture}\end{center}
\caption{Intersection of special Lagrangian D6-branes dual
to M-theory on $G_2$ holonomy cone over $SU(3)/U(1)^2$.
Resolution of the conical singularity corresponds to deforming
intersecting D6-branes into a smooth surface $L$ with two
connected components.}
\label{figh}
\end{figure}
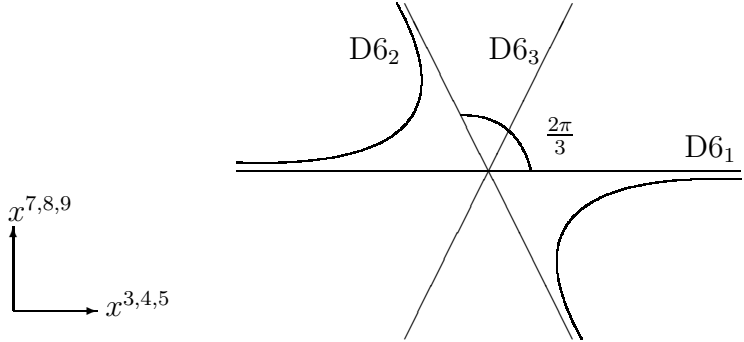

There is a natural guess for the D6-brane intersection of the 
singular conical manifold $X$. This is based on observation 
that adding an extra D6-brane oriented at angle $4 \pi / 3$
to the previous example does not break supersymmetry further.
Note, due to the special Lagrangian condition \eqn{lorientation}
a similar configuration of D6-branes rotated by angle $2 \pi / 6$
would be non-supersymmetric; the D6-branes would have opposite
orientation, {\it i.e.} correspond to anti-branes.
Summarizing, we obtain a configuration of three intersecting
D6-branes, which look like 3-planes intersecting at angles
$2 \pi / 3$ in directions $x^3 - x^7$, $x^4 - x^8$, $x^5 - x^9$,
see Figure \ref{figh}.

There is a simple way to obtain this configuration, which
makes it clear that the amount of unbroken supersymmetry is
as in the previous example.
Suppose, we start with a single D6-brane, say D6$_1$
in Figure \ref{figh}, in the flat ten-dimensional space-time.
This configuration is $1/2$ BPS. Now, let us orbifold this space
by a $\mathbb{Z}_3$ group acting as follows:
\be
\mathbb{Z}_3 \colon \quad z_k \mapsto \omega z_k, \quad
k = 1,2,3
\ee
where, as before, $\omega$ is denotes a cube root of unity,
$\omega = \exp (2 \pi i /3)$.
Since this $\mathbb{Z}_3$ is a subgroup of $SU(3)$,
this orbifold preserves $1/4$ of the original supersymmetry,
i.e. the resulting configuration is $1/8$-BPS
(equivalent to $\mathcal{N}=1$ supersymmetry in four dimensions).
We can think of the resulting configuration as an intersection
of three D6-branes on the covering space (which is again just
$\mathbb{C}^3$), corresponding to the $\mathbb{Z}_3$ mirror
images of the original D6$_1$ brane. Thus, we find
the proposed configuration of D6-branes, shown on Figure \ref{figh}.

Explicitly, we can describe the $i$-th D6-brane
by the set of linear equations:
\be
Im (\omega^{i-1} z_k)=0, \quad \forall k=1,2,3
\ee

We can write these equations in terms
of real coordinates $x^m$.
The equations for the first two D6-branes are
exactly the same as in the previous example:
\be
D6_1 : \quad x^7 = x^8 = x^9 = 0
\ee
\be
D6_2 : \quad
{1 \over 2} x^7 + {\sqrt{3} \over 2} x^3 =
{1 \over 2} x^8 + {\sqrt{3} \over 2} x^4 =
{1 \over 2} x^9 + {\sqrt{3} \over 2} x^5 = 0
\ee
and for the new D6$_3$ brane we have a similar
set of three linear equations:
\be
D6_3 : \quad
{1 \over 2} x^7 - {\sqrt{3} \over 2} x^3 =
{1 \over 2} x^8 - {\sqrt{3} \over 2} x^4 =
{1 \over 2} x^9 - {\sqrt{3} \over 2} x^5 = 0
\ee
Let us compare the symmetries of this D6-brane configuration 
with those expected from the quotient of the cone over the 
flag manifold $Y=SU(3)/U(1)^2$. The symmetry group of 
M-theory in the singular conical limit is,
\be
K_Y = SU(3) \times U(1)^2_C
\ee
The $U(1)_C^2$ symmetry coincides with the (axial) gauge symmetries 
of the D6-branes, while the $SU(3)$ geometrical symmetry is 
reduced upon squashing to 
\be
SU(3)\rightarrow U(1)_J\times SU(2)
\ee
As usual, the $U(1)_J$ corresponds to the M-theory circle, 
while the $SU(2)$ must survive as a geometrical symmetry 
of the IIA configuration. Indeed, our three intersecting 
D6-branes have precisely such an invariance:
\be
SO(3) \colon \quad \vec{\phi}_1 \mapsto a \cdot \vec{\phi}_1,
\quad \vec{\phi_2} \mapsto a \cdot \vec{\phi}_2
\ee
We would now like to discuss the resolution of the singularity 
in this model. From equations \eqn{genhom}, we expect the 
resolved locus $L$ to have $h_0(L)=h_2(\mathbb{C}{\bf P}^2)+1=2$ 
and $H_2(L,\mathbb{Z})\cong H_4(\mathbb{C}\bf{P}^2)\cong 
\mathbb{Z}$, so that topologically
\be 
L\cong\mathbb{R}\times S^2 \cup \mathbb{R}^3
\label{ub40}\ee 
In the previous section we noticed that 
the intersecting D6$_1$-brane and D6$_2$-brane
can be continuously deformed into a single smooth D6-brane
described by the equations:
\be
\vec{\phi}_1\cdot\vec{\phi}_2=-|\vec{\phi}_1||\vec{\phi}_2|, 
\quad\quad\quad
|\vec \phi_1| \cdot (3 |\vec \phi_2|^2 - |\vec \phi_1|^2) = \rho
\label{although}\ee
In fact, it is easy to see that the same deformation also resolves
singularity in this model. Namely, it deforms three intersecting
special Lagrangian planes into two components of the smooth
locus $L$:
\be
L = \left\{ \vec{\phi}_1\cdot\vec{\phi}_2=-|\vec{\phi}_1||\vec{\phi}_2| 
\ ; \quad|\vec \phi_1|(3 |\vec \phi_2|^2 - |\vec \phi_1|^2) = \rho 
\right\}
\cup \left\{ |\vec{\phi}_1-\sqrt{3}\vec{\phi}_2|=0 \right\}
\ee
The first connected component of $L$ is what one finds from
the deformation of D6$_1$ and D6$_2$, as in the previous section,
whereas the second component is just the original plane D6$_3$-brane.
Note, that equations, which parameterize the first and the second
component of $L$ have no common solutions. Deforming the intersection 
of two of the D6-branes creates a hole through which the third 
passes. Therefore, the above
deformation completely removes the singularity, and we obtain
a completely smooth D6-brane locus, diffeomorphic to the locus 
\eqn{ub40} as expected.

Of course, instead of deforming the D6$_1$ and D6$_2$ branes, 
we could choose any pair to resolve the singularity. There are three
choices, related by triality permutation symmetry. Each such 
resolution has a complex line of moduli space. These three lines 
meet at the origin. Therefore, the model has three branches, with a 
singular point at the origin. This is as predicted in \cite{aw}. 

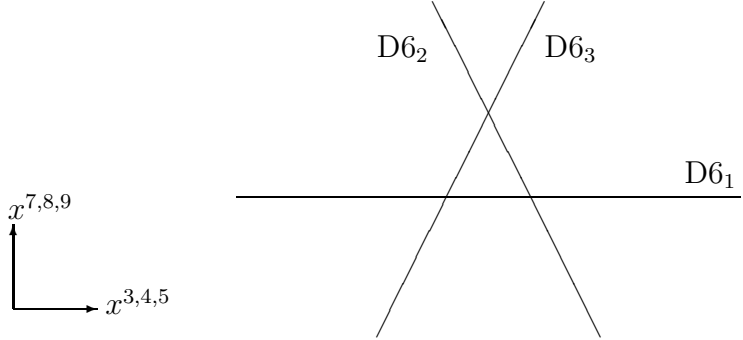
\begin{figure}

\setlength{\unitlength}{0.9em}
\begin{center}
\begin{picture}(22,11)
\put(7,4){\line(1,0){18}}\put(23,4.5){D6$_1$}
\put(12,-1){\line(1,2){6}}\put(18,9){D6$_3$}
\put(20,-1){\line(-1,2){6}}\put(12,9){D6$_2$}

%\put(17,7){${2 \pi \over 3}$}
%\qbezier(15,7)(17,7)(17.5,5)
%\qbezier(15.2,3.4)(17,3)(17.3,5)
%\qbezier(17,7)(18,6)(18,5)
%\put(18,6){${2 \pi \over 3}$}
\put(-1,0){\vector(1,0){3}}\put(-1,0){\vector(0,1){3}}
\put(2.3,-0.2){$x^{3,4,5}$}\put(-1.2,3.2){$x^{7,8,9}$}

\end{picture}\end{center}
\caption{Partially resolved singularity in which the 
D6$_3$-brane is translated.}
\label{figi}
\end{figure}

Perhaps more interestingly, the L-picture suggests that
there are further non normalizable deformations, which
must be included in the complete moduli space of this model.
To see this, note that we may simply translate one of the branes, 
let us say D6$_3$, to soften the singularity,
\ba
D6_3&:& \ft12\vec{\phi}_1-\ft{\sqrt{3}}{2}\vec{\phi}_2=\ft12\vec{m}
\nn\ea
This deformation breaks the  $SU(2)$ rotation symmetry of the 
IIA background to $U(1)$, so every point with $\vec{m} \ne 0$
corresponds to a different branch. The L-picture is drawn 
in Figure \ref{figi}. Moreover, we see that if we now 
resolve D6$_1$ and D6$_2$ onto the curve \eqn{although}, 
then D6$_3$ does not intersect this curve as long as,
\be
|\vec{m}|^2<\rho
\label{bighole}\ee
Note that, unlike the deformation \eqn{although}, the 
translation of the D6-brane is non-normalizable.

\subsubsection*{Probe Theory}

Now, let us introduce a probe D2-brane in this background,
and look at the $\mathcal{N}=1$ gauge theory on its world-volume.
In the singular limit of flat intersecting D6-branes, the 
addition of the extra D6-brane simply ensures the addition
of an extra hypermultiplet,
\begin{center}
{\bf Theory A:} $U(1)$ with 6 scalars and 3 hypermultiplets
\end{center}
As in the case of two intersecting D6-branes, the 6 scalar 
multiplets combine into two triplets whose interactions 
with the hypermultiplets are again of the form \eqn{yoyomar},
\be
f=\sum_{i=1}^3\sum_{a=1}^2 {T}^a_i\,W_i^\dagger\,\vec{\tau}\cdot\,
\vec{\Phi}_aW_i
\ee
where the Yukawa matrix is given by,
\be
T_i^a=\left(\begin{array}{cc} 1 & 0 \\ -\ft12 & -\ft{\sqrt{3}}2 \\
-\ft12 & \ft{\sqrt{3}}2\end{array}\right)
\ee
%Again, the superpotentials may be determined from the geometry,
%The corresponding interactions are encoded in the (super)potential:
%\ba
%f&=&\Phi_1\left(Q_1^\dagger Q_1 -
%\tilde{Q}_1\tilde{Q}^\dagger_1\right) + \left(\frac{1}{2}\Phi_1
%+\frac{\sqrt{3}}{2}\Phi_2\right)\left(Q_2^\dagger Q_2 -
%\tilde{Q}_2\tilde{Q}^\dagger_2\right)
%\nn\\ && + \left(\frac{1}{2}\Phi_1
%-\frac{\sqrt{3}}{2}\Phi_2\right)\left(Q_3^\dagger Q_3 -
%\tilde{Q}_3\tilde{Q}^\dagger_3\right)
%\\
%{\cal W}&=&\tilde{Q}_1\Psi_1 Q_1+\tilde{Q}_2\left(\frac{1}{2}
%\Psi_1+\frac{\sqrt{3}}{2}\Psi_2\right)Q_2+\tilde{Q}_3\left(\frac{1}{2}
%\Psi_1-\frac{\sqrt{3}}{2}\Psi_2\right)Q_3
%\%nn\ea
The D2-brane world-volume theory enjoys a
$U(1)_J\times SU(2) \times U(1)^2_F$ global symmetry. 
The first three factors act in the same manner as the 
previous case, while the two
$U(1)$ flavor symmetries again act orthogonally
to the gauge group; we may take $w_i$ to have
charges $(+1,0)$, $(-1,+1)$ and $(0,-1)$ respectively.
As the D2-brane world-volume theory flows to the infra-red,
we expect the $SU(2)\times U(1)$ isometry group of the 
Coulomb branch to be enhanced to the full $SU(3)$ isometry 
group of the flag manifold $Y$.

As discussed above, there are two deformations of the 
manifold. The first, in which two of the D6-branes 
combine to form a smooth locus $L$, induces the 
same deformation operator \eqn{defg2} as in the 
previous section. The second deformation, in which 
one of the D6-branes is moved away from the others, 
is even easier: it corresponds to a triplet of mass 
parameters for one of the hypermultiplets.

\subsubsection*{IIB Brane Construction and the Mirror Theory}

As in the previous section, we may try to determine an algebraic 
expression for the $G_2$ manifold, $X$, by realizing it as  the 
Higgs branch of a mirror theory. Once again, we may bring to bear 
either the field theoretic or brane construction techniques. We 
start with the field theory approach. Following the prescription 
of Section \ref{qftsec}, we find the mirror theory,
\begin{center}
{\bf Theory B:} $U(1)^2$ with 3 scalar and 3 hypermultiplets
\end{center}
The charges of the three hypermultiplets under the $U(1)^2$ gauge group 
are $(+1,-1,0)$ and $(0,+1,-1)$. The three scalar multiplets form a 
triplet, $\vec{\phi}$, and couple to the hypermultiplets through 
the real superpotential,
\be
f_B=\sum_{i=1}^3W^\dagger_i\vec{\tau}W_i\cdot\vec{\phi}
\nn\ee
This provides 3 real constraints on the 12 real scalar fields contained 
in the hypermultiplets. After dividing by the gauge group, we are 
left with a Higgs branch of dimension 7, as required. It is given 
by the constraints,
\ba
\sum_{i=1}^3|q_i|^2-|\tilde{q}_i|^2&=&0
\label{itsnicea}\\
\sum_{i=1}^3\tilde{q}_iq_i&=&0
\label{itsniceb}
\ea
This space has a manifest $SU(3)$ isometry group, in agreement with our 
expectations that this is the cone over $SU(3)/U(1)^2$.
Given that the Higgs branch is already a quotient by a $U(1)^2$ action,
it is natural to conjecture that the constraints
\eqn{itsnicea} - \eqn{itsniceb} taken alone
(i.e. before we divide by the gauge group)
yield the cone over $SU(3)$, at least topologically.

In order to see that this is indeed the case, we must find the base 
of the cone described by the eqs. \eqn{itsnicea} - \eqn{itsniceb}. 
To do this, we intersect this space with a sphere of a given
radius (it is convenient to choose the radius to be $\sqrt{2}$),
\be
\sum_{i=1}^3|q_i|^2 + |\tilde{q}_i|^2=2
\label{bigsphere}
\ee
Then, we can rewrite conditions \eqn{itsnicea} - \eqn{bigsphere} as:
\ba
\sum_{i=1}^3|q_i|^2 &=& 1
\label{itsnicec}\\
\sum_{i=1}^3|\tilde{q}_i|^2 &=& 1
\label{itsniced}\\
\sum_{i=1}^3\tilde{q}_iq_i&=&0
\label{itsnicef}
\ea
These equations define a codimension four submanifold
in $\mathbb{C}^3 \times \mathbb{C}^3$,
parameterized by $q_i$ and $\tilde{q}_i$,
the topology of which is to be determined.
The first two equations, \eqn{itsnicec} and \eqn{itsniced},
restrict $q$ and $\tilde{q}$ to be unit vectors in $\mathbb{C}^3$,
and \eqn{itsnicef} further implies that they are orthogonal.
Therefore, the manifold we are looking for can be
viewed\footnote{We thank James Sparks for pointing this out to us.}
as a set of orthonormal 2-frames in $\mathbb{C}^3$.
By definition, this is a complex Stiefel manifold:
\be
V_{3,2} (\mathbb{C}) = U(3)/U(1) = SU(3)
\ee
%The first equation, eq. \eqn{itsnicec}, restricts us to:
%\be
%{\bf S}^5 \times \mathbb{R}^6 \subset \mathbb{R}^6 \times \mathbb{R}^6
%\ee
%Suppose, we fix the values of $q_i$ and, in doing so, pick 
%a point on the ${\bf S}^5$. Then, 
%the last equation \eqn{itsnicef} defines a four-plane
%suitably oriented in the second $\mathbb{R}^6$.
%In other words, eqs. \eqn{itsnicec} and \eqn{itsnicef} alone
%define a 9-dimensional manifold:
%\be
%{\bf S}^5 \times \mathbb{R}^4 \subset \mathbb{R}^6 \times \mathbb{R}^6
%\ee
%where the orientation of $\mathbb{R}^4$ depends on a point
%in the ${\bf S}^5$.
%More precisely, it is an $\mathbb{R}^4$ bundle over the ${\bf S}^5$.
%Finally, implementing the remaining
%condition \eqn{itsniced} cuts a 3-sphere in the $\mathbb{R}^4$.
%So, the resulting manifold we find
%is an ${\bf S}^3$ bundle over ${\bf S}^5$.
%Using, for example, the spectral sequence for this fiber bundle
%one easily finds the following Betti numbers (note, the Euler number
%of this bundle vanishes, since $H^4 ({\bf S}^5; \mathbb{Z})=0$):
%$$
%h_3 = 1, \quad h_5 = 1
%$$
This proves that the base of the cone defined by the constraints
\eqn{itsnicea} - \eqn{itsniceb} is isomorphic to $SU(3)$.

The constraints \eqn{itsnicea} - \eqn{itsniceb}
have a natural deformation which partially 
resolves the singularity. This is obtained by simply adding constants to 
the right-hand side of each equation, and corresponds to moving the 
D6$_3$-brane as described earlier in this section.

One can also attempt to re-derive the mirror theory using brane techniques. 
As in the previous example, by compactifying the $x^6$ direction and 
performing subsequent T- and S-dualities, 
we can find dual D-brane configurations in type IIB string theory: 
\ba
D3 && 126 \nn\\
NS5_1 && 12789 \nn\\
NS5_2 && 12 [37]_{\theta_1} [48]_{\theta_2} [59]_{\theta_3} \nn\\
NS5_3 && 126 [37]_{2 \theta_1} [48]_{2 \theta_2} [59]_{2 \theta_3}
\nn\ea
where
\be
\theta_1 = \theta_2 = \theta_3 = {2 \pi \over 3}
\ee
This brane configuration is drawn in Figure \ref{figj}. 
\begin{figure}

\setlength{\unitlength}{0.9em}
\begin{center}
\begin{picture}(22,11)
\qbezier(8,6)(13.7,8)(19.5,6)\put(9.5,7){D3}
\qbezier(8,6)(13.7,4)(19.5,6)
\put(17.5,2){\line(1,2){4}}\put(22,10){NS5$_2$}
\put(8,2){\line(0,1){8}}\put(8.5,10){NS5$_1$}
\put(16,2){\line(-1,2){4}}\put(12.5,10){NS5$_3$}
%\put(14.6,5.5){$\updownarrow$}
%\put(19,4.8){$\updownarrow$}
\put(-1,0){\vector(1,0){3}}\put(-1,0){\vector(0,1){3}}
\put(-1,0){\vector(1,1){3}}
\put(2.3,-0.2){$x^{6}$}\put(-1.2,3.2){$x^{7,8,9}$}\put(2.3,3.2){$x^{3,4,5}$}

\end{picture}\end{center}
\caption{IIB Brane model for Theory B$^\prime$:
$U(1)^2$ with 3 hypermultiplets.}
\label{figj}
\end{figure}
Each segment of D3-brane carries a $U(1)$ gauge field, 
while each intersection yields a hypermultiplet, charged 
oppositely under the adjacent gauge fields. The overall 
$U(1)$ decouples, leaving only $U(1)^2$ gauge group to act 
on the hypermultiplets. However, there are no further 
massless, neutral scalar multiplets as in Theory B above. We 
thus have
\begin{center}
{\bf Theory B$^\prime$:} $U(1)^2$ with 3 hypermultiplets
\end{center}
Naively this theory has a Higgs branch of real dimension 
$12-2=10$. This clearly cannot be the case. What we 
have missed is the contribution from the massive neutral scalar 
fields. As in the case of the generalized conifold \cite{u}, 
these mediate interactions which result in a superpotential for the 
hypermultiplets. However, as for the Calabi-Yau example, determining 
the exact superpotential is somewhat more complicated than simply 
integrating the massive fields at the tree-level. 
One must follow the anomalous dimension of the various fields under 
renormalization group flow. This seems out of our reach in the present 
case, so instead we conjecture the simplest coupling consistent 
with all symmetries of the theory which still yields a seven dimensional 
Higgs branch. It is the quartic superpotential, 
\be
f^\prime_B=\left(\sum_{i=1}^3W_i^\dagger\vec{\tau}W_i-\vec{m}\right)^2
\nn\ee
For $\vec{m}=0$, the Higgs branch of this theory coincides with that of 
Theory B given in \eqn{itsnicea} - \eqn{itsniceb}.
Non-zero $\vec{m}$ corresponds 
to moving the D6$_3$-brane. In this fashion, the brane picture yields 
the same results as the field theory analysis.

%%%%%%%%%%%%%%%%%%%%%%%%%%%%%%%%%%%%%%%%%%%%%%%%%%%%%%%%%%%%%%%%%%%%%%%%%

\subsection{A New Model with 3 D6 Branes}
\label{gsecd}

We now turn to new models, which after reduction to Type IIA theory
can be described by three, flat D6-branes lying in the world-volume directions
\ba
D6_1 && 123456 \nn\\
D6_2 && 123689 \nn\\
D6_3 && 124679
\nn\ea
It is easy to check that this configuration of D6-brane is supersymmetric.
Indeed, the locus $L$, which is a union of three special Lagrangian
3-planes, is calibrated with respect to
\ba
\mathrm{Re} (\Omega) 
&=& dx^3 \wedge dx^4 \wedge dx^5 + dx^3 \wedge dx^8 \wedge dx^9
+ dx^4 \wedge dx^7 \wedge dx^9 + dx^5 \wedge dx^7 \wedge dx^8 \nn\\ 
&=& \mathrm{Re} (dz_1 \wedge dz_2 \wedge dz_3) 
\nn\ea
and the form $\mathrm{Re} (\Omega)$ restricts to volume form
on each 3-plane. This locus $L$ has no continuous isometries 
although, upon lifting to M-theory the resulting $G_2$ 
manifold is expected to have at least a $U(1)$ isometry. 

Let us discuss the possible resolutions of the singularity. 
Notice that, in the absence of any one of the three branes, the 
remaining two lift to the singular Calabi-Yau conifold discussed in 
Section \ref{susec}. Ignoring the $x^1$ and $x^2$ directions, 
any pair of branes intersect over a line. All three branes 
intersect at a point: $x^p=0$ for $p=3,\ldots,9$. 

The simplest resolution of the singularity involves
separating the branes.  There are three parameters, $a$, $b$ and $c$, 
corresponding to the relative separations of the branes which we 
choose as
\ba
D6_1: && x^7=m_1,\quad x^8=0,\quad x^9=0 \nn\\
D6_2: && x^4=0,\quad x^5=0,\quad x^7=0 \nn\\
D6_3: && x^3=0,\quad x^5=m_2,\quad x^8=m_3 
\nn\ea
If all three parameters are non-zero, then the branes do not intersect. 
This is analogous to the small resolution of the conifold singularity. 
The resulting locus has topology,
\be
L=\mathbb{R}^3\cup\mathbb{R}^3\cup\mathbb{R}^3
\ee
From equation \eqn{genhom}, we see that this IIA background lifts 
to a manifold $X$ of $G_2$ holonomy with 
$h_2(X)=2$. We will give an algebraic description of this manifold 
when we come to discuss the IIB brane construction.

There is another way to remove the singularity, which uses the 
methods of the deformed and the resolved conifold. In 
this approach, we set $m_a=0$, and pick two D6-branes, say 
$D6_1$ and $D6_2$, which intersect over the 
$x^7$ 
line. Defining the complex variables,
\be
\psi_1=x^8+ix^9,\quad\quad \psi_2=x^4+ix^5
\nn\ee
the position of the first two D6-branes is described as $\psi_1\psi_2=0$ 
and $x^7=0$. We now deform the first of these equations to 
\be
\psi_1\psi_2=\rho\equiv\rho_1+i\rho_2
\label{myfirstslag}\ee
Together with the equation for the flat D6${}_3$-brane, this 
defines a special Lagrangian curve. To see this, it suffices 
to note that D6-branes lying on the curve \eqn{myfirstslag} 
preserve the same supersymmetry as the flat, intersecting 
branes. The D6$_3$-brane intersects the complex curve 
\eqn{myfirstslag} only asymptotically, ensuring that the 
singularity is indeed removed. The smooth locus of D6-branes has 
topology
\be
L=S^1\times\mathbb{R}^2\cup \mathbb{R}^3
\nn\ee
Once more employing equations \eqn{genhom}, we see that 
this D6-brane configuration lifts to a manifold $X$ of 
$G_2$-holonomy with Betti numbers:
$$
h_2(X)=1, \quad h_3 (X)=1
$$
This manifold therefore 
admits a geometric transition. From the L-picture, it is 
clear that this is entirely analogous to the 
conifold transition, in which an ${\bf S}^2$ shrinks,
and an ${\bf S}^3$ grows. The L-picture suggests that, as for 
the conifold, this process necessarily involves passing through 
a singular point. Also, since each deformation is equivalent 
to that in the conifold, they are non-normalizable. 

Finally we note that, with three D6-branes, we may perform both 
deformation and resolution simultaneously. This involves deforming,
say, the D6$_1$ and D6$_2$-brane as in \eqn{myfirstslag}, while 
simply moving the D6$_3$-brane in the $(x^5=m_2)-(x^8=m_3)$-plane. 
If the D6$_3$-brane moves a short distance away from the 
origin, it will intersect the complex curve \eqn{myfirstslag}, 
resulting in a singular manifold. However, if we move them into 
the region defined by,
\be
m_2^2m_3^2-m_2m_3\rho_2>\ft14\rho_1^2
\ee
then the $G_2$ lift becomes smooth once again. It is interesting 
to note that this condition is opposite to the corresponding 
condition \eqn{bighole} for the cone over $SU(3)/U(1)^2$ 
discussed in the previous section. In the latter case, 
deforming two of the D6-branes created a hole through which 
the third could pass. In the present case, however, deforming 
two of the D6-branes does not create a hole. Rather, 
the third D6-brane must be moved sufficiently far from the 
initial two in order to avoid them. (Alternatively, as mentioned 
above, if it is left at the origin, it intersects only asymptotically).

The various phases of the manifold $X$ we just discussed are listed
in Table 3. The dots in the last row indicate
that our analysis is by no means complete, and there might be
more phases to be discovered.

\begin{table}\begin{center}
\begin{tabular}{|c|c|c|}
\hline
Betti Numbers & Number of Phases & Number of Deformations \\
\hline
\hline
$h_2=2$, $h_3=0$ & 1 & 3 \\
\cline{1-3}
$h_2=1$, $h_3=1$ & 3 & 2 \\
\cline{1-3}
$h_2=1$, $h_3=1$ & 3 & 4 \\
\cline{1-3}
$\ldots$ & $\ldots$ & $\ldots$ \\
\hline
\end{tabular}\end{center}
\caption{Some topological phases of a manifold $X$ with $G_2$ holonomy.}
\end{table}

\subsubsection*{Probe Theory}

As usual, we probe this brane set-up with a D2-brane extended
in the $1-2$ plane. As in previous sections, the probe 
theory has ${\cal N}=1$ supersymmetry, and is simple to 
write down,
\begin{center}
{\bf Theory A:} $U(1)$ with 6 scalar and 3 hypermultiplets
\end{center}
If $\phi_a$, $a=1,\ldots,6$, denote the fields
corresponding to D2-brane motion along $x^{3,4,5,7,8,9}$ 
then we have the real superpotential,
\be
f=\sum_{i=1}^3 \vec{A}_i\cdot W_i^\dagger \vec{\tau} W_i
\ee
where the triplets $\vec{A}_i$ are suitable combinations 
of the $\Phi_i$,
\be
\vec{A}_1=(\Phi_7,\Phi_8,\Phi_9)\ \ \ \ ,\ \ \ \ 
\vec{A}_2=(\Phi_7,\Phi_4,\Phi_5)\ \ \ \ ,\ \ \ \ 
\vec{A}_3=(\Phi_3,\Phi_8,\Phi_5)
\label{itsthea}\ee
The deformations of this model are the same as those 
described in the section on the conifold. The 
deformation \eqn{myfirstslag} is given by the addition 
of the superpotential \eqn{defcon1}. The translations 
of the D6-branes correspond to mass parameters. For 
example, if we set $\rho=0$ in \eqn{myfirstslag}, but include 
the translation parameters $m_a$, then the scalar potential is,
\ba
V_A&=& e^2\left(\re(\tilde{q}_1q_1)+\re(\tilde{q}_2q_2)\right)^2 + 
e^2\re(\tilde{q}_3q_3)^2+\left((\phi_7-m_1)^2+\phi_8^2+\phi_9^2\right)
w_1^\dagger w_1 \nn\\ 
&& +e^2\left(\im(\tilde{q}_1q_1)+\im(\tilde{q}_3q_3)\right)^2 
+e^2\im(\tilde{q}_2q_2)^2 +\left(\phi_4^2+\phi_5^2+\phi_7^2\right)
w_2^\dagger w_2 \nn\\
&& + e^2\left(|q_1|^2-|\tilde{q}_1|^2\right)^2+e^2\left(|q_2|^2-
|\tilde{q}_2|^2+|q_3|^2-|\tilde{q}_3|^2\right)^2 \nn\\ 
&& + \left((\phi_3^2+(\phi_5-m_2)^2+(\phi_8-m_3)^2\right)w_3^\dagger 
w_3
\nn\ea
Upon the lift to M-theory, this configuration of D6-branes
yields a $G_2$-manifold $X$ with a single $U(1)$ isometry
that comes from $U(1)_J$ symmetry and $U(1)^2$ flavor symmetry that 
is inherited from gauge symmetries on D6-branes. As usual, the diagonal
$U(1)$ decouples. Hence, we expect that when $X$ develops
a conical singularity the total global symmetry group is:
\be
K_Y = U(1)_J \times U(1)_C^2
\ee
We will return to the geometry of spaces $X$ and $Y$ in a moment.

%The one-loop metric on the Coulomb branch of Theory A is given by,
%\be
%ds^2=\sum_{a=1}^6 H_a d\phi_a^2 + H^{-1}\left( d\sigma^2
%+\sum_{i=1}^4\vec{\omega}_i\cdot d\vec{\phi}_i\right)^2
%\ee
%{\bf not too sure about this} where $\vec{\phi}_i$ denote
%the three directions transverse to the $i^{\rm th}$ D6-brane
%(e.g. $\vec{\phi}_1=(\phi_4,\phi_5,\phi_6)$). The harmonic
%functions are given by,
%\be
%H=\frac{1}{e^2}+\sum_{i=1}^4\frac{1}{|\phi_i|}\,\quad
%H_1=\frac{1}{e^2} + \frac{1}{|\vec{\phi}_3|}+
%\frac{1}{|\vec{\phi}_4|}
%\ee
%with similar expressions for the other $H_a$; in general
%$H_a$ receives a contribution of the form $1/|\vec{\phi}_i|$
%if $\phi_a$ is a component of $\vec{\phi}_i$.

\subsubsection*{IIB Brane Construction and the Mirror Theory} 

In the previous section we have determined the topology 
of the manifold $X$. Here we shall give an algebraic 
description of the manifold using mirror symmetry. As 
we have many times, we perform a T-duality along the 
$x^6$ direction, followed by an S-duality. We end up 
with a IIB brane construction with three NS5-branes, 
and a D3-brane wrapped on the compact $x^6$ direction:
\ba
NS5_1: && 12345 \nn\\
NS5_2: && 12389 \nn\\
NS5_3: && 12479 \nn\\ 
D3: && 126 
\nn\ea
At this stage, the reader with a (very) good memory will 
recognize this as the model considered as Example 2 in 
Section \ref{branesec}, where we derived the mirror theory to 
be 
\begin{center}
{\bf Theory B:} $U(1)^2$ with 3 scalar and 3 hypermultiplets
\end{center}
where we have ignored the overall, free $U(1)$ gauge symmetry. 
The hypermultiplets may be taken to have charges $(+1,-1,0)$ and 
$(0,-1,+1)$ under the gauge group. In Section \ref{branesec} we 
further found both Yukawa 
and quartic superpotential interactions, given in 
equations \eqn{corblimey} and \eqn{ffour} respectively. 
These yield the constraints,
\ba
|q_1|^2 - |\tilde{q}_1|^2 - |q_2|^2 + | \tilde{q}_2|^2 &=& 0 \nn\\
\re(\tilde{q}_2q_2-\tilde{q}_3q_3) &=& 0 \nn\\ 
\im(\tilde{q}_3q_3-\tilde{q}_1q_1) &=& 0
\nn\ea
These give 3 real constraints on the 12 real parameters $w_i$. 
After dividing by the $U(1)^2$ gauge action, we arrive at 
the Higgs branch. Since the above set of equations is invariant 
under rescaling of the fields $q$, it describes a conical space $X$.
%In order to find the base of the cone, $Y$, we fix the `radius':
%\be
%|q_1|^2 + |\tilde{q}_1|^2 + |q_2|^2 + | \tilde{q}_2|^2
%+ |q_3|^2 + | \tilde{q}_3|^2 = {\rm const}
%\ee
%{\bf A natural guess is that the 6-manifold $Y$ is:}
%\be
%Y = SU(2) \times SU(2) \times SU(2) / U(1)^3
%\ee

As we have seen, the locus $L$ has a rich moduli space 
of deformations. The deformations which blow up a 
${\bf S}^3$, given by \eqn{myfirstslag} are realized by the 
mirror operator \eqn{defcon17}, relevant to the deformed 
conifold. This is poorly understood. More simple are the 
translations of the D6-branes which blow up ${\bf S}^2$'s.
{}From the brane picture, we see that these correspond to FI terms. 
Specifically, we have the additional real superpotential,
\be
\Delta f = \sum_{a=1}^3 m_a\Phi_a
\nn\ee
However, this cannot be the full answer because this 
interaction partially lifts the Higgs branch, setting 
$W_i^\dagger\tau^iW_i=0$ for $i=1,2,3$. In order to 
compensate for this, we must further add the bare mass 
terms,
\be
\Delta f = \sum_{i=1}^3 m_i W_i^\dagger \tau^i W_i
\nn\ee
The theory once again has a seven dimensional Higgs branch, 
now given by the constraints, 
\ba
|q_1|^2 - |\tilde{q}_1|^2 - |q_2|^2 + | \tilde{q}_2|^2 &=& m_1 \nn\\
\re(\tilde{q}_2q_2-\tilde{q}_3q_3) &=& m_2 \nn\\ 
\im(\tilde{q}_3q_3-\tilde{q}_1q_1) &=& m_3
\nn\ea
modulo the $U(1)^2$ gauge action.

Finally, we comment that the above Higgs branch description of 
$X$ may also be recovered using the field theory method 
of describing ${\cal N}=1$ duals presented in Section \ref{qftsec}. 
In this case, one finds the mirror theory to have the same 
field content as Theory B, but without the 
quartic superpotential \eqn{ffour}. However, as discussed at length 
in Section \ref{branesec}, 
the quartic superpotential does not add any further constraints 
to determine the Higgs branch, and the two techniques therefore 
agree. 

\subsubsection*{Generalizations}

This model may be extended to consist of many, mutually 
orthogonal D6-branes:
\ba
N_1~~ D6_1 && 123456 \nn\\
N_2~~ D6_2 && 123689 \nn\\
N_3~~ D6_3 && 124679
\nn\ea
The moduli space of deformations is a simple generalization of 
that considered above, where the singularity appearing on the 
intersection of any pair of D6-branes may be resolved either
through translation or complex deformation. At least part of the physics 
associated with compactification of M-theory on such a $G_2$ manifold $X$ 
is obvious: as $X$ develops a conical singularity, non-abelian degrees 
of freedom appear with gauge group $U(N_1) \times U(N_2) \times U(N_3)$. 
It would be 
interesting to understand the resulting $G_2$ manifolds 
further -- they seem to be analogous to the generalized 
conifold studied, for example, in \cite{aklm}. It would 
also be interesting to see if these manifolds can be 
obtained from partial resolution of the orbifold 
$X = \mathbb{R}^7 /\mathbb{Z}_{k_1} \times \mathbb{Z}_{k_2} 
\times \mathbb{Z}_{k_3}$, where the orbifold group is 
generated by three elements
$\alpha$, $\beta$, and $\gamma$ of order $k_1$, $k_2$, and $k_3$,
respectively:
\ba
\alpha &:& (x^7 + i x^8) \mapsto e^{2 \pi i / k_1} (x^7 + i x^8), \quad
(x^9 + i x^{11}) \mapsto e^{- 2 \pi i / k_1} (x^9 + i x^{11}) \nn\\
\beta &:& (x^4 + i x^5) \mapsto e^{2 \pi i / k_2} (x^4 + i x^5), \quad
(x^7 + i x^{11}) \mapsto e^{- 2 \pi i / k_2} (x^7 + i x^{11}) \nn\\
\gamma &:& (x^3 + i x^5) \mapsto e^{2 \pi i / k_3} (x^3 + i x^5), \quad
(x^8 + i x^{11}) \mapsto e^{- 2 \pi i / k_3} (x^8 + i x^{11})
\nn\ea
An example of such space, with $k_i=2$, was recently studied in 
\cite{tsimpis}. Since $X$ is orbifold, the base of the cone is simply 
$Y = {\bf S}^6 / \mathbb{Z}_{k_1} \times \mathbb{Z}_{k_2} \times 
\mathbb{Z}_{k_3}$. Note, however, that orbifold action has fixed 
points (of codimension 4) on ${\bf S}^6$.

%%%%%%%%%%%%%%%%%%%%%%%%%%%%%%%%%%%%%%%%%%%%%%%%%%%%%%%%%%%%%%%%%%%%%%%%%

\subsection{A New Model with 4 D6 Branes}
\label{gsece}

A careful reader might notice that one can add an additional 
D6-brane to the configuration of 3 orthogonally intersection 
D6-branes, discussed in the previous section:
\ba
D6_1 && 123456 \nn\\
D6_2 && 123689 \nn\\ 
D6_3 && 124679 \nn\\ 
D6_4 && 125678
\nn\ea 
The D6$_4$ brane does not break supersymmetry further since the 
special Lagrangian calibration ${\rm Re} (\Omega)$ restricts 
to the volume form on the $x^{5,7,8}$-plane. 
All of the four D6-branes are on an equal footing. 

As in the previous section, we may resolve the singularity  
using the two deformations of the conifold. Firstly,  
let us consider translating the D6-branes. There are six such 
parameters, 
\ba
D6_1: && x^7=m_1,\quad x^8=0,\quad x^9=0 \nn\\ 
D6_2: && x^4=0,\quad x^5=0,\quad x^7=0 \nn\\ 
D6_3: && x^3=0,\quad x^5=m_2,\quad x^8=m_3 \nn\\ 
D6_4: && x^3=n_1,\quad x^4=n_2,\quad x^9=n_3 
\nn\ea
If $m_i,n_i\neq 0$ for all $i=1,2,3$, then the D6-branes do  
not intersect and locus $L$ has topology, 
\be
L=\mathbb{R}^3\cup\mathbb{R}^3\cup\mathbb{R}^3\cup\mathbb{R}^3 
\ee 
so, from \eqn{genhom}, we see that the lift to M-theory 
results in a smooth manifold $X$ of $G_2$ holonomy with topology 
$h_2(X)=3$, with no other non-trivial cycles. As we shall now 
show, there are geometrical transitions which involve blowing 
down up to two of these ${\bf S}^2$'s and replacing them with 
${\bf S}^3$'s. There are therefore three topologically distinct 
manifolds. 

Let us first consider blowing up a single 
${\bf S}^3$. This requires us to pick a pair of D6-branes, and there are 
therefore six independent ways of doing this. Here we give an 
example. Let us define the two complex variables:
\ba
\psi_1=x^8+ix^9,&&\quad\quad \psi_2=x^4+ix^5 \nn\\ 
\tilde{\psi}_1=x^5+ix^8,&&\quad\quad \tilde{\psi}_2=x^4+ix^9 
\nn\ea
If we set $m_i=n_2=n_3=0$, then we can deform the first two 
D6-branes into the complex curve $\psi_1\psi_2=\rho$, in 
which case the singularity is removed providing $n_1\neq 0$. 
As in the previous example, we may move the D6$_3$ and D6$_4$ 
brane by turning on $m_2,m_3$ and $n_2,n_3$ respectively. The 
lift to M-theory is again non-singular providing we move them 
far enough. Alternatively, we may set $n_i=m_2=m_3=0$, and then
deform the D6$_3$ and D6$_4$ branes on the curve 
$\tilde{\psi}_1\tilde{\psi}_2=\tilde{\rho}$. Again, the 
special Lagrangian manifold is smooth if $m_1\neq 0$. 
In each of these cases, the locus $L$ has topology:
\be
L=S^1\times\mathbb{R}^2\cup\mathbb{R}^3\cup\mathbb{R}^3
\ee
which ensures that the lift to M-theory results in a manifold $X$ 
with $h_2(X)=2$, and $h_3 (X)=1$.

Finally, we may resolve two pairs of D6-branes. There are three ways of 
choosing two pairs. Let us examine the deformation of the 
D6$_1$ -- D6$_2$ pair, and the D6$_3$ -- D6$_4$ pair:
\be
\psi_1\psi_2=\rho,\quad\quad \tilde{\psi}_1\tilde{\psi}_2=\tilde{\rho}
\ee
For suitable choices of the parameters, the two curves do not 
intersect (e.g. $\re\rho=\im\tilde{\rho}=0$). The locus $L$ 
has topology 
\be
L=S^1\times\mathbb{R}^2\cup S^1\times\mathbb{R}^2
\ee
so the smooth manifold $X$ has Betti numbers:
$$
h_2 (X)=1, \quad h_3 (X)=2
$$
As in the previous example, since each of these 
deformations is inherited from the conifold, they are non-normalizable. 

\subsubsection*{Probe Theory}

The ${\cal N}=1$ theory living on a probe D2-brane
is again easy to determine:
\begin{center}
{\bf Theory A:} $U(1)$ with 6 scalar and 4 hypermultiplets
\end{center}
with a superpotential given by:
\be
f=\sum_{i=1}^4 \vec{A}_i\cdot W_i^\dagger \vec{\tau} W_i
\ee
where the triplets $\vec{A}_i$ where given in equation \eqn{itsthea} 
for $i=1,2,3$, and the fourth is,
\be
\vec{A}_4=(\Phi_3,\Phi_4,\Phi_9) 
\ee
As before, we write the scalar potential only for the model 
with $h_2(X)=3$, which consists of four, separated flat D6-branes:
\ba
V_A&=& e^2\left(\re(\tilde{q}_1q_1)+\re(\tilde{q}_2q_2)\right)^2 + 
e^2\left(\re(\tilde{q}_3q_3)+\re(\tilde{q}_4q_4)\right)^2 \nn\\ 
&& +e^2\left(\im(\tilde{q}_1q_1)+\im(\tilde{q}_3q_3)\right)^2 
+e^2\left(\im(\tilde{q}_2q_2)^2+\im(\tilde{q}_4q_4)\right)^2 \nn\\ 
&& + e^2\left(|q_1|^2-|\tilde{q}_1|^2+|q_4|^2-|\tilde{q}_4|^2
\right)^2+e^2\left(|q_2|^2-
|\tilde{q}_2|^2+|q_3|^2-|\tilde{q}_3|^2\right)^2 \nn\\ 
&& +\left((\phi_7-m_1)^2+\phi_8^2+\phi_9^2\right)
w_1^\dagger w_1 +\left(\phi_4^2+\phi_5^2+\phi_7^2\right)
w_2^\dagger w_2 \nn\\
&& + \left((\phi_3^2+(\phi_5-m_2)^2+(\phi_8-m_3)^2\right)w_3^\dagger 
w_3 \nn\\ 
&& + \left((\phi_3-n_1)^2+(\phi_4-n_2)^2+(\phi_9-n_3)^2\right)
w_4^\dagger w_4 
\nn\ea
Upon the lift to M-theory, the manifold $X$ has a single 
$U(1)$ isometry arising from $U(1)_J$ symmetry. In the singular 
conical case, there is a 
further $U(1)^3_C$ that comes from
the non-diagonal $U(1)$ gauge symmetries on D6-branes. 
\be
K_Y = U(1)_J \times U(1)_C^3
\ee

\subsubsection*{IIB Brane Construction and the Mirror Theory}

As in the previous section, we shall endeavor to find an 
algebraic expression for $X$ Upon compactifying the $x^6$ 
direction, and performing both T- and S-dualities, we 
end up with a Hanany-Witten type periodic model, consisting of:
\ba
NS5_1: && 12345 \nn\\
NS5_2: && 12389 \nn\\
NS5_3: && 12479 \nn\\ 
NS5_4: && 12578 \nn\\ 
D3: && 126 
\nn\ea
It has a low-energy ${\cal N}=1$ three-dimensional description 
given by:
\begin{center}
{\bf Theory B:} $U(1)^3$ with 4 scalar and 4 hypermultiplets
\end{center}
where we have ignored the overall, free $U(1)$ gauge symmetry. 
The hypermultiplets may be taken to have charges $(+1,-1,0,0)$, 
$(0,+1,-1,0)$, and $(0,0,+1,-1)$ under the three gauge groups. 
The Yukawa couplings are simple to write down following the 
prescription given in the previous section
\ba
f_{\rm Yuk}&=&\Phi_1(W_1^\dagger\tau^3W_1-W_2^\dagger
\tau^3W_2)+ \Phi_2(W_2^\dagger\tau^1W_2-W_3^\dagger\tau^1W_3) 
\nn \\ && +\Phi_3(W_3^\dagger \tau^3W_3-W_4^\dagger\tau^3W_4)
+ \Phi_4(W_4^\dagger\tau^1W_4-W_1^\dagger\tau^1W_1)
\label{corblimeyguv}\ea
where we have taken the liberty of performing a field redefinition 
of $W_3$ relative to the previous section to make the symmetries 
more manifest. Notice that this time there is an important 
difference from the three D6-brane case discussed in the 
previous section. In that case, the Yukawa terms alone were 
enough to result in a Higgs branch of real dimension seven; 
any quartic superpotential was required to be of a form that 
didn't impose any further constraints, which indeed was what 
we found. In the present case with four 
D6-branes however, this is no longer the case. The Yukawa 
couplings above, together with the $U(1)^3$ gauge action, 
result in a Higgs branch of real dimension nine. We therefore 
expect the quartic superpotential to yield two further 
real constraints. This is an important check on the answer below. 

The quartic superpotential can be easily written down following 
the discussion of Section 2.2, and \cite{u}:
\ba
f_4&=&(W^\dagger_1\tau^1W_1\,W^\dagger_2\tau^1W_2-
W^\dagger_1\tau^2W_1\,W^\dagger_2\tau^2W_2) \nn\\ && + 
(W^\dagger_2\tau^2W_2\, W^\dagger_3\tau^2W_3-
W^\dagger_2\tau^3W_2\, W^\dagger_3\tau^3W_3) \nn\\ 
&& - (W^\dagger_3\tau^1W_3\,W^\dagger_4\tau^1W_4-
W^\dagger_3\tau^2W_3\,W^\dagger_4\tau^2W_4) \nn\\ && - 
(W^\dagger_4\tau^2W_4\, W^\dagger_1\tau^2W_1-
W^\dagger_1\tau^3W_1\, W^\dagger_4\tau^3W_4)
\ea 
Reassuringly, when combined with the Yukawa couplings 
above, this does indeed lead to 6 real constraints 
on the 16 hypermultiplet degrees of freedom,
\ba
|q_1|^2 - |\tilde{q}_1|^2 - |q_2|^2 + | \tilde{q}_2|^2 &=& 0 \nn\\
|q_3|^2 - |\tilde{q}_3|^2 - |q_4|^2 + | \tilde{q}_4|^2 &=& 0 \nn\\
\re(\tilde{q}_2q_2-\tilde{q}_3q_3) &=& 0 \nn\\ 
\re(\tilde{q}_4q_4-\tilde{q}_1q_1) &=& 0 \label{ohyeah}\\
\im(\tilde{q}_1q_1-\tilde{q}_3q_3) &=& 0 \nn\\ 
\im(\tilde{q}_2q_2+\tilde{q}_4q_4) &=& 0
\nn\ea
where the first four constraints arise from the Yukawa 
couplings alone, and the last two come from the quartic 
superpotential. 
After dividing by the $U(1)^3$ gauge action, we arrive at 
the seven dimensional Higgs branch. Once again, the above set 
of equations is invariant under rescaling of the fields $q$, and 
so describes a conical singularity $X$. 

We now turn to the deformations corresponding to separating 
the D6-branes. As before, we expect these to correspond to 
FI parameters. In fact, only four of them are FI parameters: 
$m_1, m_2, n_1$ and $n_3$ each correspond to separating 
adjacent NS5-branes, and so give rise to FI parameters. As in 
the previous section, it is necessary that these are accompanied 
by bare mass parameters for hypermultiplets so as not to lift 
the Higgs branch. In contrast, $m_3$ and $n_2$ correspond to separating 
opposite NS5-branes. These induce only the mass terms for the 
hypermultiplets, 
\be
\Delta f = m_3(W_2^\dagger\tau^2W_2+W_4^\dagger\tau^2 W_4)
+n_2(W_3^\dagger\tau^2W_3-W_1^\dagger\tau^2 W_1)
\ee
The net result of these deformations is to add constant 
terms to the right-hand-side of \eqn{ohyeah},
\ba
|q_1|^2 - |\tilde{q}_1|^2 - |q_2|^2 + | \tilde{q}_2|^2 &=& m_1 \nn\\
|q_3|^2 - |\tilde{q}_3|^2 - |q_4|^2 + | \tilde{q}_4|^2 &=& n_1 \nn\\
\re(\tilde{q}_2q_2-\tilde{q}_3q_3) &=& m_2 \nn\\ 
\re(\tilde{q}_4q_4-\tilde{q}_1q_1) &=& n_3 \label{dadio}\\
\im(\tilde{q}_1q_1-\tilde{q}_3q_3) &=& m_3 \nn\\ 
\im(\tilde{q}_2q_2+\tilde{q}_4q_4) &=& n_2 \nn
\ea
Again, this model is but the simplest example of a family of models 
consisting of $N_i$ of the D6$_i$-brane. It would be of interest 
to examine if these too can be thought of as partial resolutions 
of a $\mathbb{R}^7$ orbifold. 

Finally, let us mention that we can also derive the same result 
for the Higgs branch using the field theory techniques 
explained in Section \ref{qftsec}. In this case, however, the 
mirror theory differs somewhat from that obtained using 
brane techniques. Specifically, we find:
\begin{center}
{\bf Theory B$^\prime$:} $U(1)^3$ with 6 scalar and 4 hypermultiplets
\end{center}
The gauge multiplets and the first four hypermultiplets have 
the same couplings as in Theory B above. However, Theory 
B$^\prime$ has 
only Yukawa couplings, and no quartic superpotential. 
The two extra scalars in Theory B$^\prime$ play the role of 
the quartic superpotential by imposing the two extra constraints 
(the imaginary equations in \eqn{dadio}). The two methods therefore 
yield the same Higgs branch and, indeed, the same physics at 
scales below those set by the mass terms $m_3$ and $n_2$.

%In a more general model of this kind, one could also find
%a non-abelian gauge symmetry enhancement. Such models
%correspond to configurations of multiple intersecting D6-branes:
%\ba
%D2 && 12 \nn\\
%N_1~~ D6_1 && 123456 \nn\\
%N_2~~ D6_2 && 123689 \nn\\
%N_3~~ D6_3 && 124679 \nn\\
%N_4~~ D6_4 && 125678
%\nn\ea
%In M-theory, the local geometry near the isolated singularity
%is that of the $G_2$ orbifold:
%\be
%X = \mathbb{R}^7 /
%\mathbb{Z}_{N_1} \times \mathbb{Z}_{N_2} \times \mathbb{Z}_{N_3}
%\times \mathbb{Z}_{N_4}
%\ee
%where the orbifold group is generated by elements
%$\alpha$, $\beta$, $\gamma$, and $\delta$:
%\ba
%\alpha &:& (x^7 + i x^8) \mapsto e^{2 \pi i / N_1} (x^7 + i x^8), \quad
%(x^9 + i x^{11}) \mapsto e^{- 2 \pi i / N_1} (x^9 + i x^{11}) \nn\\
%\beta &:& (x^4 + i x^5) \mapsto e^{2 \pi i / N_2} (x^4 + i x^5), \quad
%(x^7 + i x^{11}) \mapsto e^{- 2 \pi i / N_2} (x^7 + i x^{11}) \nn\\
%\gamma &:& (x^3 + i x^5) \mapsto e^{2 \pi i / N_3} (x^3 + i x^5), \quad
%(x^8 + i x^{11}) \mapsto e^{- 2 \pi i / N_3} (x^8 + i x^{11}) \nn\\
%\delta &:& (x^3 + i x^4) \mapsto e^{2 \pi i / N_4} (x^3 + i x^4), \quad
%(x^9 + i x^{11}) \mapsto e^{- 2 \pi i / N_4} (x^9 + i x^{11})
%\nn\ea
%Again, we can understand this family of orbifold models as
%cones on:
%\be
%Y = {\bf S}^6 /
%\mathbb{Z}_{N_1} \times \mathbb{Z}_{N_2} \times \mathbb{Z}_{N_3}
%%times \mathbb{Z}_{N_4}
%\ee

%%%%%%%%%%%%%%%%%%%%%%%%%%%%%%%%%%%%%%%%%%%%%%%%%%%%%%%%%%%%%%%%%%

\subsection{The cone over ${\bf S}^3\times {\bf S}^3$}
\label{gsecf}

Let us now turn to the most subtle example discussed in \cite{aw},
that of a cone over $Y={\bf S}^3\times {\bf S}^3$. This has proven to 
be an extremely interesting manifold, providing a new 
handle on four dimensional ${\cal N}=1$ super Yang-Mills 
\cite{bob,amv,aw}. Algebraically this 
dimension 7 space $X$ is described by the simple equation:
\ba
\sum_{i=1}^2 |q_i|^2-|\tilde{q}_i|^2=m
\label{g2}\ea
for real $m$, and is topologically an 
$\mathbb{R}^4$ bundle over ${\bf S}^3$.  This space has at least 
four interesting reductions to IIA string theory. As well as the 
two discussed above --- namely, reduction on a transverse circle, 
and the L-picture --- we could also choose to embed the M-theory 
circle either in the ${\mathbb R}^4$ fiber, or the ${\bf S}^3$ 
base. These reductions result in the deformed conifold with a 
wrapped D6-brane or the resolved conifold with RR flux, respectively 
\cite{bob,amv,aw}. D2-brane probes on the IIA background with flux 
were previously discussed by Aganagic and Vafa \cite{av}. Their 
method also yields the theory on the probe of $X$ and, as we shall 
show, mirror symmetry gives the probe of the L-picture. The 
theory on a D2-brane probe of a D6-brane wrapping the deformed 
conifold remains elusive. 

Let us start
by reducing on an ${\bf S}^1\subset{\bf S}^3$. The resulting 
IIA conifiguration is the resolved conifold with a single unit of 
RR two-form flux through the two-cycle.
The theory on a transverse D2-brane probe of the resolved conifold
was reviewed in Section \ref{susec}.
It has ${\cal N}=2$ supersymmetry and a Higgs branch,
which reproduces the geometry of the conifold. 
Recall that it comes complete with a
free ${\cal N}=2$ vector multiplet, including a gauge field 
$\tilde{A}$ corresponding to the motion of D2-brane along the M-theory
circle, as well as a scalar describing the direction transverse
to the conifold. The addition of the RR flux through the
two-cycle breaks supersymmetry on the D2-brane to ${\cal N}=1$
by inducing a CS-coupling, $\tilde{A}\wedge F$,  between the
two ${\cal N}=1$ vector multiplets of the theory \cite{av}.
The simplest way to see this is to examine the Wess-Zumino
terms on a fractional D4-brane wrapped around the ${\bf S}^2$.
The resulting ${\cal N}=1$ theory is,
\begin{center}
{\bf Theory A:} $U(1)\times\tilde{U(1)}$ with 1 real scalar 
and 2 hypermultiplets
\end{center}
with a further, free scalar multiplet which we ignore.
The hypermultiplets are charged only under the $U(1)$ gauge group, 
with the $\tilde{U(1)}$ coupling only through the CS term, 
\be
\frac{1}{e^2}F^2+\frac{1}{\tilde{e}^2}\tilde{F}^2+A\wedge\tilde{F}
\ee
Upon exchanging $\tilde{A}$ for its dual photon $\tilde{\sigma}$, 
we may replace the above interaction with
\be
\frac{1}{e^2}F^2+\tilde{e}^2\left(\partial \tilde{\sigma} 
+A\right)^2
\nn\ee
so that $\tilde{\sigma}$ transforms transitively under the 
$U(1)$ gauge action. The scalar potential is the same as 
for the conifold \eqn{apot}
\ba
V_A=e^2\left(|q_1|^2+|q_2|^2-|\tilde{q}_1|^2-|\tilde{q}_2|^2-m\right)^2
+\phi^2\left(w_1^\dagger w_1+w_2^\dagger w_2\right)
\label{apotagain}\ea
whose Higgs branch reproduces the resolved conifold 
$O(-1)+O(-1)\rightarrow {\bf \mathbb{C} P}^1$. The 
transformation of the dual photon $\tilde{\sigma}$ ensures 
it has unit Hopf fibration over the zero section ${\bf P}^1$, 
yielding a space of the requisite topology \cite{av}. To see 
this, note that the asymptotic radius of the M-theory circle is 
$\tilde{e}^2$. In the strong coupling limit
$\tilde{e}^2\rightarrow\infty$, the dual photon removes the $U(1)$ 
gauge invariance, and the effect of the CS coupling is 
simply to ungauge the original $U(1)$ . We are then left with:
\begin{center}
{\bf Theory A${}^\prime$:} 1 scalar multiplet and 2 hypermultiplets
\end{center}
which has the scalar potential \eqn{apotagain},
but no gauge fields. The vacuum moduli space coincides with the
equation \eqn{g2} for $X$.

Both Theories A and A${}^\prime$ 
acquire a new branch of vacua as the three-cycle collapses, 
$m\rightarrow 0$. For the conifold, the interpretation is clear: 
this branch describes fractional D2 branes, corresponding
to D4-anti-D4 branes wrapping the collapsed 2-cycle.
Lifting this picture to M-theory, and including the twist
of the dual-photon, the interpretation is equally clear
in the $G_2$ case: the fractional D2-branes correspond
to M5-anti-M5 branes wrapping the collapsed three-cycle.

In \cite{av}, it was further argued that including the Chern-Simons 
term $kA\wedge\tilde{F}$ results in the $\mathbb{Z}_k$ quotient 
of $X$ with topology ${\bf S}^3/\mathbb{Z}_k\times\mathbb{R}^4$. 
In this case, we see that the $U(1)$ action on $\tilde{\sigma}$ 
leaves a remanant $\mathbb{Z}_k$. The resulting vacuum moduli 
space is therefore \eqn{g2}, with the identification,
\be
q_i\rightarrow \omega q_i,\quad\quad 
\tilde{q}_i\rightarrow -\omega\tilde{q}_i
\ee
where $\omega^k=1$. This theory therefore describes the space
$(S^3\times \mathbb{R}^4)/\mathbb{Z}_k$,
which is isomophic to ${\bf S}^3/\mathbb{Z}_k\times\mathbb{R}^4$ 
in agreement with \cite{av}. 

Let us return to the $k=1$ case, and reduce to IIA on a circle 
transverse to both $X$ and the M2-brane. The resulting probe theory 
remains Theory A$^\prime$ above, now with a further free ${\cal N}=1$ 
vector multiplet describing the motion of the D2-brane in the M-theory 
direction. In other words, this theory describes IIA string theory 
on $X$: we have travelled clockwise around the circle of dualities 
in Figure \ref{figa}.

Our final type IIA dual can be obtained by embedding M-theory
circle in the ${\bf S}^3$ inside $X$, such that \cite{aw}:
$$
X / {\bf S}^1 \cong \mathbb{R}^6
$$
Unlike the other two cases, here the circle has fixed points
on a codimention four locus $L \subset \mathbb{R}^6$,
which gets identified with the position of D6-branes.
Let us determine what $L$ looks like.
{}From \eqn{genhom}, we see that it has topology:
\be
L\cong {\bf S}^1\times\mathbb{R}^2
\ee
More explicitly, if we parameterise
$\mathbb{R}^6 \cong \mathbb{C}^3$
by complex coordinates $z_i$, $i=1,2,3$,
we can write $L$ as \cite{joycecount,avfirst}:
\be
L = \{ \vert z_1 \vert^2 - m^2 = \vert z_2 \vert^2 = \vert z_3 \vert^2,
\im (z_1 z_2 z_3) =0 , \re (z_1 z_2 z_3) \ge 0 \}
\ee
It is easy to check that it indeed has the right topology.
In order to see that $L$ is (special) Lagrangian,
it is convenient to rewrite the K\"ahler form as:
$$
J = \sum_{i=1}^3 d \vert z_i \vert^2 \wedge d \theta_i
$$
The defining equations for $L$ can be expressed in the form:
\ba
\vert z_1 \vert^2 - \vert z_2 \vert^2 &=& m^2 \nn\\
\vert z_2 \vert^2 - \vert z_3 \vert^2 &=& 0 \nn\\
\theta_1 + \theta_2 + \theta_3 &=& 0
\ea
Since all of these relations are linear, it is straighforward
to check that $J$ restricts to zero on $L$. A little bit more
work shows that $L$ is {\it special} Lagrangian.
Furthermore, in the limit $t \to 0$ it degenerates into a cone
over ${\bf S}^1 \times {\bf S}^1$:
\be
L_0 = \{ \vert z_1 \vert = \vert z_2 \vert = \vert z_3 \vert,
\im (z_1 z_2 z_3) =0 , \re (z_1 z_2 z_3) \ge 0 \}
\ee
Our next task is to reproduce $L$ as the locus of massless 
hypermultiplets in the mirror theory. Using the standard 
mirror symmetry techniques, the mirror ${\cal N}=1$ 
theory is:
\begin{center}
{\bf Theory B:} $U(1)^2$ with 5 scalar and 2 hypermultiplets
\end{center}
The hypermultiplets have charge $(+1,-1)$ under the first
$U(1)$, and charge $(+1,+1)$ under the second. If we were 
to neglect one of the $U(1)$ vector multiplets, this would 
be precisely the matter content which realises the 
resolved conifold on the Coulomb branch. Indeed, 
the coupling of the scalars agrees with that of the conifold 
theory. In the notation of a real superpotential, we have:
\be
f=\sum_{i=1}^2\vec{A}_i\cdot W_i^\dagger \vec{\tau}W_i
\ee
where the triplets of scalar multiplets are:
\be
\vec{A}_1=(\Phi_1,\Phi_2,\Phi_3),\quad\quad\vec{A}_2=
(\Phi_1-m,\Phi_4,\Phi_5)
\ee
which results in the scalar potential \eqn{bpot}. Note that 
the size of the ${\bf S}^3$ no plays the role of a mass parameter. 

The Coulomb branch of this theory is described by the two dual
photons and the real scalar, together with the two
chiral multiplets, completing a seven real dimensional
manifold. To determine the locus $L$, we must first decide 
which of the $U(1)$ factors to nominate as the M-theory circle. 
For simplicity, we choose the diagonal combination. Only the first 
hypermultiplet is charged under this, and becomes massless 
at $\phi_i=0$ for $i=1,2,3$. At one-loop level, the dual photon 
degenerates at these points, leaving a locus of 
fixed points parameterised by $\phi_4$, $\phi_5$ and the linear 
combination of dual photons, $\sigma_1-\sigma_2$. This 
latter scalar has non-vanishing period on ${L}$ as long as $m\neq 0$. 
This ensures that the locus $L$ does indeed have topology 
${\bf S}^1\times \mathbb{R}^2$ as required. 

Although Theory B correctly reproduces the topology of $L$, several 
puzzles remain. Firstly, it is unclear why a D-brane probe in 
flat space would have two $U(1)$ gauge fields on its world-volume, 
as suggested by the above analysis. Secondly, we have been unable 
to reproduce the famous three-legged topology for the moduli space 
of $X$ from the probe theory. This remains an open problem.

\section{$Sp(2)$ Holonomy}
\label{spsec}

Starting with this section, we turn to compactifications 
of M-theory on eight dimensional manifolds $X$ of special
holonomy. We shall start with a discussion of hyperk\"ahler 
manifolds with  $Sp(2)$ holonomy group since here we have
much more control than the $Spin(7)$ models to be discussed
in the following section.
D2-brane probes of hyperK\"ahler 8-folds have
${\cal N}=3$ supersymmetry on their worldvolume.

There are several features that distinguish M-theory on 
manifolds $X$ of dimension eight (or greater). For example, 
suppose that $X$ is a non-compact manifold
of dimension eight (at this point we make no specific
assumptions about holonomy of $X$). Suppose, further,
that $X$ develops an isolated conical singularity.
Motivated by the analysis in the previous sections,
one might be interested in a dual type IIA description 
that involves intersecting D6-branes in topologically
flat space-time.
If it exists, such a description can be obtained from M-theory
via reduction on a circle, such that:
\be
X/U(1) \cong \mathbb{R}^7
\ee
The fixed point set, $L$, of the circle action has codimension
four and can be understood as a D6-brane locus \cite{aw,GS}.
It turns out, that if $X$ develops an isolated 
conical singularity, $L$ is also conical.
What is peculiar about compactification on eight-manifolds
is that $L$ can not be represented by intersection of two
coassociative planes in $\mathbb{R}^7$;
generically, two 4-planes in a 7-dimensional
space intersect over a set of dimension one. 
In this sense, models with $X$ of dimension eight are hard.

Another problem with compactifications on 8-manifolds
is that we no longer have extra spatial dimension to
perform the M-theory flip. Therefore, there is no 
string theory method to derive a mirror three-dimensional 
gauge theory, and we must resort to the field theory 
techniques of Section \ref{mirrorsec}.

Keeping these peculiar subtleties in mind, let's begin
with the simplest example of a hyperk\"ahler manifold $X$
given by cotangent bundle of ${\bf \mathbb{C}P}^2$:
$$
X= T^* {\bf \mathbb{C}P}^2
$$
M-theory on this manifold has  very interesting dynamics.
For example, a membrane anomaly requires the introduction 
of non-zero $G$-flux in order to obey the shifted quantization
condition:
$$
\int_{{\bf \mathbb{C}P}^2} {G \over 2 \pi} \in \mathbb{Z} + {1 \over 2}
$$
This flux generates a Chern-Simons coupling in the effective
three-dimensional theory, thus breaking the parity symmetry \cite{GS,AGdO}.
Its value becomes an additional parameter, and for certain
values of this parameter the model has two vacua.
Otherwise, $\mathcal{N}=3$ effective field theory has
a single massive vacuum \cite{gvw}.
Below we extend these results to compactifications of
M-theory on cotangent bundle of a general two-dimensional
Fano surface\footnote{A Fano variety is a projective variety whose
anticanonical class is ample. In complex dimension two,
a Fano variety is also called a del Pezzo surface.}.

To obtain a three-dimensional field theory with a vacuum moduli
space, we can consider adding extra membranes to this configuration.
By analogy with the other models, we shall mainly think of
a type IIA dual, which is much easier to analyze.
In type IIA theory a membrane become a D2-brane.
It preserves the same amount of supersymmetry as
the original bulk theory,
so that the effective theory on the D2-brane is
an $\mathcal{N}=3$ abelian gauge theory in 2+1 dimensions. 
What is this theory? There is an obvious ${\cal N}=3$ candidate 
with $X = T^* {\bf \mathbb{C}P}^2$ as the Higgs branch \cite{gvw}:
\begin{center}
{\bf Theory:} $\mathcal{N}=4$ vector multiplet
with Chern-Simons coupling and 3 hypermultiplets
\end{center}
The sole role of the Chern-Simons coupling is to break
supersymmetry from $\mathcal{N}=4$ to $\mathcal{N}=3$.
Since all hypermultiplets have charge $+1$,
the Higgs branch of this theory is described by
the following constraints:
\be
\sum_{i=1}^3 \vert q_i \vert^2 - \vert \tilde q_i \vert^2 = m_3, 
\quad\quad \sum_{i=1}^3 q_i \tilde q_i = m_1+im_2
\label{thisisbol}\ee
From the perspective of the gauge theory, $\vec m$ is a FI parameter. 
The Higgs branch, as written in this form, has a manifest 
$SU(3)_F\times U(1)_R$ 
symmetry. The simplest way to see that this space is indeed 
$T^\star{\bf \mathbb{C} P}^2$ is to set, without loss of generality, 
$m_1+im_2=0$, so that the variables 
\be
y_i = {q_i \over \sqrt {\vert \tilde q_i \vert^2 + m}}
\ee
takes values in a 5-sphere, ${\bf S}^5$.
Its quotient by the $U(1)$ gauge action gives us ${\bf \mathbb{C}P}^2$. 
The fields $\tilde q_i$, constrained by the second equation in 
\eqn{thisisbol}, provide the fibre to give us 
$X = T^* {\bf \mathbb{C}P}^2$. 
The metric on the Higgs branch may be computed explicitly 
using the hyperkahler quotient construction and is of the 
toric hyperkahler form,
\be
ds^2 = H_{ab} d \vec \phi_a \cdot d \vec \phi_b 
+ H^{ab} (d \sigma_a + \vec{\omega}_{ac} d\vec{\phi}^c)(d \sigma_b + 
\vec{\omega}_{bd}\vec{\phi}^d)
\label{hktoric}\ee
where the matrix of harmonic functions is given by,
\be
H_{ab}= {1 \over \vert \vec \phi_1 \vert} 
\pmatrix{1 & 0 \cr 0 & 0}+ {1 \over \vert \vec \phi_2 \vert} 
\pmatrix{0 & 0 \cr 0 & 1}
+ {1 \over \vert \vec \phi_1 + \vec \phi_2 + \vec m \vert}
\pmatrix{1 & 1 \cr 1 & 1}
\ee
These coordinates make manifest only $U(1)^2_F\subset SU(3)_F$, 
which acts by shifts of $\sigma_a$. 

While this theory certainly gives the correct moduli space,
it needs a little modification to describe correctly a theory
on the D2-brane probe, where the dual photon is identified
with the M-theory circle.
In fact, we expect the theory on the D2-brane probe to realise 
$X$ as a ``mixed'' branch: partially Higgs, partially Coulomb. 
To make this issue more apparant, let us deform $X$ 
by squashing the $T^2$ fibers at infinity. This is acheived by 
gauging the $U(1)_J$ symmetries \cite{ks} as described
in Section \ref{mirrorsec}.   
The net result is simply to add to a constant term to the harmonic 
function,
\be
H_{ab}= \pmatrix{e_1^{-2} & 0 \cr 0 & e_2^{-2}}
+{1 \over \vert \vec \phi_1 \vert} 
\pmatrix{1 & 0 \cr 0 & 0}+ {1 \over \vert \vec \phi_2 \vert} 
\pmatrix{0 & 0 \cr 0 & 1}
+ {1 \over \vert \vec \phi_1 + \vec \phi_2 + \vec m \vert}
\pmatrix{1 & 1 \cr 1 & 1}
\label{tcp}\ee
If we now blow-up the ${\bf \mathbb{C} P}^2$ zero section to 
become very large, $\vec m\rightarrow \infty$, 
the last term in $H$ decouples and we are 
left with the product of two Taub-NUT spaces. The 
theory described above indeed produces this target 
space in the limit of large FI parameter. However,  
both copies of this space are in terms of Higgs branch 
variables. From the discussion above, one of the 
${\bf S}^1$ fibers is the M-theory circle, we know that one 
of these copies must arise in a Coulomb branch description. 
We therefore expect the following two decoupled theories in 
the $\vec{m}\rightarrow \infty$ limit, each of which 
realizes the Taub-NUT manifold as a different branch of 
vacuum moduli space
\ba
{\mbox{\bf Coulomb Branch}}:&& {\cal N}=4\ U(1)_1\ \mbox{with 1 
hypermultiplet} \nn\\ 
{\mbox{\bf Higgs branch}}:&& {\cal N}=4\ U(1)_2\times \hat{U(1)}\ 
\mbox{with 1 hypermultiplet}
\nn\ea
In the Higgs branch theory, the hypermultiplet is charged 
only under $\hat{U(1)}$, which couples to $U(1)_2$ via 
a CS term. 
To rediscover $T^\star{\bf \mathbb{C} P}^2$, it is natural to allow the 
above two theories to interact in a manner which decouples 
in the limit $\vec{m}\rightarrow \infty$. There is a very 
natural candidate deformation: we couple a further hypermultiplet 
to $U(1)_1$ and $U(1)_2$, with bare mass $\vec{m}$. Moreover, 
to break supersymmetry to ${\cal N}=3$, we add a CS coupling for 
$\hat{U(1)}$ 
\begin{center}
{\bf Theory}: $U(1)_1\times U(1)_2\times \hat{U(1)}$ with 3 hypermultiplets 
\end{center}
The three hypermultiplets have charges $(1,0,0)$, $(0,0,1)$ and 
$(1,1,0)$ respectively under the gauge factors, and the third 
hypermultiplet is assigned a bare mass $\vec{m}$. Moreover, we include 
a Chern-Simons couplings
\be
\hat{A}\wedge(F_2+\hat{F})
\ee
The latter term breaks supersymmetry to ${\cal N}=3$. Note 
that further self-CS couplings would lift the moduli space of 
interest and so are disallowed. The scalar potential of this 
theory is,
\ba
V&=&|\vec{\phi}_1|^2w_1^\dagger w_1 + |\vec{\hat{\phi}}|^2
w_2^\dagger w_2 
+|\vec{\phi}_1+\vec{\phi}_2+\vec{m}|^2w_3^\dagger w_3 + \nn\\ && 
e_1^2|w_1^\dagger \vec{\tau}w_1+w_3^\dagger \vec{\tau}w_3|^2 
+e_2^2|w_3^\dagger \vec{\tau}w_3+\vec{\hat{\phi}}|^2 
+\hat{e}^2|w_2^\dagger \vec{\tau}w_2-\vec{\phi}_2|^2 
\nn\ea
The vacuum moduli space $V=0$ is given by $\hat{\phi}=w_1=w_3=0$, 
while $\phi_1,\phi_2$ and $w_2$ satisfy only the constraint 
$w_2^\dagger \vec{\tau}w_2=\vec{\phi}_2$. This branch of 
vacua is indeed a mixed branch as we anticipated. Classically, 
the metric is of the form \eqn{hktoric} with,
\be
H^{(0)}_{ab}=\pmatrix{e_1^{-2} & 0 \cr 0 & e_2^{-2}}
+ {1 \over \vert \vec \phi_2 \vert} 
\pmatrix{0 & 0 \cr 0 & 1}
\ee
However, this receives quantum corrections upon integrating 
out the massive hypermultiplets $w_1$ and $w_3$. The 
final metric is given by the harmonic function \eqn{tcp}, 
which indeed reduces to the hyperK\"ahler metric on 
$T^* {\bf \mathbb{C} P}^2$ in the limit $e_1^2,e_2^2\rightarrow \infty$.

What of mirror symmetry in our model? There is a version 
of the ``M-theory'' flip in the present situation which 
involves exchanging the two dual photons $\sigma_1$ and $\sigma_2$. 
This $Z_2$ symmetry, far from obvious from the classical 
lagrangian, is expected to hold at the quantum level whenever 
$e_1^2=e_2^2$. Moreover, in the 
$e_a^2\rightarrow \infty$ limit, it combines with the two 
$U(1)_J$ symmetries to yield the full $SU(3)$ isometry 
of the target space.

\subsubsection*{Generalizations and IIB Models}

One may consider more general hyperK\"ahler manifolds of the form:
$$
X = T^* B
$$
where $B$ is a smooth, compact Fano surface, {\it i.e.}
$B ={\bf \mathbb{C}P}^1 \times {\bf \mathbb{C}P}^1$
or a del Pezzo surface $\mathcal{B}_n$.
We will be particularly interested in models which admit
a toric description of the form \eqn{hktoric}:
\be
H_{ab}=\pmatrix{e_1^{-2} & 0 \cr 0 & e_2^{-2}}+
\frac{1}{|\vec{\phi}_2|} \pmatrix{0 & 0 \cr 0 & 1}
+\sum_{i=1}^{n+2}
\frac{1}{|p_i\vec{\phi}_1+q_i\vec{\phi}_2+\vec{m}_i|}
\pmatrix{ p_i^2 & p_iq_i \cr p_iq_i & q_i^2}
\label{ghk}\ee
The topology of the hyperK\"ahler toric manifold $X$
is encoded in the toric data; that is, in the pairs of
integers $(p,q)$. The space $X$ is a ${\bf T}^2$ fibration 
over $\mathbb{R}^6$, parametrised by two 3-vectors: $\vec \phi_1$ 
and $\vec \phi_2$. A 1-cycle in $T^2$ degenerates at loci in 
which $det(H^{-1})=0$ or, alternatively, on the loci in which 
$H$ diverges. The class of the cycle is determined by the values of $(p,q)$.
%(in fact, the class is just
%$(p,q) \in H_1 (T^2, \mathbb{Z}) = \mathbb{Z} \oplus \mathbb{Z}$).
%So, if there are no five-branes, we have:
%\be
%X \cong \mathbb{R}^6 \times T^2
%\ee
%The next simple case is when there is one five-brane,
%say, of type $(p,q)$. Then, $T^2$ fiber degenerates
%over a 3-plane:
%\be
%p \vec \phi_1 + q \vec \phi_2 = \vec a
%\ee
%Namely, a 1-cycle in $T^2$ degenerates at these points,
%and the class of the cycle is determined by the values
%of $(p,q)$ (in fact, the class is just
%$(p,q) \in H_1 (T^2, \mathbb{Z}) = \mathbb{Z} \oplus \mathbb{Z}$).
%Specifically, we have:
%\be
%H_{ij} = \pmatrix{1 & 0 \cr 0 & 1}
%+ {1 \over \vert p \vec \phi_1 + q \vec \phi_2 - \vec a \vert}
%\pmatrix{p^2 & pq \cr pq & q^2}
%\ee
%and:
%\be
%{\rm det} (H_{ij}) = 1 + {p^2 + q^2
%\over \vert p \vec \phi_1 + q \vec \phi_2 - \vec a \vert}
%\ee
%Therefore, ${\rm det} (H_{ij}) \to \infty$
%on the five-brane locus, {\it i.e.} ${\rm det} (H^{-1}) \to 0$.
%It follows that the corresponding $X$ has topology:
%\be
%X \cong \mathbb{R}^7 \times {\bf S}^1
%\ee
This story is very similar to toric geometry
in complex dimension 2, except that the base is 6-dimensional,
rather than 2-dimensional. Note, however, that there
are only two integers $(p,q)$ which specify
3-planes in the base. In this sense, we lose no
information if we think of $\vec \phi_{1,2}$
as real numbers, rather than vectors.
Then we get a usual toric diagram for the 4-manifold $B$,
which is a ${\bf T}^2$ fibration over $\mathbb{R}^2$
parametrised by $\phi_1$ and $\phi_2$.

Recall that 
a del Pezzo surface $\mathcal{B}_n$ can be constructed by blowing up
$n$ points on ${\bf \mathbb{C}P}^2$, where $0 \le n \le 8$.
If we denote $\ell$ the class of a line in the original
complex projective space ${\bf \mathbb{C}P}^2$,
and $E_i$ ($i = 1, \ldots, n$) the exceptional divisors
of the blown up points, then the canonical class of
$\mathcal{B}_n$ is given
by $K_{\mathcal{B}_n} = -3 \ell + \sum_i E_i$.
For the first Chern class of $\mathcal{B}_n$ we have:
\be
c_1 = - K_{\mathcal{B}_n} = 3 \ell - \sum_i E_i
\label{fcdelp}
\ee
By definition, $c_1$ is ample, {\it i.e.} it has positive
intersection with every effective curve in $\mathcal{B}_n$.
It is also useful to know that $\mathcal{B}_n$ for $n \ge 1$ has
a description in terms of a fiber space over ${\bf \mathbb{C}P}^1$. 
The intersection numbers of the basis elements
$\{ \ell, E_1, \ldots, E_n \} \in H_2 (B)$ are:
\be
\ell \cdot \ell = 1, \quad
E_i \cdot E_j = - \delta_{ij}, \quad \ell \cdot E_i = 0
\label{interll}
\ee
Since non-zero Betti numbers of $B = \mathcal{B}_n$ look like:
\be
h^0 (B) = h^4 (B) =1, \quad h^2 (B) = n+1
\label{delph}
\ee
the models based on cotangent bundles of del Pezzo
surfaces are labeled by a number $n$, which essentially
determines the topology of $B$.
Since $X$ is contractible to $B=\mathcal{B}_n$, one has
$H^4(X;\mathbb{R}) =H^4(\mathcal{B}_n;\mathbb{R})$,
and the non-zero Betti numbers are given by (\ref{delph}):
\be
h^0 (X) = h^4 (X) =1, \quad h^2 (X) = n+1
\ee
As in the case of $T^* {\bf \mathbb{C}P}^2$ model \cite{gvw},
dynamics of M-theory on these hyperK\"ahler manifolds has many
interesting aspects. For example, some of the models are
inconsistent unless we introduce background $G$-fields.
Indeed, due to a membrane anomaly, the flux quantization condition
requires the period of $G$-field to be congruent to $c_2 (X)$
modulo integers. On the other hand, it is easy to show that
$c_2 (X)$ for manifolds $X$ of the form $T^* B$ is determined
by the topology of $B$:
$$
\int_B {c_2 (X) \over 2} \cong \int_B c_1^2 (B)~~ {\rm mod}~~ \mathbb{Z}
$$
For our examples, the right-hand side can be easily evaluated
using \eqn{fcdelp} and \eqn{interll}. The result is:
$$
\int_B {G \over 2 \pi} = k_0 + {9-n \over 2}, \quad k_0 \in \mathbb{Z}
$$
and, in particular, $G$-flux can not be zero for $B = \mathcal{B}_n$
with $n$ even. The shift in the $G$-flux quantization condition
is expected to be related to the shift in the Chern-Simons coefficient
in the effective three-dimensional theory \cite{GS}. One would
expect a violation of parity symmetry in such theories.

Models with different values of the $G$-flux can be connected by
domain walls, obtained from five-branes wrapped on the 4-cycle $B$.
Hence, they are classified by
$H_4 (X, \mathbb{Z}) \cong H^4_{cpct} (X, \mathbb{Z})$.
Since possible values of the $G$-flux are classified
by $H^4 (X, \mathbb{Z})$, the number of models which cannot
be connected by domain walls is given by the quotient of
these two groups,  
$H^4 (X, \mathbb{Z}) / H^4_{cpct} (X, \mathbb{Z})$.
By Poincar\'e duality, the cohomology with compact
support is generated by the class $[B]$. On the other hand,
$B \cdot B = \chi (B) = n+3$, so that $H^4 (X, \mathbb{Z})$
is generated by $[B]/(n+3)$. Hence,
$$
H^4 (X, \mathbb{Z}) / H^4_{cpct} (X, \mathbb{Z}) = \mathbb{Z}_{n+3}
$$
Summarizing, different $\mathcal{N}=3$ theories obtained from
compactification on $X = T^* B$ are labeled by the value of
``flux at infinity'' \cite{gvw}:
\be
\Phi_{\infty} = N + {1 \over 2} \int_X {G \wedge G \over (2 \pi)^2}
= N + {1 \over 2 (n+3)} \Big( k_0 + {9-n \over 2} \Big)^2
\label{infflux}
\ee
where $N$ is the number of space-filling membranes.
Given the value of $k_0$ in a mod $(n+3)$ coset,
a vacuum is then found by choosing a non-negative $N$ and
an integer $k_0$ in the given mod $(n+3)$ coset, such that
the anomaly condition \eqn{infflux} is satisfied.
For all del Pezzo surfaces $\mathcal{B}_n$, there is one
special case of a model having more than one vacuum (branch of vacua).
They appear for $\Phi_{\infty} = N + (n+3)/8$, with:
$$
k_0 = {(n-9) \pm (n+3) \over 2}
$$
These results generalise \cite{gvw} to a more general class
of hyperK\"ahler manifolds of the form $T^* B$, where $B$ is
a smooth, compact Fano surface.
Similarly, it is easy to generalize our $\mathcal{N}=3$
D2-brane probe theory for $T^* {\bf \mathbb{C}P}^2$ model
to reproduce a more general metric like \eqn{ghk} on
the moduli space:
\begin{center}
{\bf Theory}: $U(1)_1\times U(1)_2\times \hat{U(1)}$
with $(n+3)$ hypermultiplets 
\end{center}
The first hypermultiplet is charged only under $\hat{U(1)}$, while 
the remaining $(n+2)$ have charges $(p_i,q_i,0)$ under 
$U(1)_1\times U(1)_2\times \hat{U(1)}$. As before the theory 
has CS terms $\hat{A}\wedge(F_2+\hat{F})$, the latter 
of which breaks supersymmetry to the desired ${\cal N}=3$. 
It is simple to check that the (mixed) vacuum moduli space of this 
theory indeed reproduces the metric with harmonic function \eqn{ghk}.

For a related discussion of M-theory on these (and more general)
toric hyperK\"ahler manifolds see \cite{jerome}, where it 
was shown that these spaces are T-dual to intersecting 
five-brane configurations in IIB theory which preserve 
3/16 fraction of supersymmetry. The constant term 
in $H$ encodes the asymptotic IIB coupling constant, 
\be
\frac{1}{g_s}=\frac{e_1 e_2}{e_1^2}
\nn\ee
The second term in  
\eqn{ghk} describes TN space in IIA theory, and is 
T-dual to an NS5-brane in 12345 directions. The remaining 
terms are dual to $(p_i,q_i)5$-branes. We thus have the 
configuration of intersecting branes,
\ba
NS5&& 12345 \nn\\
(p,q)5 && 12[37]_\theta [48]_\theta [59]_\theta,\quad\quad 
\tan\theta=p/q
\nn\ea
%The matrix $H^{\infty}_{ij}$ encodes the asymptotic value
%of type IIB coupling constant via relation:
%\be
%\tau = - {H_{12} \over H_{11}} + i {\sqrt{\rm{det} H} \over H_11}
%\ee
%For the sake of simplicity, we can choose:
%\be
%H^{\infty}_{ij} = \pmatrix{1 & 0 \cr 0 & 1}
%\ee

%%%%%%%%%%%%%%%%%%%%%%%%%%%%%%%%%%%%%%%%%%%%%%%%%%%%%%%%%%%%%%%%%%%%%%

\section{$Spin(7)$ Holonomy}
\label{spinsec}

The most subtle, and also most interesting class of models
corresponds to compactification of M-theory on manifold $X$
of $Spin(7)$ holonomy. Such theories use the rich dynamical structure
of $\mathcal{N}=1$ three-dimensional gauge theories to the fullest.
In this section we shall examine some of only the simplest models, 
arising from mututally orthogonal intersecting D6-branes. 
D-brane probes of these models have a world-volume theory in 
the same class as those described in Section \ref{mirrorsec} and, 
in particular, do not have any Chern-Simons couplings. We shall 
describe the moduli space of these compactifications and show 
that there are the geometrical transitions between branches. 
However, as in the case of the orthogonal $G_2$ branes, 
each of these geometrical transitions is inherited 
from those of the Calabi-Yau manifold discussed in Section \ref{susec}. 
We shall further use our mirror symmetry result to conjecture 
an algebraic quotient construction of these $Spin(7)$ manifolds.

%\subsection{Orthogonal D6 Branes}
As we pointed out in the previous section, IIA D-brane picture
dual to M-theory on eight-dimensional manifolds is typically
more involved since the D6-brane locus $L$ often has a conical
singularity. There is, however, a simple analog of the 
orthogonal D6-brane models discussed in sections \ref{gsecd}. 
To see this, we calibrate the locus of D6-branes with the 
following coassociative 4-form,
\be
\Psi^{(4)}
= * \Psi^{(3)} 
= e^{4679} + e^{4589} + e^{5678} + e^{3478}
+ e^{3698} + e^{3579} + e^{3456}
\label{caf}\ee
which restricts to the volume form on any of the seven 4-planes,
corresponding to different terms in \eqn{caf}.
These four-planes have the 
property that any pair of them intersect over a two-plane. 
If we pick two D6-branes with corresponding world-volumes 
(also filling the ubiquitous $x^1-x^2$ direction), then 
we reproduce the conifold model of Section \ref{susec}.

For three D6-branes, there are two possibilities: the four-planes 
may intersect either over a line, or alternatively over a point. 
In the former case, we return to the $G_2$ model discussed in 
the Section \ref{gsecd}. In the latter case, where three four-planes 
intersect over a point, the configuration breaks $(1/2)^3=1/8$ of 
supersymmetry. The lift to M-theory therefore results in a 
Calabi-Yau four-fold. We will not discuss this possibility 
further here.

Therefore the simplest $Spin(7)$ case 
consists of four D6-branes with the following world-volume 
directions:
\ba
D6_1 && 123456 \nn\\
D6_2 && 123689 \nn\\
D6_3 && 124679 \nn\\
D6_4 && 123579
\nn\ea
Since all four D6-branes intersect only at a point
(the origin, in our notation)
in  $\mathbb{R}^7$, which is parameterized by 
$x^3,x^4,x^5,x^6,x^7,x^8,x^9$, the lift of this configuration to 
M-theory gives
a $Spin(7)$ manifold $X$ with isolated conical singularity.
Moreover, since the D6-brane locus $L$ is a collection of four
coassociative 4-planes in the flat $\mathbb{R}^7$, 
it is natural to expect that $L$ can be continuously deformed
into a smooth coassociative 4-manifold $L \subset \mathbb{R}^7$. 
In fact, it is simple to see that we may once again make use 
of the two deformations of the conifold to smoothen the 
singularity. Firstly let us consider translating the D6-branes. 
There are five such parameters, 
\ba
D6_1: && x^7=m_1,\quad x^8=0,\quad x^9=0 \nn\\ 
D6_2: && x^4=0,\quad x^5=0,\quad x^7=0 \nn\\ 
D6_3: && x^3=0,\quad x^5=m_2,\quad x^8=m_3 \nn\\ 
D6_4: && x^4=n_1,\quad x^6=0,\quad x^8=n_2 
\nn\ea
If $m_i\neq 0$ for $i=1,2,3$ and $n_i\neq 0$ for $i=1,2$, then 
the D6-branes do  not intersect and locus $L$ has topology, 
\be
L=\mathbb{R}^4\cup\mathbb{R}^4\cup\mathbb{R}^4\cup\mathbb{R}^4 
\ee 
So, from \eqn{genhom}, we see that the lift to M-theory 
results in a smooth manifold $X$ of $Spin(7)$ holonomy
with three 2-cycles:
$$
h_2(X)=3,
$$
and all other Betti numbers vanishing.

Exactly as in 
the $G_2$ case, we shall see that there are geometric transitions,
which involve replacing these 2-cycles with with ${\bf S}^3$'s, 
resulting once again in three topologically distinct 
manifolds. However, unlike the $G_2$ case we shall see that 
we can only blow up a single ${\bf S}^3$ if the manifold 
is to remain non-singular. 
The analysis is identical to the $G_2$ case, the 
only difference being the two pairs of complex structures. If 
we single out the D6$_1$ and D6$_2$ pair, as well as the 
D6$_3$ and D6$_4$ pair, it is useful to define the complex 
combinations, 
\ba
\psi_1=x^8+ix^9,&&\quad\quad \psi_2=x^4+ix^5 \nn\\ 
\tilde{\psi}_1=x^3+ix^5,&&\quad\quad \tilde{\psi}_2=x^4+ix^6 
\nn\ea
We may set $m_1=0$, to allow us to deform the first pair 
on the curve $\psi_1\psi_2=\rho$. This curve fails 
to intersect with the D6$_3$ brane if
either\footnote{In the first case, if $\rho_1 \neq 0$,
the intersection occurs only asymptotically,
as in Section \ref{gsecd}.}
$m_2=m_3=0$ or, alternatively, if we move the 
D6$_3$ far enough,
\be
m_2^2m_3^2-m_2m_3\rho_2-\ft14\rho_1^2 > 0
\ee
The same analysis holds for D6$_4$: it fails to intersect the 
complex curve if either $n_1=n_2=0$ (and $\rho_2\neq 0$) or 
\be
n_1^2n_2^2-n_1n_2\rho_1-\ft14\rho_2^2 > 0
\ee
If we also want no intersection between 
D6$_3$ and D6$_4$, this requires us to turn on 
at least, say, $n_2$. We must therefore move at 
least one of the branes away from the origin. 
Of course, one may pick any pair of D6-branes and 
perform a similar deformations, leading to six branches 
in each of which the locus $L$ has topology,
\be
L=S^1\times\mathbb{R}^3\cup\mathbb{R}^4\cup\mathbb{R}^4 
\ee
which ensures that the lift to M-theory results in
a non-compact manifold $X$ with non-trivial Betti numbers:
$$
h_2(X)=2, \quad h_3(X)=1
$$
Finally, as in the $G_2$ case, we may attempt to resolve the 
two pairs simultaneously:
\be
\psi_1\psi_2=\rho,\quad\quad \tilde{\psi}_1\tilde{\psi}_2=\tilde{\rho}
\ee
However, in this case there are no choice of the parameters for 
which these two complex curves fail to intersect: this therefore 
always results only in a partial resolution of the $Spin(7)$ singularity.

We catalog different topological phases we found in Table 4.
However, there might be other phases of our $Spin(7)$ manifold $X$,
and it would be interesting to complete this quest.

\begin{table}\begin{center}
\begin{tabular}{|c|c|c|}
\hline
Betti Numbers & Number of Phases & Number of Deformations \\
\hline
\hline
$h_2=3$, $h_3=0$ & 1 & 5 \\
\cline{1-3}
$h_2=2$, $h_3=1$ & 6 & 4 \\
\cline{1-3}
$h_2=2$, $h_3=1$ & 6 & 6 \\
\cline{1-3}
$\ldots$ & $\ldots$ & $\ldots$ \\
\hline
\end{tabular}\end{center}
\caption{Some topological phases of a manifold $X$ with $Spin(7)$ holonomy.}
\end{table}

\subsubsection*{Probe Theory and the Mirror Theory}

As in the previous sections, it is straightforward to describe 
the theory on a D2-brane probe of this D6-brane, oriented in 
the $x^1-x^2$ plane. We have,
\begin{center}
{\bf Theory A:} $U(1)$ with 7 scalars and 4 hypermultiplets
\end{center}
The charges of hypermultiplets are all equal to $+1$.
If we denote the scalar fields as $\phi_i = x^{i+2}$, 
the coupling to the hypermultiplets is described by the 
real superpotential,
\be
f=\sum_{i=1}^4 \left(\vec{A}_i-\vec{M}_i\right)
\cdot W_i^\dagger \vec{\tau} W_i
\ee
where the triplets $\vec{A}_i$ and $\vec{M}_i$ are suitable 
combinations of the $\Phi_i$ and deformation parameters 
respectively,
\be
\vec{A}_1=(\Phi_9,\Phi_8,\Phi_7)\ ,\ \vec{A}_2=(\Phi_5,\Phi_4,\Phi_7)
\ ,\ \vec{A}_3=(\Phi_5,\Phi_8,\Phi_3)\ ,\ 
\vec{A}_4=(\Phi_6,\Phi_8,\Phi_4)
\label{itstheb}\ee
and
\be
\vec{M}_1=(0,0,m_1)\ ,\ 
\vec{M}_2=(0,0,0)\ ,\ 
\vec{M}_3=(m_2,m_3,0)\ ,\ 
\vec{M}_4=(0,n_2,n_1)
\ee
As in previous sections, we have included only the translational 
deformations corresponding to blowing up ${\bf S}^2$'s. Further 
deformations, corresponding to blowing up ${\bf S}^3$'s are 
given by the usual terms \eqn{defcon1}. At the singular point, 
$m_i=n_i=0$, the theory has 
three global $U(1)_C$ flavor symmetries coming from gauge symmetries 
on the D6-branes (not including the diagonal one). 
As usual, there is also a $U(1)_J$ symmetry corresponding to shift
of the dual photon. Therefore, upon lift to M-theory we expect a 
$Spin(7)$ conical
singularity with the global symmetry group 
$K_Y = U(1)_J \times U(1)^3_C$. 
Since the flavor symmetries are unbroken by the mass parameters 
$m_i,n_i$, this is expected to survive a deformation to a smooth manifold
$X$, with all singular points replaced by ${\bf S}^2$'s. The same 
is not true on the other branches where we blow up some ${\bf S}^3$'s. 

Recall that in the previous sections, we were able to derive 
an algebraic expression for the manifold $X$ by dualising 
to a configuration of NS5-branes in IIB. In the present 
case, this is only possible if we T-dualise along $x^1$ or 
$x^2$ directions. In this case, the D2-brane turns into 
a D-string as opposed to a D3-brane, and we are unable to 
read off the mirror theory. Nevertheless, it is still possible 
to derive the mirror theory using the field theory techniques 
of Section \ref{qftsec}. As we mentioned above, these reproduce 
the same Higgs branch as the string picture for the $G_2$ case. 
Here we must rely on these techniques for want of another method. 
We find,
\begin{center}
{\bf Theory B:} $U(1)^3$ with 5 scalars and 4 hypermultiplets
\end{center}
The charges of the hypermultiplets are $(+1,-1,0,0)$, $(0,+1,-1,0)$,
and $(0,0,+1,-1)$. The theory consists of only Yukawa couplings, 
and no quartic (or higher) superpotentials. These Yukawa 
couplings are given by,
\ba
f_{\rm Yuk}&=&\Phi_1(W_1^\dagger\tau^3W_1-W_2^\dagger
\tau^3W_2-m_1)+ \Phi_2(W_2^\dagger\tau^1W_2-W_3^\dagger\tau^1W_3-m_2) 
\nn \\ && +\Phi_3(W_3^\dagger \tau^2W_3-W_1^\dagger\tau^2W_1-m_3)
+\Phi_4(W_2^\dagger \tau^2W_2-W_4^\dagger\tau^3W_4-n_1) \nn\\
 && +\Phi_5(W_1^\dagger \tau^2W_1-W_4^\dagger\tau^2W_4-n_2)
\ea
Note that the first three terms coincide with the $G_2$ 
Yukawa couplings \eqn{corblimey}. The last two terms are 
unique to this $Spin(7)$ example. The Higgs branch of Theory B 
is parameterized by the 16 real variables in the hypermultiplets, 
subject to 3 $U(1)$ gauge orbits and the 5 constraints,
\ba
|q_1|^2 - |\tilde{q}_1|^2 - |q_2|^2 + | \tilde{q}_2|^2 &=& m_1 \nn\\
\re(\tilde{q}_2q_2-\tilde{q}_3q_3) &=& m_2 \nn\\ 
\im(\tilde{q}_3q_3-\tilde{q}_1q_1) &=& m_3 \nn\\ 
\im(\tilde{q}_1q_1-\tilde{q}_4q_4) &=& n_2 \nn\\
\im(\tilde{q}_2q_2)-|q_4|^2+|\tilde{q}_4|^2&=&n_1 
\nn\ea
Note that, unlike the $G_2$ cases, the final equation 
relates different components of the 3-vectors  
$W^\dagger \vec{\tau}W$. 

\subsubsection*{Further D6-Branes}

As is clear from the coassociative form \eqn{caf}, there 
exist further loci $L$ consisting of $N=5,6$ or 7 
D6-branes, each of which lifts to M-theory on a 
$Spin(7)$ manifold. Here we describe briefly only the case 
of $N=7$, with other cases (including the $N=4$ model 
described above) arising as limiting cases. The D6-branes 
have world-volumes, 
\ba
D6_1 && 123456 \nn\\
D6_2 && 123689 \nn\\
D6_3 && 124679 \nn\\
D6_4 && 123579 \nn\\
D6_5 && 125678 \nn\\
D6_6 && 124589 \nn\\
D6_7 && 123478 
\nn\ea
It is a simple matter to determine the possible resolutions 
of this singularity by implementing the resolution and 
deformation of the conifold pairwise. For example, there 
are 14 parameters arising from moving the D6-branes 
away from the origin. If all D6-branes remain flat, but 
non-intersecting, then the locus $L$ lifts to a $Spin(7)$ 
manifold $X$ with homology $h_2(X)=6$, and no further 
cycles. As in the previous case of four D6-branes, we 
may blow down any one of these ${\bf S}^2$'s and replace it 
with a ${\bf S}^3$, by picking any pair of D6-branes and deforming  
them on a complex curve. If the other D6-branes are moved far 
enough from the origin, the M-theory lift is once again smooth. 
After passing through this geometrical 
transition, $X$ has Betti numbers:
$$
h_2(X)=5, \quad h_3(X)=1
$$
Finally, we saw in the case of 
four D6-branes that it was not possible to deform two pairs 
of D6-branes simultaneously without the complex curves intersecting. 
The same is true here.

It is straightforward to write down the theory on a probe 
D2-brane
\begin{center}
{\bf Theory A:} $U(1)$ with 7 scalars and 7 hypermultiplets
\end{center}
with the Yukawa couplings determined, as usual, by the orientation 
of the D6-branes. Once again, we are forced to use field theoretic 
techniques to determine the mirror theory, 
\begin{center}
{\bf Theory B:} $U(1)^6$ with 14 scalars and 7 hypermultiplets
\end{center}
where the hypermultiplets have the usual alternating $+1,-1$ 
charges under the gauge group, and the superpotential 
consists only of Yukawa couplings, which result in 14 
real constraints on the 14 complex degrees of freedom in
the hypermultiplets. These constraints may be written 
in compact form,
\be
\re(\tilde{q}_iq_i)=\im(\tilde{q}_{i+4}q_{i+4})
=|q_{i+6}|^2-|\tilde{q}_{i+6}|^2 \ \ \ \ i=1,\ldots,7
\ee
%\ba
%\re(\tilde{q}_1q_1)&=&\im(\tilde{q}_5q_5)=|q_7|^2-|\tilde{q}_7|^2 \nn\\ 
%\re(\tilde{q}_2q_2)&=&\im(\tilde{q}_6q_6)=|q_1|^2-|\tilde{q}_1|^2 \nn\\
%\re(\tilde{q}_3q_3)&=&\im(\tilde{q}_7q_7)=|q_2|^2-|\tilde{q}_2|^2 \nn\\
%\re(\tilde{q}_4q_4)&=&\im(\tilde{q}_1q_1)=|q_3|^2-|\tilde{q}_3|^2    \\
%\re(\tilde{q}_5q_5)&=&\im(\tilde{q}_2q_2)=|q_4|^2-|\tilde{q}_4|^2 \nn\\
%\re(\tilde{q}_6q_6)&=&\im(\tilde{q}_3q_3)=|q_5|^2-|\tilde{q}_5|^2 \nn\\
%\re(\tilde{q}_7q_7)&=&\im(\tilde{q}_4q_4)=|q_6|^2-|\tilde{q}_6|^2 
%\nn\ea
where $i$ is defined modulo $7$; i.e. $q_{i+7}\equiv q_i$. The 
14 translational parameters equate each of these terms 
up to a constant.

As in previous cases, we may easily generalize this configuration 
by having $N_i$ of each branes, so that when the resulting $Spin(7)$ 
manifold $X$ develops a conical singularity, there is an 
enhanced $\prod_iU(N_i)$ gauge symmetry.

\section*{Acknowledgements}

We are grateful to
B.~Acharya, M.~Aganagic, N.~Constable, A.~Hanany, 
J.~Sparks, N.~Seiberg, M.~Strassler, C.~Vafa and E.~Witten
for useful discussions.
This research was conducted during the period S.G.
served as a Clay Mathematics Institute Long-Term Prize Fellow.
The work of S.G. is also supported in part by grant RFBR No. 01-02-17488,
and the Russian President's grant No. 00-15-99296.
D.T. is a Pappalardo fellow and would like to thank the 
Pappalardo family for their largesse. The work of D.T. also supported 
in part by funds provided by the U.S. Department of Energy 
(D.O.E.) under cooperative research agreement \#DF-FC02-94ER40818.

\appendix{${\cal N}=1$ Chern-Simons Mirrors}

In Section \ref{mirrorsec} of this paper we derived a large class of abelian 
mirror pairs with ${\cal N}=1$ supersymmetry. However, in 
each case all charged matter fields lie within 
hypermultiplets, ensuring invariance under charge conjugation. 
From the field theory perspective, this restriction arose 
because we derived the mirror pairs from deforming ${\cal N}=4$
flavor symmetries which act symmetrically upon the
two chiral multiplets $Q$ and $\tilde{Q}$. 

In this appendix, we show how to relax this constraint 
and obtain ${\cal N}=1$ three-dimensional mirror pairs with 
only chiral (as opposed to hyper-) multiplets. 
However, we have not yet found a place 
for these theories in our probe set-ups, and they play 
no role in the bulk of this paper. We suspect they may 
be important in more complicated $Spin(7)$ compactifications, 
and  include them here only for completeness.

The prescription we use was given in  \cite{tong}, where 
the ${\cal N}=4$ mirror pairs were
deformed by gauging the R-symmetry currents. Specifically,
an abelian R-symmetry contained within the diagonal
group of $SU(2)_R\times SU(2)_N$ may be gauged
preserving ${\cal N}=2$ supersymmetry. This gives a mass
splitting not only to the ${\cal N}=4$ vector multiplets,
as in the previous subsection, but also to the ${\cal N}=4$ 
hypermultiplets. As a result, certain charged chiral multiplets 
become heavy and decouple, leaving behind a reminder of their
presence in the form of CS couplings
$\kappa^{ab}A^a\wedge F^b$. The resulting ${\cal N}=2$ mirror pairs
had been previously discovered in \cite{dt},
\ba
{\bf Theory\ A:} && U(1)^r\ \mbox{with $N$ chiral multiplets} \nn\\
{\bf Theory\ B:} && \hat{U(1)}^{N-r}\ \mbox{with $N$ chiral multiplets}
\nn\ea
As in the Section \ref{mirrorsec}, the chiral multiplets of the two 
theories carry charges $R$ and
$\hat{R}$ respectively, satisfying
\be
\sum_{i=1}^NR_i^a\hat{R}_i^p=0
\ee
The new element in these theories is the presence of CS couplings, 
$\kappa$ and $\hat{\kappa}$ for the Theories A and B respectively,
\ba
\kappa^{ab}=\ft12\sum_{i=1}^NR_i^aR_i^b\ \ \ \ ;\ \ \ \ \
\hat{\kappa}^{ab}=\ft12\sum_{i=1}^N\hat{R}_i^a\hat{R}_i^b
\nn\ea
Including all possible mass and FI parameters, the scalar
potential for Theory A takes the form,
\ba
V_A=\sum_{a=1}^re_a^2\left(R_i^a|q_i|^2
-\ft12R_i^aR_i^b\phi^b-\zeta^a-\kappa_{ab}\phi^b\right)^2
+\sum_{i=1}^N(R_i^a\phi^a+m_i)^2|q_i|^2
\nn\ea
with a similar tale for Theory B, with parameters $\hat{\zeta}$
and $\hat{m}$. These are determined by the mirror map \cite{ahkt},
\ba
\zeta^a-\ft12 R_i^am_i = R^a_i\hat{m}_i\ \ \ \ ;\ \ \ \ \
\hat{\zeta}^p+\ft12 \hat{R}_i^p\hat{m}_i=-\hat{R}^p_i{m}_i
\label{newmirror}\ea
The difference between this and the ${\cal N}=4$ mirror
map \eqn{massfimap} may be traced to a finite renormalization
of the FI parameters. As in previous
cases, the two theories as stated are mirror only in the
strong coupling limit $e_a^2\rightarrow \infty$. However,
there exists a deformation to ``Theory \Bprime'' which
is conjectured to be valid at all energy scales \cite{tong},
thus justifying further attempts to deform these theories. 
In fact, these deformations proceed as in Section \ref{mirrorsec}, 
resulting in ${\cal N}=1$ Maxwell-Chern-Simons mirror pairs,
\ba
{\bf Theory\ A:} && U(1)^r\ \mbox{with $k$ scalars and $N$ chiral
multiplets} \nn\\
{\bf Theory\ B:} && \hat{U(1)}^{N-r}\ \mbox{with $N-k$ scalars
and $N$ chiral multiplets}
\nn\ea
The charges and CS couplings are as above. We
further have a real superpotential leading to a Yukawa
coupling and real masses for Theory A,
\ba
f= S_i^\alpha \Phi_\alpha Q_i^\dagger Q_i + M^{\alpha\beta}
\Phi_\alpha\Phi_\beta
\nn\ea
with coupling constants given by the maximal rank matrix
$S$. The Yukawa couplings and real masses for Theory
B are denoted by $\hat{S}$ and $\hat{M}$ respectively. The 
former satisfy equation \eqn{ss} from Section 2,
\be
\sum_{i=1}^NS_i^\alpha S_i^\rho = 0
\ee
while the real masses for both Theory A and B are determined in 
terms of the Yukawa couplings, 
\ba
M^{\alpha\beta} = \ft12 S_i^\alpha S_i^\beta\ \ \ \ \ ;\ \ \ \ \
\hat{M}^{\rho\lambda} = \ft12\hat{S}_i^\rho \hat{S}_i^\lambda
\nn\ea
Since the both FI parameters and mass parameters are associated
with scalar, rather than vector, multiplets in ${\cal N}=1$
theories, the renormalization of $\zeta$ depends on 
$S$ rather than $R$, and the mirror map \eqn{newmirror} becomes,
\ba
\zeta^\alpha-\ft12 S_i^\alpha m_i = R^\alpha_i\hat{m}_i\ \ \ \ ;\ \ \ \ \
\hat{\zeta}^\rho+\ft12 \hat{R}_i^\rho\hat{m}_i=-\hat{R}^\rho_i{m}_i
\nn\ea
Let us focus on the Coulomb branch of Theory A, parameterized by
$k$ real scalars and $r$ dual photons. It naturally described by
a $k$ real dimensional base space $Q$, fibered with the torus
${\bf T}^r$. It exists only if
the CS couplings $\kappa$, the real masses $M$
and the FI parameters $\zeta$ all vanish. However, unlike
the situation with hypermultiplets, each of these
quantities receives a correction at one-loop, and we require
this quantum corrected parameter to vanish. The shifts are,
\ba
\kappa^{ab} &\rightarrow& \kappa^{ab}-\ft12\sum_i R_i^aR_i^b
\,\sign\,M_i \nn\\
M^{\alpha\beta} &\rightarrow& M^{\alpha\beta} -\ft12
\sum_i S_i^\alpha S_i^\beta\,\sign\, M_i \nn\\
\zeta^\alpha &\rightarrow& \zeta^a -\ft12\sum_i S_i^\alpha m_i
\,\sign\, M_i
\nn\ea
where 
\be
M_i=S_i^\alpha\phi_\alpha+m_i
\ee 
is the effective mass
of the $i^{\rm th}$ chiral multiplet. We see that the Coulomb
branch exists for $\zeta = \ft12S_i^\alpha m_i$ and for
$\phi_\alpha$ restricted to lie within the range $M_i\geq 0$.
This describes the base $Q$.
At the boundaries of $Q$, given by $M_i=0$, a cycle of
${\bf T}^r$, corresponding to the linear combination
$R_i^a\sigma_a$, degenerates. In contrast, the Higgs branch
of Theory B is given by the vanishing of the D-term,
$\hat{S}_i^\rho|q_i|^2=\hat{\zeta}^\rho=\hat{S}^\rho_im_i$,
modulo the gauge action, $q_i\rightarrow \exp(\hat{R}_i^p c_p)q_i$.
Comparing to the Coulomb branch, we have the mirror map between
fields,
\ba
|q_i|^2=M_i\ \ \ \ \ ;\ \ \ \ \ 2{\rm arg}\ (q_i)=R_i^a\sigma_a
\nn\ea
As explained in previous sections, any attempt to derive mirror 
symmetry for theories with only ${\cal N}=1$ supersymmetry 
is necessarily conjectural, since there are few quantitative 
tests available. Nevertheless, here we present a simple example 
which illustrates how mirror symmetry works in this case.

\subsubsection*{\rm{\em An example: ${\bf S}^3$}}

One advantage of using chiral multiplets is that we may
engineer the compact Coulomb branches. As an example
we examine the case of ${\bf S}^3$ in detail. As a Higgs branch of 
Theory B, this is extremely easy to construct,
\begin{center}
{\bf Theory B:} 1 scalar and 2 chiral multiplets
\end{center}
Each chiral multiplet has Yukawa coupling $+1$ with the
real scalar. A FI parameter $2\zeta$ yields the D-term
constraint,
\ba
|q_1|^2+|q_2|^2=2\zeta
\nn\ea
which is the ${\bf S}^3$ of interest. The theory has
a global $SO(4)$ symmetry group, which protects the
metric from quantum corrections. It will prove useful 
to change coordinates to
$q_i=r_i\exp(i\alpha_i)$. Then defining
$\phi=r_1^2-\zeta=\zeta-r_2^2$, we see that this
coordinate is restricted to the interval
\ba
-\zeta\leq \phi\leq\zeta
\label{interval}\ea
We may now view ${\bf S}^3$
as a fibration of ${\bf T}^2$ over this interval, with the
round metric given by,
\ba
2ds^2=\frac{\zeta}{\zeta^2-\phi^2}d\phi^2
+2(\zeta+\phi)d\alpha_1^2+2(\zeta-\phi)d\alpha_2^2
\label{round}\ea
Our task is to reconstruct ${\bf S}^3$ as the Coulomb branch.
The mirror theory is,
\begin{center}
{\bf Theory A:} $U(1)^2$ with 1 scalar and 2 chiral multiplets.
\end{center}
where each chiral has charge $+1$ under only a single
$U(1)$ factor, and the Yukawa coupling is $S=(+1,-1)$. This
theory has vanishing FI coupling, but each chiral is
assigned a real mass $m_1=m_2=\zeta$. Moreover, there
is a real mass $M=-1$ for $\phi$ and a CS coupling
$\kappa^{ab}=-\ft12\delta^{ab}$. Following the
prescription above, we see that the Coulomb branch
exists for $M_i\geq 0$, which translates precisely to
the interval \eqn{interval}. We can begin to calculate 
the metric perturbatively. The relevant dimensionless 
expansion parameters are $e^2/|\zeta\pm m|$. At one-loop 
we find,
\ba
ds^2=\left(\frac{1}{e^2}+\frac{\ft12}{\zeta-\phi}
+\frac{\ft12}{\zeta+\phi}\right)d\phi^2
+\left(\frac{1}{e^2}+\frac{\ft12}{\zeta-\phi}\right)^{-1}
d\sigma_1^2+\left(\frac{1}{e^2}+\frac{\ft12}{\zeta+\phi}\right)^{-1}
d\alpha_2^2
\nn\ea
With ${\cal N}=1$ supersymmetry, one would expect this
metric to receive many more quantum corrections, especially in
the neighborhood $\phi\rightarrow \pm\zeta$, where we
have integrated out light matter fields. Nonetheless, it
is amusing to note that in the strong coupling limit
$e^2\rightarrow \infty$, this metric reproduces the
round metric on ${\bf S}^3$ \eqn{round}. In this limit
the manifest $U(1)_J^2$ symmetry of the Coulomb branch
is enhanced to $SO(4)$. It is tempting to conjecture that
this non-abelian infra-red symmetry protects the
Coulomb branch from further corrections.

\end{document}